\documentclass[%
    aip,
     amsmath,amssymb,
    reprint,%
    twocolumn
]{revtex4-1}

\usepackage{graphicx}
\usepackage{dcolumn}
\usepackage{bm}

\usepackage{dcolumn}

\usepackage[utf8]{inputenc}
\usepackage[T1]{fontenc}
\usepackage{mathptmx}
\usepackage{mathtools}
\usepackage{etoolbox}

\makeatletter
\def\@email#1#2{%
 \endgroup
 \patchcmd{\titleblock@produce}
  {\frontmatter@RRAPformat}
  {\frontmatter@RRAPformat{\produce@RRAP{*#1\href{mailto:#2}{#2}}}\frontmatter@RRAPformat}
  {}{}
}%
\makeatother

\usepackage[margin=2cm]{geometry}
\allowdisplaybreaks

\usepackage{subfig}

\usepackage{hyperref}
\usepackage[dvipsnames]{xcolor}
\usepackage{amsmath}
\usepackage{amsfonts,amssymb}
\usepackage{array}

\usepackage{tikz}
\usetikzlibrary{quantikz2}

\usepackage{soul}  
\usepackage{pgfplots}
\pgfplotsset{compat=1.18}

\usepackage{mathbbol}
\usepackage{xifthen}

\usepackage[titletoc,title]{appendix}

\usepackage{multirow}

\usepgfplotslibrary{fillbetween}

\usepackage{amsthm}

\theoremstyle{definition}

\colorlet{ygray}{gray!20}

\def\yk{k}

\newcommand{\oO}{\mathcal{O}}

\newcommand*\diff{\mathop{}\!\mathrm{d}}

\newcommand{\yA}[2]{A_{#1}^{[#2]}}
\newcommand{\yIN}[1]{I_{N}^{\otimes #1}}

\def\dim{{N_x}}
\def\yNnl{{N_{\rm nl}}}
\def\yNca{{N_{\rm ca}}}
\def\yNu{{N_u}}

\def\dim{n}
\def\yNnl{p}
\def\yNca{P}
\def\yNu{N}

\renewcommand{\yIN}[1]{I_{\dim}^{\otimes #1}}

\def\Binomial{C}

\def\yFc{\mathcal{F}_c}
\def\ySc{\mathcal{S}_c}
\def\ytol{\epsilon_{\rm tol}}
\def\yCE{\mathcal{E}}
\def\yAR{\mathcal{A}}
\def\yDE{\mathcal{D}}

\def\ContFunc{\mathcal{C}}
\def\Manifold{M}
\def\Reals{\mathbb{R}}

\newcommand\abs[1]{\left| #1 \right|}

\newcommand\FuncOne[2]{\mathrm{#1}\,#2}
\newcommand\LScat[1]{\FuncOne{cat}\,#1}
\newcommand\dimRm[1]{\FuncOne{dim}\,#1}
\newcommand\Fix[1]{\FuncOne{Fix}\,#1}

\def\xinit{x_0}

\def\dx{\Delta x}

\def\Binomial{C}
\def\Mpoly{\yNnl}
\def\Npoly{\yNca}
\def\dPoly{\dim}
\def\Velocity{F}

\begin{document}

\preprint{AIP/123-QED}

\title[]{Globalizing the Carleman linear embedding method for nonlinear dynamics}
\author{I. Novikau}
\affiliation{Lawrence Livermore National Laboratory, Livermore, California 94550, USA}
\author{I. Joseph}
\email{joseph5@llnl.gov}
\affiliation{Lawrence Livermore National Laboratory, Livermore, California 94550, USA}

\date{\today}

\begin{abstract}
The Carleman embedding method is a well-known
technique for linearizing a system of nonlinear differential equations, but fails to converge in regions where there are multiple fixed points.
We prove that the need for the solution to be expressible as a convergent Taylor series of the initial conditions leads to restrictions on the stability and convergence of the Carleman algorithm. Inspired by this realization,
we propose and test three different versions of a global piecewise
Carleman embedding technique, based on partitioning space into multiple regions where the center and size of the embedding region are chosen to control convergence.
The first method switches between local linearization regions of fixed size once the
trajectory reaches the boundary of the current linearization chart.
During the transition, the embedding is reconstructed within the newly created chart, centered at the transition point. 
The second method also adapts the chart size dynamically, enhancing accuracy in regions where multiple fixed points are located.
The third method partitions the state space using a static grid with precomputed linearization charts of fixed size, making it more suitable for applications that require high speed.
All techniques are numerically tested on multiple integrable and chaotic nonlinear dynamical systems demonstrating their applicability for problems that are completely intractable for the standard Carleman embedding method.
Simulations of chaotic dynamical systems such as various types of strange attractors demonstrate the power of the adaptive methods, if a sufficiently low tolerance is imposed. 
The success of the global Carleman methods imply fundamental limitations on the ability of quantum computers to accelerate the simulation of the trajectories of a nonlinear dynamical system. The need to perform nonlinear operations at the remapping stage when a trajectory crosses domain boundaries appears to be easiest to implement as a hybrid algorithm and implies that, for generic nonlinear chaotic dynamical systems, the gains in compression and work that are possible for integrable dynamics are lost at the remapping stage.
\end{abstract}
\maketitle

\section{Introduction}\label{sec:introduction}

\subsection{Motivation}

Carleman linearization~\cite{Carleman32} is a widely used technique for embedding nonlinear dynamics within an infinite dimensional linear system of differential equations \cite{Steeb83, Kowalski87}.
The embedding of nonlinear ordinary differential equations (ODE) into an infinite-dimensional linear system is performed by augmenting the simulated state with monomials of the coordinates with ever increasing polynomial degree.
Extensions of the standard Carleman technique use other libraries of basis functions such as Fourier series or orthogonal polynomials, like the Chebyshev polynomials.
If the Carleman method converges, then one can substitute a nonlinear system for a much larger linear system, which can lead to practical advantages for solution speed and accuracy.

However, a serious drawback is that the convergence of the Carleman method has severe limitations that limit its ability to handle truly nonlinear dynamics as well as vast domains of initial conditions.~\cite{Lin22Koopman}
For example, the Carleman method must normalize the equations so that the solution remains in the interior of the unit disk in the complex plane and will fail if the solution tries to cross the boundary of the unit disk.
More disconcerting is that the Carleman method will fail whenever the system has more than one fixed point inside the unit disk and the solution is too far from the linearization point at the center of the disk.
A simple proof of this fact is presented for autonomous 1D flows in Appendix~\ref{sec:1dflows}.
Since a linear system can only ever have a single fixed point, it should not be surprising that the Carleman linearizing transformation typically fails when the nonlinear system has multiple fixed points.
It also fails if the fixed point is of higher order, unless a transformation of the type given in Appendix~\ref{sec:higher-order-fixed-points} is used. 
As proven in Appendix~\ref{sec:higher-dflows}, a similar but more complex set of issues arise for nonlinear systems of ODEs in higher dimensions.
For example, for completely integrable system of ODEs, there are special coordinates in which the system reduces to integration by quadratures, i.e. a tensor product of autonomous 1D systems.
Hence, completely integrable system of ODEs suffer from a multi-dimensional generalization of the same issues experienced in 1D.
This clearly limits the domain of applicability of the Carleman method to \emph{weakly nonlinear dynamic}s and to initial conditions that are \emph{attracted to a region within the unit disk.}
In contrast, for many of the examples considered below, one of the key issues is that the radius of Carleman convergence is smaller than the radius of the trajectory, and, hence, more than one linearization region is necessary to cover the entire trajectory.
This obstruction appears for all of the strange attractors that we have studied.

The goal of this work is to \emph{globalize} the Carleman method so that it can be applied to more general classes of nonlinear dynamical systems and much wider domains for the initial conditions.
Our ultimate motivation is to enable the use of a globalized Carleman method for advanced applications like learning the Koopman mode decomposition (KMD),~\cite{Mezic05, Brunton22, Colbrook2024} for prediction and  control~\cite{Korda18, Korda20} and for emerging computational paradigms that are optimized for linear systems such as tensor networks and quantum algorithms.~\cite{Joseph23}

Newton's method is another example of a nonlinear solution method that uses the assumption of linearizability as an essential part of the solution strategy. 
Yet, Newton's method fails when the initial guess is too far from the fixed point because the linear approximation is no longer accurate. 
Newton's method can be rescued with a globalization method like line search, dog leg search, or trust region search that ensures that the proposed step is limited to a size that actually improves the residual.

The finite element and spectral element methods were developed in order to handle similar globalization issues. 
Generic series expansions only have a finite radius of convergence that depends on the location of the basepoint of the expansion.
Thus, in order to represent the solution in a manner that converges everywhere, it is necessary to cover the region of interest with multiple local domains (called elements) that have different series expansions within each local domain.

Here, we develop a globalized linearization method with a similar strategy of using multiple linearization regions to ensure that the trajectory is accurately represented over the entire time interval of interest.
Each valid linearization region is called a \emph{linearization chart}.
In many cases of practical interest, the entire solution manifold can be covered with a countable number of linearization charts, which form a \emph{linearization atlas} for the solution manifold.
Each linearizing transformation is assumed to be invertible within each chart, which allows one to uniquely transform between charts.
Unlike the case of a piecewise linear manifold, here, the transition functions between linearization charts are allowed to either be smooth nonlinear functions of class $\ContFunc^k$ or piecewise continuous functions of class $\ContFunc^0$. 
This allows one to define the linearization atlas as either being smooth of class $\ContFunc^k$ or piecewise continuous of class $\ContFunc^0$, respectively.

Depending on how much is known about the system of ODEs, the linearization charts can either be precomputed in advance or, if the vector field is assumed to be given by a black-box oracle, computed on the fly.
Clearly, for the Carleman method, a new chart must be used each time the solution leaves the Carleman unit disk.
Moreover, the radius of the unit disk may need to be rescaled as the next chart is accessed in order to optimize accuracy and ensure convergence.
Thus, a well defined algorithmic approach is needed to guarantee that a target error budget can be achieved.

This work explores these ideas and issues with a proposed series of algorithms tested on realistic examples of integrable and chaotic dynamical systems. We numerically demonstrate that the globalized Carleman method can be extended to a wide range of nonlinear dynamical systems that cannot be treated with the standard approach.  Examples where the globalized version succeeds but the standard approach fails include 1D systems with multiple fixed points, stable limit cycles in the 2D plane, and 3D systems with strange attractors, including the R\"ossler, Lorentz, and Chen's attractors. 
While this demonstrates the great power of the globalized Carleman method, our understanding of how to develop optimal Carleman atlases has only just begun.

To emphasize the piecewise use of linearization charts, and, in line with prior terminology in ~\onlinecite{Weber16, Forets17, Sanchez25}, we refer to these methods as \emph{piecewise Carleman embedding} (PCE).
We note that a rigorous convergence analysis was performed in ~\onlinecite{Weber16, Forets17, Sanchez25}, proving that re-expanding the solution around a new basepoint doubles the convergence rate.

Motivated by the desire to build a convergent quantum Carleman algorithm, this piecewise embedding technique was recently explored in Ref.~\onlinecite{Endo2026} under the name of \textit{pivot-switching} Carleman embedding.
An important contribution is that they demonstrated that this methodology works well for PDEs that have have an attracting fixed point.
Motivated by our work and that of \onlinecite{Endo2026}, the recent  Ref.~\onlinecite{Wang2026quantum} also provides a convergence analysis for a single step of the shifted Carleman approach, which they refer to as the \textit{pivot-shifted} Carleman embedding.

Our work goes beyond prior art in applying the method to truly chaotic dynamical systems, explaining the importance of rescaling in addition to shifting, proposing and demonstrating the convergence of adaptive algorithms, and proving the generalized Carleman stability and convergence results in Appendices \ref{sec:1dflows}-\ref{sec:higher-dflows}.
One of the key points that we are making is that the linearization of a flow on a manifold does not require a complete KMD.
Rather, \emph{the flow can be linearized using any appropriate set of basis functions if and only if this set of basis functions can capture the correct topological structure needed to represent the flow on the manifold.}
While a complete generalized KMD, including singular Dirac-delta-like eigendistributions, clearly provides a complete and simple representation for the Koopman evolution operator, it is not analytic and in fact becomes singular at the boundaries of independent dynamically invariant sets.
Thus, if the goal is to discover the KMD through numerical computation, then it is important to first linearize the flow with a numerically well-behaved basis set from which one can then determine the KMD.

\subsection{The Carleman and Koopman embedding methods}
The Carleman method was derived just after the work of Koopman \cite{Koopman31} and von Neumann \cite{Koopman32} on the linearization of Hamiltonian dynamics.
The general linearizing transformation for a set of ordinary differential equations (ODEs) can always be understood as a linear partial differential equation (PDE) for the conservation of the probability distribution function (PDF) of the set of solutions.~\cite{Joseph20, Joseph23}
This linear conservation law can always be expressed as an infinite dimensional linear matrix equation, by computing its components over a set of basis functions that is complete over phase space.
In fact, the Carleman method has been proven to simply be a complex analytic version of the Koopman-von Neumann (KvN) approach (see  \onlinecite{Joseph23} Appendix B). 

A few of the main limitations of the Carleman embedding technique are that that it is inherently local, has a limited radius of convergence, and only operates correctly in the vicinity of a single fixed point.
Importantly, it cannot globally represent a system with multiple distinct fixed points.
If the problem is initialized near two or more fixed points, or if the trajectory leaves the initial linearization region, the Carleman method fails to accurately simulate the system's evolution and may become unstable.

Several techniques were recently proposed to provide global embedding capturing dynamics near multiple equilibrium points.
One option is to use a basis other than monomials for linearization.
For instance, one can use Fourier basis functions to enable the embedding to capture periodic or oscillatory dynamics.~\cite{Motee25}
In the so-called super-linearization approach,~\cite{Arathoon23, Belabbas23} additional variables beyond simple monomials are introduced using, in general, non-polynomial coordinate transformations.
The disadvantage of this approach is that it is still not guaranteed to converge, and, in fact, the linear system becomes more poorly conditioned as the library of functions increases in expressivity. 
Moreover, the conditions under which a general dynamical system can be super-linearized remain an open question.
Apart from this, machine learning techniques also can be used for constructing global linearization of nonlinear systems,~\cite{Breunung24} again, with some improvement in performance but without guarantees of convergence. 

The key idea behind the global Carleman embedding technique, which is more suitable for truly nonlinear applications, 
is piecewise linearization, where the method switches to a new local linearization chart once the convergence radius is exceeded.
While this idea was mentioned as the motivation for~\onlinecite{Forets17}, the authors did not describe an algorithm for switching between charts, and, hence, an explicit switching procedure was not proposed, implemented, or tested. 

The authors of~\onlinecite{Weber16, Sanchez25} demonstrated that the precision of Carleman embedding can be improved by linearizing the nonlinear system around its initial conditions.
They proved that, in general, this doubles the order of the polynomial for which the Carleman method can exactly represent the solution.
The recent work of \onlinecite{Wang2026quantum}, which considers a single shift, appears to agree with those analyses.
If one partitions the space into separate linearization regions, which we call charts, and applies a local Carleman embedding within each chart, the overall precision of the method can be improved.
However, such piecewise embeddings require careful detection of the moment when it is necessary to switch between charts along the evolution of the trajectory. 

In the following, we prove that rescaling of the solution vector can be essential for achieving convergence of the Carleman algorithm, a fact that was not fully appreciated by other authors.
Assuming that there is a smallest radius that is required and assuming that one has a good estimate for this radius, then one can consider using the piecewise strategy alone, which leads to a much simpler algorithm. 
However, we show that rescaling the radius has the potential to be more efficient in practice.
Finally, the Carleman stability and convergence criteria that we derive in Appendices~\ref{sec:1dflows} and \ref{sec:higher-dflows} are more stringent than previously been recognized. 
Understanding how our new criteria and the various convergence criteria that appear in the literature are related is an important subject for future work.

\subsection{Connection to Koopman theory}

Carleman linearization can be regarded as a special case of Koopman theory\cite{Koopman31, Brunton22}.
In the latter, one seeks a linear operator, namely, the Koopman operator, that acts on a set of observables such that the underlying nonlinear dynamics become linear on the span of the chosen observables.
The Carleman technique is simply 
an explicit construction of the Koopman operator using monomials of the state as observables\cite{Mauroy2020koopman, Dongwei24}.
Koopman theory is more general than Carleman embedding and can potentially extend the regions where the linear representation accurately emulates the nonlinear dynamics. 
However, for an arbitrary nonlinear system, Koopman theory  
does not  
guarantee that a smooth global transformation exists that converges everywhere because any dynamically invariant linearizing transformation must become singular near fixed points and strange attractors \cite{Kvalheim24flows}.
For example, cases with multiple isolated equilibria are particularly challenging for many 
embedding methods\cite{Liu23Koopman, Liu25Koopman}.
However, even in such cases, counterexamples exist where a smooth global embedding can be constructed\cite{Arathoon23}.

In the KMD,~\cite{Mezic05} each observable is represented as an expansion in a basis of the Koopman operator's eigenfunctions, where the coefficients in the expansion are called Koopman modes.
Numerical algorithms such as Dynamic Mode Decomposition (DMD)\cite{Schmid10} approximate KMD in real-world applications using experimental or numerically computed data.
The finite-dimensional linear model produced by DMD can be interpreted as a projection of the true Koopman operator onto the span of the observed snapshots.
While Carleman requires explicit nonlinear equations to construct the linear embedding, the DMD technique relies on measured data, such as time-series data, to approximate observed nonlinear dynamics with a finite linear model.
Ultimately, both Carleman linearization and DMD/KMD methodologies yield linear representations that serve as approximations for the simulation and analysis of inherently nonlinear dynamical systems.

Koopman theory and DMD techniques are widely employed in the analysis of complex nonlinear problems,~\cite{Budisic12, Brunton16, Brunton17} especially for fluid dynamics.~\cite{Brunton20, Brunton22}
In fact, the well known Sparse Identification of Nonlinear Dynamics (SINDy) technique \cite{Brunton16} is a data-driven method for discovering a sparse library of functions that are suitable for describing the dynamics of interest.

For example, in the field of plasma physics, these techniques have been applied to the analysis of both numerical 
simulations and experiments.~\cite{Taylor18, Sasaki19, Nayak21, Kusaba22, DiGrazia24}
In Ref.~\onlinecite{Kaptanoglu20}, the authors utilized DMD to characterize the spatial and temporal behavior of plasma instabilities in a spheromak plasma device.
In Refs.~\onlinecite{Faraji24Part1, Faraji24Part2}, the authors investigated DMD-based dimensionality reduction for high-dimensional plasma simulations.
In Refs.~\onlinecite{Pascuale22, Indranil24}, DMD was employed to accelerate plasma simulations.
This technique was further applied in Ref.~\onlinecite{LoVerso25} to construct a reduced numerical magnetohydrodynamics (MHD) model.
In Ref.~\onlinecite{Dudkovskaia25}, DMD was used to extract dominant drift-wave modes from gyrokinetic simulations and was benchmarked against results obtained from gyrokinetic eigensolver simulations.

Linear embeddings have also found widespread use in machine learning~\cite{Brunton20, Ji23} and robotics,~\cite{Taylor21} where high-dimensional and nonlinear dynamics are common.
A good example is the forecasting of the time evolution of nonlinear problems using Koopman-based autoencoder networks.~\cite{Azencot20, Fukami20}
Another example is the use of the Koopman operator in reinforcement learning to estimate the value function, thereby effectively eliminating the dependence on future system states.~\cite{Rozwood24}

One of the key advantages of applying the Koopman operator in machine learning is better interpretability of input-output relationships, in contrast to neural networks, which are often regarded as black-box models.~\cite{Carrier24, Shi25koopman}
Linear embeddings are also beneficial in the context of optimal control problems.~\cite{Rauh09, Abraham17, Korda20, Otto21, Bevanda21}
In particular, linear Koopman Model Predictive Control (MPC)~\cite{Korda18} leads to convex optimization problems with a unique global solution.
In this context, the Koopman operator can be approximated using deep neural networks, enabling its subsequent application in the optimal control of dynamical systems~\cite{Han20}.
However, one should be careful when attempting to learn linear embeddings blindly, as there are fundamental limitations which constrain the applicability of linearization techniques.~\cite{Kvalheim21, Liu23Koopman, Liu25Koopman, Kvalheim25}


\subsection{Quantum Carleman algorithms}

Quantum computers have been predicted to be quite powerful because they can  efficiently apply exponentially large dense unitary matrices to dense state-vectors, and, hence, have the ability to speed up many calculations of interest to scientific computing.
For example, they can speed up the solution of large linear differential equations.
However, because their great power relies on the ability to form superposition states, even quantum computers are limited in their ability to speed up nonlinear operations, which are fundamentally incompatible with superposition.~\cite{Joseph23}

While nonlinear operations can be performed by working with a tensor product of copies of states, the need to hold tensor product states in quantum memory precludes the ability to compress the information.
Moreover, the so-called \textit{No-Cloning Theorem},~\cite{Wootters82} which states that an unknown quantum state cannot be copied, implies that the only way to make such tensor products is to independently prepare copies from scratch.~\cite{Joseph23}
This implies that performing iterative nonlinear operations becomes exponentially costly in quantum memory.~\cite{Leyton08,Brustle2025quantum}
This fundamentally limits the complexity of na\"ive quantum approaches to iterative nonlinear solvers or optimizers, e.g. that rely on Newton's method. 

Thus, the quantum algorithms community has focused on linear embedding methods, such as the Koopman-von Neummann (KvN) approach~\cite{Joseph20,Joseph23,Novikau24KvN} and Carleman linearization,~\cite{Liu21, Krovi23, Jennings25} to solve nonlinear differential equations.
Quantum algorithms were developed that efficiently implement the Carleman embedding technique~\cite{Liu21, Krovi23, Vaszary25} and enable the efficient simulation of \textit{weakly nonlinear dissipative} dynamical systems on quantum computers.
An important advancement was the ability to rigorously prove the convergence of the quantum Carleman algorithm for the first time.~\cite{Liu21}
However, convergence can only be proven (1) if the nonlinearity is sufficiently weak and (2) given the promise that the initial condition is sufficiently close to an attracting fixed point (e.g. see \onlinecite{Liu21}).
Assuming the linearly embedded system converges to a fixed point, it can be solved efficiently on a quantum computer using quantum Hamiltonian simulation methods such as quantum linear system solvers\cite{Ambainis12, Lin20, Low24QLSA, Dalzell24}
or the more recent Linear Combination of Hamiltonian Simulation~\cite{Jin22Sch, An23, An23impr} method.

Motivated by the desire to build a convergent quantum Carleman algorithm, the recent work of \onlinecite{Endo2026} also explored the effect of shifting the solution and \onlinecite{Wang2026quantum} provided a convergence analysis for a single step of the shifted Carleman approach.
However, because of the concern that multiple shifts might be difficult to apply in a quantum algorithm, their analysis only applies to a single step of the algorithms explored here and in ~\onlinecite{Endo2026}.
For example, they prove that a single shift is sufficient to improve the Carleman solution of the Lotka-Volterra equations for a specific finite time interval but cannot provide the solution for all times.
However, these authors did not consider chaotic dynamical systems.

In fact, the first proposed quantum embedding technique for nonlinear dynamics was the KvN method~\cite{Joseph20} and it is not restricted to dissipative or weakly nonlinear dynamics.
The KvN method has been successfully employed to represent nonlinear dynamics through augmented linear systems in several specific applications.~\cite{Joseph20, Joseph23JPA, Lin22Koopman, Novikau24KvN}
Explicit quantum algorithms with rigorous convergence guarantees were recently developed for the first time in \onlinecite{Novikau24KvN, Novikau25Opt}.
Today, both embedding techniques are widely discussed as promising candidates for designing quantum algorithms for fluid dynamics,~\cite{Succi23, Sanavio24Three, Sanavio24, Conde25, Li25} kinetic theory,~\cite{Joseph23, Akiba23, Vaszary24, Vaszary25} and, with certain caveats, general nonlinear differential equations.~\cite{Liu21, Lin22Koopman, Liu23, Krovi23, Novikau24KvN, Jennings25}

\subsection{What is the optimal linearization atlas? }

The number of linearization charts needed to cover the manifold with a linearization atlas are determined by at least two separate considerations: \emph{topology} and \emph{analyticity}.
Koopman charts are only constrained by topology, because they can use any library of functions to define the linearization map.
In contrast, Carleman charts make a specific choice for the basis functions, namely polynomials in a specific coordinate system that are centered at a particular basepoint of the manifold, and, hence, the Carleman embedding procedure will only converge if the linearization map converges when written in terms of this basis.
In DMD terminology, one would refer to a convergent Koopman linearization process as an extended-- or super--Carleman embedding.
However, a better description is simply that the Carleman embedding is a specific type of Koopman embedding. 

When using Koopman theory, one should also distinguish between two separate notions of linearizability.
The first notion of linearizability is given by the  \emph{flow-box rectification} theorem, which states that a vector field can always be rectified in any region where it has no singular points, i.e. fixed-points where it vanishes or truly singular points where it blows up. This is the notion used here to design a global linearization atlas.
In contrast, determining a \emph{dynamically invariant linearization} which defines a complete set of generalized Koopman eigendistributions requires determining all distinct dynamically invariant sets on the manifold, which is clearly a much more challenging problem.
While the Koopman eigenproblem can always be solved for integrable vector fields, it is generically intractable for chaotic dynamics, which is the most common setting where one would like to design a linearization atlas.

The question of how to design an optimal linearization atlas with a minimal number of charts is an important question. 
It is possible that the overall number of Koopman linearization charts that are needed may be quite small, perhaps as small as the number of fixed points of the vector field, $\Fix{(V)}$.
Due to the \emph{Poincar\'e-Hopf theorem}, this number must be greater than zero if the Euler characteristic of the manifold does not vanish.

However, another bound on this number is given by the minimum number of open charts needed to cover a manifold, $\Manifold$, which is determined by the Lustnerik-Schnirelman category,~  \cite{Cornea2003Lusternik-Schnirelmann} $\LScat{(\Manifold)}+1$.
For example, only one set is needed for a contractible space, two for the $\dim$-sphere, $S^\dim$, and $\dim+1$ for the $\dim$-torus, $(\times S^1)^\dim$.
Ostrand's theorem \cite{Ostrand1971covering} on covering dimension implies that an upper bound on the number of charts needed is $\dimRm{(\Manifold)}+1$, so that $\LScat{(\Manifold)}\leq \dimRm{(\Manifold)}$.

However, due to the requirement of analyticity, the number of Carleman linearization charts needed could still be larger than this.
Hence, in this work, we will not attempt to find an optimal nor minimal Carleman atlas, i.e. an atlas that only requires the minimum number of Carleman charts for a given vector field necessary to cover a given manifold.
Instead, we will seek to find robust numerical procedures for ensuring that the global Carleman embedding is convergent.
We leave it as an open question to find an explicit algorithm for determining an optimal and/or minimal Carleman atlas.

\subsection{Main results}

We propose a globalized Carleman linearization method that enables the simulation  of dynamical systems with multiple fixed points, including chaotic dynamical systems, and that allows one to track trajectories far from any fixed points, including those diverging from unstable equilibria.
This technique divides the space into charts, each with a convergence radius $\xi$.
If enough information is given about the vector field, then the required atlas of charts can be precomputed before integrating the trajectories in time.
Here, one constructs an atlas of linearization charts that covers the entire solution manifold and allows the Carleman embedding to remain convergent over the entire manifold.

Alternatively, linearization charts can be constructed dynamically along the trajectory.
Here, the goal is only to generate a series of charts that cover a neighborhood of the trajectory of interest.
The simple piecewise embedding method, defined below, follows the trajectory by transitioning from one chart to the next each time the trajectory reaches the boundary of the current chart.
We also propose an adaptive version of the piecewise embedding, that varies the linearization chart radius $\xi$ on the fly depending on the difference between two trajectories simulated within charts of different sizes.
In this case, a larger or smaller radius is selected depending on whether a rapid decline in convergence is detected.

This paper is organized as follows.
In Sec.~\ref{sec:carleman-std}, we review the standard Carleman technique.
In Sec.~\ref{sec:carleman-gl}, we extend the technique to 
two
globalized piecewise adaptive versions.
In Sec.~\ref{sec:carleman-fixed}, we propose an alternative globalized embedding method with a static predefined atlas.
In Sec.~\ref{sec:numetical-tests}, the piecewise technique is tested on various nonlinear problems progressing from 1D and 2D systems, where good decompositions are known, to 3D systems that harbor completely chaotic dynamics.  For example, we study the ability to capture dynamics on strange attractors, such as the R\"ossler, Lorenz, and Chen's attractor.  
We also study integrable and non-integrable Lotka-Volterra predator-prey systems.
Section~\ref{sec:complexity} discusses the complexity of the global Carleman embedding methods.
Section~\ref{sec:classical-global-carleman} discusses the classical complexity based on a combination of empirical results and the theoretical analysis in Appendix~\ref{sec:complexity-remap}. Section~\ref{sec:quantum-global-carleman} considers the prospects for developing efficient quantum versions of the global Carleman algorithms and discusses the implied limitations on the ability to speed up calculations of individual trajectories.
Finally, Section~\ref{sec:conclusion} 
summarizes our conclusions.

The theory behind the need for the globalized Carleman methods and their associated complexity is given in the appendices. 
Appendix~\ref{sec:1dflows} proves that multiple  Carleman charts are needed even for 1D flows. 
Appendix~\ref{sec:higher-order-fixed-points} discusses the multibranched linearizing transformations required for higher order fixed points in 1D.
Appendix~\ref{sec:higher-dflows} proves that multiple Carleman charts are required for multidimensional flows due to the presence of fixed and moveable singularities of the flow in complex time, $t$, and initial conditions, $x_0$.
Appendix~\ref{sec:complexity-remap} estimates the complexity of global Carleman algorithms, including the shifting and rescaling operations.
\section{Globalized Carleman embedding}

\subsection{Standard Carleman embedding (SCE) method}\label{sec:carleman-std}

Let us assume that we need to solve a set of nonlinear differential equations. Assume the solution $X=\{X^i\}\in \Reals^\dim$, is specified in terms of coordinates $X^i$ and that the flow is specified by a nonlinear vector field $V(X)=\{V^i(X^j)\}\in\Reals^\dim$ via:
\begin{equation}\label{eq:nl-orig}
    \diff_t X = dX/dt
    = V(X).
\end{equation}
To construct the Carleman embedding, we Taylor expand the vector field $V(X)$ around a chosen point $X_0$. 
In terms of $x=X-X_0$, this leads to the result:
\begin{equation}\label{eq:nl}
    \diff_t x 
     = F(x) \simeq F_0 + F_1 x + F_2 x^{\otimes 2} + \dots F_{\yNnl} x^{\otimes \yNnl}.
\end{equation}
When the nonlinearity is carried to order $\yNnl$, the resulting linear differential equation has order $\yNnl$. 
The vector $x \in \Reals^\dim$ is
\begin{equation}\label{eq:x}
    x = \left(x^{0}, x^{1}, \dots x^{\dim-1}\right)^T,
\end{equation}
where $x^j$ 
is the $j$-th component of the vector $x$.
In contrast, the vector $x^{\otimes 2}$ is a tensor product of two vectors $x$, i.e., $x \otimes x \in \Reals^{\dim^2}$, and so on.
The matrices $F_k \in \Reals^{\dim \times \dim^k}$ 
describe the nonlinear problem in a given linearization chart with a convergence radius $\xi$:
\begin{equation}\label{eq:emb-condition}
   |x| \leq \xi \leq 1.
\end{equation}
It should be kept in mind that the elements of the matrices $F$ depend on the point $X_0$ around which the chart is constructed.

To implement the SCE, one considers $\yNca$ monomials $x$, $x^{\otimes 2}$, $\dots$, $x^{\otimes \yNca}$ as independent components of the vector $u$:
\begin{equation}\label{eq:u}
    u = (x, x^{\otimes 2}, \dots, x^{\otimes \yNca})^T,
\end{equation} 
where $\yNca$ must be greater or equal than $\yNnl$.
The size of the vector $u$ 
is determined by the number of independent homogeneous polynomials up to degree $\yNca$ which is given by the binomial coefficient 
\begin{align}
\yNu=\Binomial_\dim^{\dim+\yNca} =\Binomial_\yNca^{\dim+\yNca} =\frac{(\dim+\yNca)!}{\dim!\yNca!}
<\frac{(\dim+\yNca)^{\dim+\yNca}}{\dim^\dim \yNca^\yNca}.
\end{align}
One can show that $\yNu$ is bounded from above by the smaller of $\yNca^\dim$ and $\dim^\yNca$.
In fact, using Stirling's approximation, when $\yNca\gg\dim$, as is the case for the low-dimensional dynamical systems studied here, $\yNu$ scales as $(\yNca/\dim)^\dim$ which is exponential in dimension $\dim$, 
but in the opposite limit, $\dim\gg\yNca$, this scales as $(\dim/\yNca)^{\yNca}$ which is exponential in the polynomial degree $\yNca$. 
For large systems of ODEs, such as those resulting from the discretization of PDEs, this leads to a significant reduction in degrees of freedom (DOFs) relative to tracking $\yNca$ polynomials for all $\dim$ variables, which would require $\yNca^\dim$ terms.

The embedding results in the linear differential equation
\begin{equation}\label{eq:carleman-emb}
    \diff_t u 
    = A u + B,
\end{equation}
where the matrices $A \in \Reals^{\yNu \times \yNu}$ and $B \in \Reals^{\yNu\times 1}$ include the matrices $F_k$ as subblocks.
For instance, for $\yNnl = 3$, the matrix $A$ is 
\begin{equation}\label{eq:A}
    A = 
    \begin{bmatrix}
    \yA{1}{1} & \yA{2}{1} & \yA{3}{1} & 0         &         0 &     0 &                 \dots &                 0 \\
    \yA{1}{2} & \yA{2}{2} & \yA{3}{2} & \yA{4}{2} &         0 &     0 &                 \dots &                 0 \\
            0 & \yA{2}{3} & \yA{3}{3} & \yA{4}{3} & \yA{5}{3} &     0 &                 \dots &                 0 \\
        \dots &     \dots &     \dots &     \dots &     \dots & \dots &                 \dots &             \dots \\
        \dots &     \dots &     \dots &     \dots &     \dots &     0 & \yA{\yNca - 1}{\yNca} & \yA{\yNca}{\yNca}
    \end{bmatrix}.
\end{equation}
Here, $\yA{k}{j} \in \Reals^{\dim^j \times \dim^{k}}$ where $j, k = 1, 2, \dots \yNca$.
The matrices $\yA{k}{j}$ can be computed in the following way:
\begin{align}\label{eq:ca-matrices-A}
    \yA{j+\yk}{j} 
    &=\sum_{i=0}^{j-1} \yIN{i}\otimes F_{\yk+1} \otimes \yIN{j-i-1}  \\
    &= F_{\yk + 1}\otimes\yIN{j-1} + I_{\dim} \otimes F_{\yk + 1}\yIN{j-2} + \dots  
    + \yIN{j-1}\otimes F_{\yk + 1},
\end{align}
where 
$I_{\dim}$ is the $\dim \times \dim$ unit matrix
and, now, $\yk = \max(-1,1-j), 0, 1, \dots,  \min( \yNnl-1, \yNca-j)$.
The matrix $B$ depends only on $F_0$:
\begin{equation}\label{eq:B}
    B = [F_0^T, 0, \dots, 0]^T.
\end{equation}
This large system of linear differential equations~\eqref{eq:carleman-emb} embeds the nonlinear system~\eqref{eq:nl}.
However, as explained above, the embedding~\eqref{eq:carleman-emb}  only works correctly within a subset of the linearization chart~\eqref{eq:emb-condition} and only in the vicinity of a single fixed point.

\begin{table}[t!]
{
\centering
\caption{Dynamic adaptive Carleman embedding (ACE) algorithm. \label{table:piecewise-adaptive}}
}
\scriptsize
\begin{tabular}{l}
    \hline
    \textbf{Given data} \\
    \hline
    \parbox{8cm}{ \raggedright
    \texttt{
    Time interval $t_{\rm max}$, 
    coefficients $c$, 
    initial conditions $X_0$,
    function $\yFc$,
    function $\ySc$, 
    tolerance $\ytol$,
    initial chart size $\xi$, 
    step $\Delta \xi$,
    minimal possible radius $\xi_{\rm min}$,
    maximum possible radius $\xi_{\rm max}$.
    }
    }\\   

    \hline
    \textbf{Embedding subroutine:} \texttt{$x(t + \Delta t),$ isUnstable $= \yCE(t, x_{\rm init}, \Xi)$} \\
    \hline
    $c' = \yFc(\Xi)$ \\
    $A, B, u_{\rm init} = \ySc(c', x_{\rm init})$ \\
    \texttt{\textcolor{gray}{// --- A single time step of a Runge-Kutta solver ---} }\\
    $u = \text{RK}(A, B, u_{\rm init}, t)$ \\
    $x = \yDE(u)$\\
    \texttt{isUnstable $=$ any$(|u| > 1)$}\\
    \texttt{return $x$, isUnstable}\\

    \hline
    \textbf{Adaptation subroutine:}
    \texttt{$t, x([t_{\rm init}, t]), \xi_{\rm new}, \Xi = \yAR(t_{\rm init}, \xi, \Xi, $flagInc$)$}\\
    \hline
    \texttt{ \textcolor{gray}{// The flagInc enables or disables the comparison} }\\
    \texttt{ \textcolor{gray}{// with enlarged charts.} }\\
    $\xi' = \xi - \Delta \xi$\\
    $x_{\rm arr}, t, x, i_t = [0], t_{\rm init}, 0, 0$\\
    \texttt{if $\xi' \geq \xi_{\rm min}$:}\\
        \texttt{$\quad$\textcolor{gray}{// --- Simulate the reduced chart ---} }\\
        \texttt{ \textcolor{gray}{// It is assumed that the trajectories embedded into} }\\
        \texttt{ \textcolor{gray}{// the concentric $\xi'$-chart and $\xi$-chart coincide.} }\\
        \texttt{$\quad$while $|x| < \xi'$:}\\
            \texttt{$\quad\quad\quad x,$ isUnstable $= \yCE(t,x,\Xi)$}\\
            $\quad\quad\quad i_t = i_t + 1$\\
            $\quad\quad\quad x_{\rm arr}[i_t], t = x, t + \Delta t$\\
        \texttt{$\quad$\textcolor{gray}{// --- Compare the reference and shifted reduced charts ---} }\\
        \texttt{ \textcolor{gray}{// The comparison starts from the point where} }\\
        \texttt{ \textcolor{gray}{// the non-shifted $\xi'$-chart ends.} }\\
        $\quad\Xi', x' = \Xi + x, 0$\\
        \texttt{$\quad$While $|x| < \xi$:} \\
            $\quad\quad \epsilon = x - x' + \Xi - \Xi'$\\
            \texttt{$\quad\quad$if $|\epsilon| \geq \ytol$ or isUnstable:}\\
                \texttt{$\quad\quad\quad$return $\yAR(t_{\rm init}, \xi', \Xi,$false$)$ \textcolor{gray}{// reduce the chart once more} } \\
            \texttt{$\quad\quad$else:}\\
                \texttt{$\quad\quad\quad x,$ isUnstable $= \yCE(t,x,\Xi)$ \textcolor{gray}{// simulate the reference chart} }\\ 
                $\quad\quad\quad x',\_ = \yCE(t,x',\Xi')$ \texttt{ \textcolor{gray}{// simulate the shifted reduced chart} }\\
                $\quad\quad\quad x_{\rm arr}[i_t], t = x, t + \Delta t$\\
    \texttt{else:}\\
        \texttt{$\quad$\textcolor{gray}{// --- Simulate just the reference chart ---} }\\
        \texttt{$\quad$while $|x| < \xi$:}\\
        $\quad\quad\quad x, \_ = \yCE(t,x,\Xi)$\\
        $\quad\quad\quad i_t = i_t + 1$\\
        $\quad\quad\quad x_{\rm arr}[i_t], t = x, t + \Delta t$\\
    \texttt{if not flagInc:}\\
        \texttt{$\quad$return $t$, $x_{\rm arr}$, $\xi$, $\Xi + x_{\rm arr}[i_t]$ \textcolor{gray}{// take the reference chart} } \\
    \texttt{\textcolor{gray}{// --- Compare the reference and enlarged charts ---} }\\
    $\xi' = \xi + \Delta \xi$\\
    \texttt{While $\xi' \leq \xi_{\rm max}$:} \\
        $\quad\Xi', x' = \Xi + x, 0$\\
        \texttt{$\quad x, $ isUnstable $ = \yCE(t,x,\Xi)$} \texttt{\textcolor{gray}{// evolution in the enlarged chart} }\\
        $\quad x', \_ = \yCE(t,x',\Xi')$ \texttt{\textcolor{gray}{// evolution in the shifted reference chart} }\\
        \texttt{$\quad$While $|x| < \xi'$:} \\
            $\quad\quad \epsilon = x - x' + \Xi - \Xi'$\\
            \texttt{$\quad\quad$if $|\epsilon| \geq \ytol$ or isUnstable:}\\
                \texttt{$\quad\quad\quad$ return $t$, $x_{\rm arr}$, $\xi$, $\Xi + x_{\rm arr}[i_t]$ \textcolor{gray}{// take the reference chart} }\\
            \texttt{$\quad\quad$else:}\\
                $\quad\quad\quad x_{\rm arr}[i_t], t = x, t + \Delta t$\\
                \texttt{$\quad\quad\quad x,\ $isUnstable$\ = \yCE(t,x,\Xi)$ }\\
                \texttt{$\quad\quad\quad x', \_ = \yCE(t,x',\Xi')$ }\\
        \texttt{$\quad\xi = \xi'$ \textcolor{gray}{// the enlarged chart becomes the new reference chart} } \\
        $\quad\xi' = \xi + \Delta \xi$\\
    \texttt{return $t$, $x_{\rm arr}$, $\xi$, $\Xi + x_{\rm arr}[i_t]$ \textcolor{gray}{// take the enlarged chart} }\\
    
    \hline
    \textbf{Overall algorithm} \\
    \hline
    $\Xi, t, X = X_0, 0, [\text{   }]$\\
    \texttt{While $t \leq t_{\rm max}$:} \\
        $\quad t_{\rm init}, \Xi_{\rm init} = t, \Xi$\\
        \texttt{$\quad t, x([t_{\rm init}, t]), \xi, \Xi = \yAR(t_{\rm init}, \xi, \Xi_{\rm init}, $true$)$}\\
        $\quad X([t_{\rm init}, t]) = x([t_{\rm init}, t]) + \Xi_{\rm init}$\\
    \texttt{return $X$}\\
    \hline
\end{tabular}
\end{table}

\begin{figure}[!t]
\centering
\includegraphics[width=0.49\textwidth]{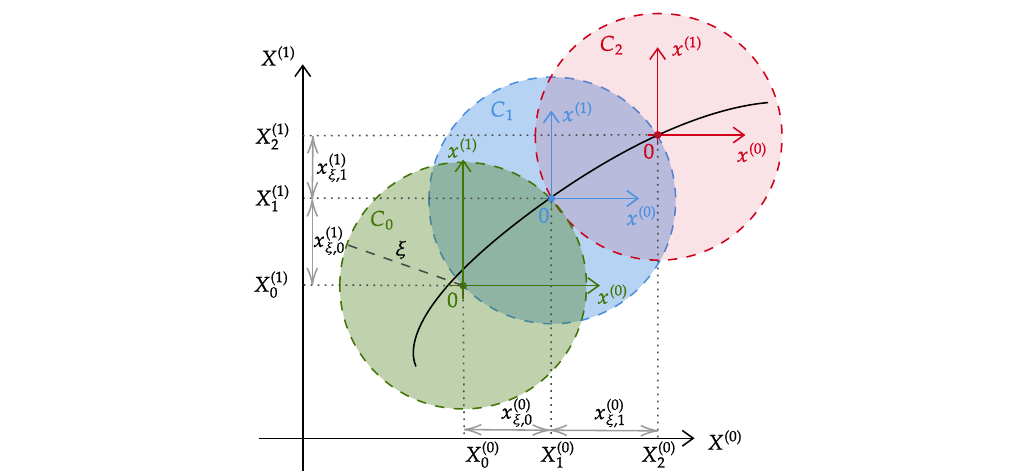}
\caption{
    \label{fig:scheme-PCE}
    Schematic depiction of the dynamic Piecewise Carleman Embedding (PCE) for a system with $\dim = 2$ degrees of freedom.
    A new linearization chart must be chosen each time the trajectory leaves the original linearization chart.
}
\end{figure}
\begin{figure*}[!t]
\centering
\includegraphics[width=0.90\textwidth]{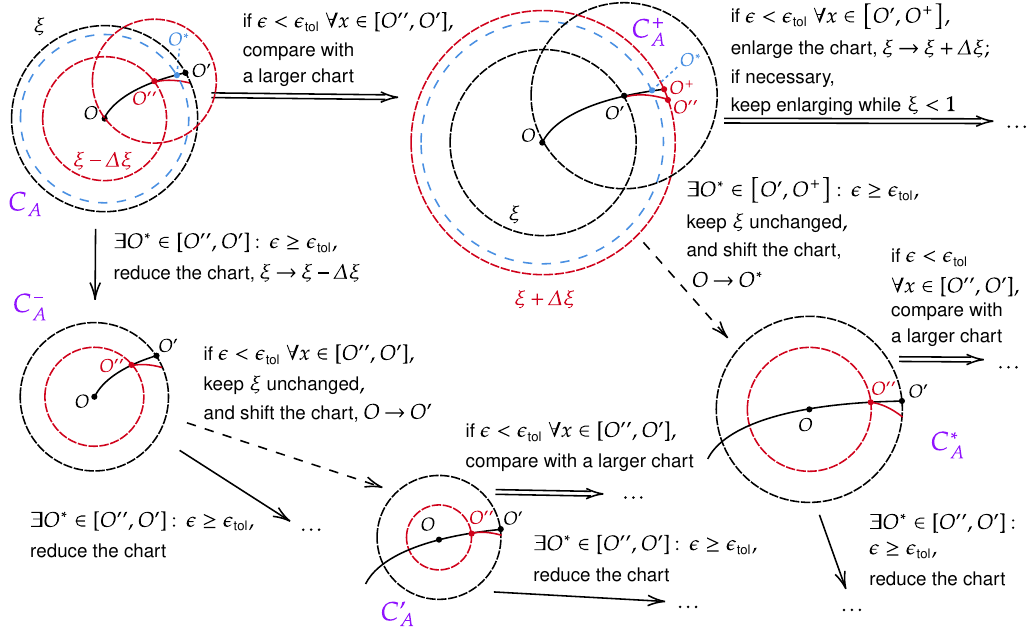}
\caption{
    \label{fig:scheme-ACE}
    The logical flow diagram for the ACE algorithm that adapts the new chart radius $\xi$ based on the need to satisfy an imposed error threshold $\ytol$.
}
\end{figure*}

\subsection{Dynamic Piecewise and Adaptive Carleman Embeddings (PCE/ACE)}\label{sec:carleman-gl}
The goal of this work is to extend the Carleman embedding method to a much larger class of nonlinear systems, including chaotic dynamical systems.
In particular, to bypass the limitation~\eqref{eq:emb-condition}, we allow the choice of embedding chart to change along the trajectory, 
as schematically shown in Fig.~\ref{fig:scheme-PCE}.
First of all, we construct an embedding chart $C_0$ (green circle) centered at the initial conditions $X_0$.
The Carleman method evolves the trajectory (solid black line) until it reaches the chart boundary, beyond which the condition~\eqref{eq:emb-condition} is no longer satisfied.
At this moment, one switches to a new chart, $C_1$ (blue circle), centered around the new point 
$X_1 = \left(X_1^{0}, X_1^{1}\right)$ 
and rewrites the embedding~\eqref{eq:carleman-emb}.
The displacement $x_{\xi,0}$ between two centers $X_0$ and $X_1$ is added to the global shift $\Xi$, 
\begin{equation}\label{eq:gl-shift}
    \Xi = \Xi + x_{\xi},
\end{equation}
where $\Xi$ is initialized as the zero vector.
Later, the vector $\Xi$ is used to reconstruct the original variable $X$ from the local variable $x$ via:
\begin{equation}\label{eq:x-reconstruction}
    X = x + \Xi.
\end{equation}
When the trajectory reaches the boundary of $C_1$, one again jumps to the new chart $C_2$ (red circle), adds the local shift $x_{\xi,1}$ to $\Xi$, reconstructs the linearization~\eqref{eq:carleman-emb} at the new point $X_2$, and models the system's evolution within the chart $C_2$, and so on.

This approach requires minimal prior knowledge about the simulated system. 
In particular, it does not require prior knowledge about the positions of the system's fixed points. 
However, it does require knowing the explicit dependence of the linearization~\eqref{eq:nl} on the point $X_j$ around which the chart $C_j$ is constructed.
For example, if $c$ represents the set of coefficients characterizing the nonlinear system~\eqref{eq:nl}, a function $\yFc(\Xi)$ can be introduced to describe the dependence of $c$ on the accumulated vector shift $\Xi$. 
The implementation of the function $\yFc(\Xi)$ strongly depends on the simulated system~\eqref{eq:nl-orig}.
Another function, $\ySc(c)$, constructs the embedding~\eqref{eq:carleman-emb} using the  
coefficients $c$.
The implementation $\ySc(c)$ follows the logic described in Sec.~\ref{sec:carleman-std}.
The function $\ySc(c)$ takes the coefficients $c$, constructs the matrices $F$, Eq.~\eqref{eq:nl}, and then builds the embedding system~\eqref{eq:carleman-emb} by preparing the matrices $A$ and $B$, and the initial vector $u$.
Once the trajectory within a single chart has been successfully simulated, e.g., by using a Runge-Kutta time integrator,  
a function $\yDE(u)$ is used to decode the embedding monomials $u$ back into the local variable $x$.
As seen from Eq.~\eqref{eq:u}, $\yDE(u)$ can be implemented just by taking the first $\dim$ elements of $u$.
After this, the original variable $X$ is reconstructed using Eq.~\eqref{eq:x-reconstruction}.

One may still encounter an unfavorable scenario in which two or more fixed points lie within a single linearization chart, i.e., within the domain~\eqref{eq:emb-condition}.
To address this case, an adaptive method that dynamically adjusts the radius of each chart, $\xi$, is required.
A schematic depiction of the logical flow diagram for an adaptive algorithm for dynamically adjusting the radius on the fly  is shown in Fig.~\ref{fig:scheme-ACE}.

One starts from a reference chart with a radius $\xi$ (black dashed circle in the case $C_A$) and compares the embedding with the linearization within a smaller chart, $\xi' = \xi - \Delta\xi$ (red dashed circle). 
Inside the $\xi'$-chart, the trajectories obtained by both linearizations coincide but can diverge outside the $\xi'$-chart starting from the point $X = O''$.
Thus, one first simulates the trajectory within $\xi'$-chart, then shifts the chart to the point $O''$ and compares the reference trajectory with the trajectory in the shifted $\xi'$-chart at each time step.
If necessary, the reduced chart must be shifted again when the simulated trajectory is found to lie outside the current reduced chart.

The difference between two trajectories is computed as
\begin{equation}\label{eq:diff-trajectory}
    \epsilon = |X-X'|,
\end{equation}
where $X$ is a point on the trajectory obtained using the $\xi$-chart, and $X'$ is a point on the trajectory obtained using the shifted $\xi'$-embedding.
It is important to note that the comparison is made between the global variables, rather than their local counterparts $x$ and $x'$, since the local variables in the $\xi$- and $\xi'$-charts differ completely.

If $\epsilon$ is less than the required tolerance $\ytol$ for all $X$ in the interval between $O''$ and $O'$, one saves the computed part of the trajectory.
Then, one compares the reference linearization with the embedding in a larger chart, $\xi' = \xi + \Delta\xi$ (red circle in the case $C_A^{+}$).
The linearizations coincide within the $\xi$-chart, but can diverge beyond the point $O'$.
To perform the comparison, one needs to shift the reference chart to the point $O'$.
If $\epsilon < \ytol$ for all points between $O'$ and $O^{+}$, the larger chart is taken as the reference, the trajectory $O'O''$ is saved, and the new chart is then compared with an even larger chart.
The linearization radius is continuously increased as long as $\epsilon < \ytol$ and $\xi \leq \xi_{\rm max} \leq 1$.
Once $\epsilon \geq \ytol$ at some point $X = O^*$ (the blue dot in the case $C_A^{+}$), the step increase is halted, and the reference chart is shifted to the point $O^{*}$ without increasing the convergence radius $\xi$.
The comparison with larger charts is necessary to accelerate simulations by avoiding the use of over-reduced charts.

If one arrives back to the case $C_A$, there is a point $X = O^*$ in the interval between the points $O''$ and $O'$ where $\epsilon \geq \ytol$, then the linearization radius is decreased, i.e. $\xi^* = \xi - \Delta \xi$.
The $\xi^*$-chart is now taken as the reference one (the case $C^{-}_A$ in Fig.~\ref{fig:scheme-ACE}).
This chart is then compared with an even smaller chart $\xi^* - \Delta \xi$.
If $\epsilon \geq \ytol$ is still observed, one keeps decreasing the linearization radius as long as $\xi$ remains larger or equal than an imposed minimal value $\xi_{\rm min}$.
On the other hand, if $\epsilon < \ytol$, the $\xi^*$-embedding is kept as the reference, the trajectory $OO'$ is saved, and the chart is shifted to the point $O'$ (the case $C'_A$).

Hence, the original chart $C_A$ is shifted either to the chart $C'_A$ or the chart $C^{*}_A$. 
After the shift, the algorithm follows the same logic as it did for the chart $C_A$.

It is worth noting that the chart size can be adjusted either by using an additive step $\Delta \xi$, or, for example, by multiplying or dividing $\xi$ by a factor of two to accelerate changes in the radius.
These two approaches can be combined: use the step $\Delta \xi$ when $\xi \geq \xi_{\rm th}$ and apply multiplication or division when $\xi < \xi_{\rm th}$ where $\xi_{\rm th}$ is a preset value.

The described piecewise Adaptive Carleman Embedding (ACE) is summarized in Table~\ref{table:piecewise-adaptive}.
We also  consider a 
simplified 
version  that we call 
Piecewise Carleman Embedding (PCE)  that does not change the radius.
For the PCE 
algorithm, one still moves the linearization chart in response to 
the simulated trajectory but keeps the convergence radius $\xi$ constant.
A comparison between the PCE and ACE approaches is given in Section~\ref{sec:ace_vs_pce}
and the complexity of the shifting and rescaling operations is discussed in Appendix~\ref{sec:complexity}.

The main input parameters of the ACE approach are the initial chart size $\xi$, the step $\Delta \xi$, the threshold $\xi_{\rm th}$, and the tolerance $\ytol$.
To prevent the chart size from becoming zero or negative, the threshold must be larger than the step $\Delta \xi$.
This also allows the chart size to be adapted independently of $\Delta \xi$ when $\xi < \xi_{\rm th}$, as discussed above, which is especially important if a large $\Delta\xi$ is chosen.

To ensure that the ACE can detect differences between trajectories embedded in charts of different sizes, the tolerance $\ytol$ should be reduced as $\Delta \xi$ decreases.
Indeed, trajectories in two charts of sizes $\xi$ and $\xi' = \xi \pm \Delta\xi$ typically differ by only a small amount $\epsilon$, Eq.~\eqref{eq:diff-trajectory}, when $\Delta\xi$ is small.
If $\ytol$ is not sufficiently small, the ACE cannot resolve the difference between these trajectories and therefore keeps the chart size unchanged.
However, a small difference between trajectories embedded in charts of similar sizes does not necessarily imply a small error in the ACE simulation of a nonlinear trajectory.
Therefore, it is recommended to begin with $\xi = \xi_{\rm max} = 1.0$, choose a relatively large step $\Delta \xi$, e.g., $\Delta \xi = 0.5$, and subsequently decrease the tolerance $\ytol$, starting from a relatively large value, e.g., $\ytol \approx 10^{-2}$.
This strategy begins with computationally inexpensive simulations of relatively low accuracy and progressively decreases $\ytol$ until an optimal value is found that provides the desired accuracy over the entire trajectory.
 
A similar procedure can be used for the PCE technique by starting with $\xi = 1$ and progressively decreasing $\xi$ in successive simulations until the desired accuracy is reached.
However, the PCE maintains a constant chart size $\xi$ along the entire trajectory.
As a result, if only a small portion of a nonlinear trajectory requires a small chart size, the PCE must nevertheless use this small value of $\xi$ throughout the simulation.
By contrast, the ACE adapts the chart size locally: it decreases $\xi$ only where necessary while keeping larger charts elsewhere, thereby potentially reducing the total number of chart shifts.

The current version of the adaptive method depends on the step size $\Delta \xi$ used to adjust the convergence radius.
To eliminate this dependence, the logic of the adaptive scheme can be reversed: 
starting from a small value of $\xi$, the convergence radius is progressively increased by applying a multiplicative factor $\eta > 1$, such that $\xi \rightarrow \eta \xi \rightarrow \eta^2 \xi \rightarrow \dots$.
Once $\epsilon \geq \epsilon_{\rm tol}$, the convergence radius is reduced by a multiplicative factor $1/\eta'$, where $\eta'=\eta/\gamma$ and $\gamma > 1$ is an irrational number.
If, after this radius reduction, i.e., $\xi \rightarrow \xi/\eta'$, the condition $\epsilon < \epsilon_{\rm tol}$ is satisfied, the radius is increased again using a modified factor $\eta'' = \eta/\gamma^2$, and this process continues accordingly.
By using this technique, it is potentially possible to eliminate the dependence on $\Delta\xi$ and to determine a more optimal value of $\xi$ for each segment of the simulated trajectory. 
The drawback of this approach is that the number of $\xi$ adjustments could become substantial.

\subsection{Static Grid-anchored 
piecewise Carleman Embedding (GCE)}\label{sec:carleman-fixed}

\begin{figure}[!t]
\centering
\includegraphics[width=0.49\textwidth]{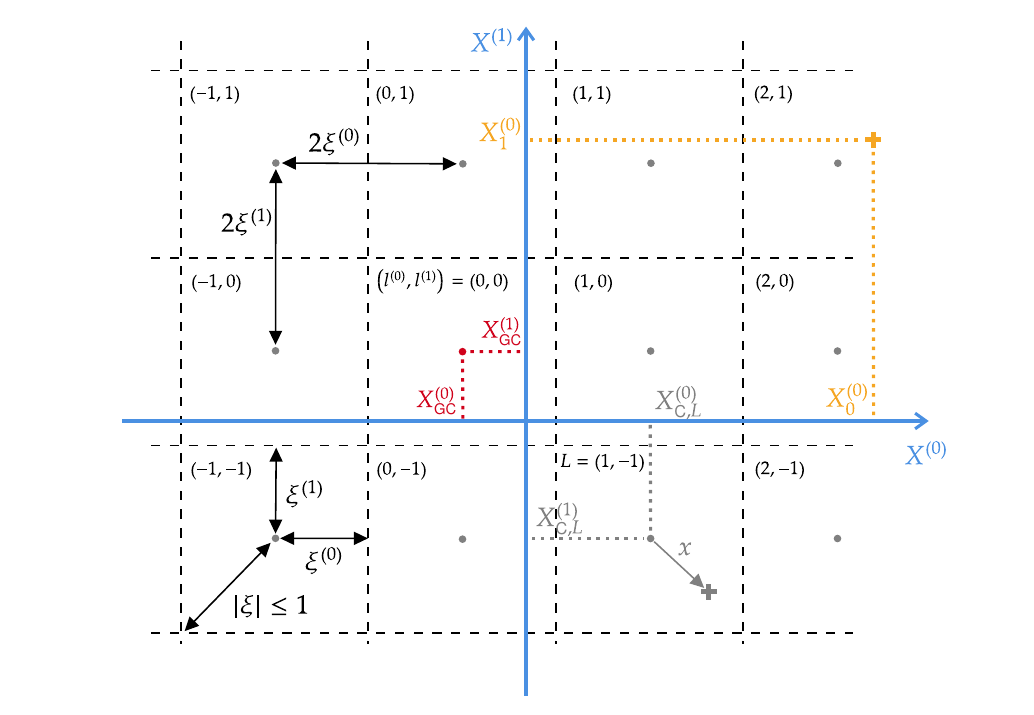}
\caption{
    \label{fig:scheme-carleman-fixed} 
    Schematic depiction of the static piecewise Grid-anchored Carleman Embedding (GCE) for the case with $\dim = 2$ variables.
    The original nonlinear variables $X^{(i)}$ are shown in blue.
    The red dot at 
    $\left(X_{\rm GC}^{0}, X_{\rm GC}^{1}\right)$ 
    marks the center of the linearization grid.
    The position of each tile-shaped element is determined by the integers 
    $l^{i}$.
    The distance between the centers of two neighboring elements is 
    $2\xi^{i}$ 
    in the $i$-th direction.
    The orange cross indicates the location of the initial condition, 
    $\left(X_{0}^{0}, X_{0}^{1}\right)$.
}
\end{figure}

The adaptive embedding approaches described in the previous section requires not only detecting the moment when the linearization chart must be switched but also comparing trajectories across charts of different sizes, or employing another method for detecting the Carleman linearization instability within a given chart.
Even the simplified PCE version, which lacks adaptivity in chart radius, results in dynamically moving linearization regions that requires high computational overhead.
These adaptive methods, which rely on many iterative nonlinear operations, 
could be difficult to implement 
on a quantum computer in a manner that achieves a quantum advantage.\cite{Joseph23}

Therefore, we propose an alternative version of the embedding using static linearization charts, which we refer to as elements or tiles.
Specifically, we perform a decomposition 
of the   
space of the variables $X^i$,  
with $i = 0, 1, \dots \dim-1$, prior to executing the Carleman simulations.
This results in a static grid of linearization regions, Fig.~\ref{fig:scheme-carleman-fixed}, whose positions are known in advance.
This grid consists of elements with fixed sizes $2\xi^{i}$  
in the $i$-th dimension and must satisfy the condition
\begin{equation}
    |\xi|\equiv\sqrt{\sum_{i = 0}^{\dim-1}\left( \xi^{i} \right)^2} \leq 1,
\end{equation}
to ensure the convergence of the Carleman linearization near the corners of the tiles.
If the tile sizes are equal in all directions than the above condition can be recast as
\begin{equation}
    \xi^{i} \leq \frac{1}{\sqrt{\dim}}.
\end{equation}

The center of the linearization grid (the red dot in Fig.~\ref{fig:scheme-carleman-fixed}) is placed at the position
\begin{equation}\label{eq:fixed-grid-center}
    X_{\rm GC} = \left(X_{\rm GC}^{0}, X_{\rm GC}^{1}, \dots X_{\rm GC}^{\dim-1} \right)^T,
\end{equation}
which does not necessarily have to be close to the initial condition (the orange cross in Fig.~\ref{fig:scheme-carleman-fixed}) of the simulated system of nonlinear equations.
The positions of the tile elements are described by the integer indices 
$l^{i}$, $l^{(i)}$, 
with $i = 0, 1, \dots, \dim-1$, where
$l^{i} = 0\ \forall i$ 
marks the center tile of the linearization grid.
The center of the tile associated with the integer vector 
\begin{equation}\label{eq:tile-L}
    L = \left(l^{0}, l^{1}, \dots, l^{\dim-1}\right)^T,
\end{equation} 
is computed as
\begin{equation}\label{eq:tile-center}
    X^{i}_{{\rm C}, L} = X_{\rm GC}^{i} + 2 \xi^{i} l^{i},\quad i = 0, 1,\dots \dim-1,
\end{equation}
or, equivalently, in vector form
\begin{equation}\label{eq:tile-center-vec}
    X_{{\rm C}, L} = X_{\rm GC} + 2\ {\rm diag}(\xi)\ L.
\end{equation}
Within each linearization chart, we work with a local variable $x$, whose components satisfy the condition
\begin{equation}\label{eq:fixed-x-cond}
    |x^{i}| \leq \xi^{i},\quad i = 0, 1, \dots, \dim-1,
\end{equation}
similarly to the local variables introduced in Sec.~\ref{sec:carleman-std}.
The local variable is equal to zero at the center of each tile.
The original variable $X$ in the $L$-th tile can be expressed as the sum of the local variable $x$ and the center of the corresponding tile:
\begin{equation}\label{eq:fixed-X}
    X = X_{{\rm C}, L} + x. 
\end{equation}
Thus, similarly to the global shift $\Xi$ introduced in Sec.~\ref{sec:carleman-gl}, the tile center $X_{{\rm C}, L}$ is used to construct the local Carleman embedding:
\begin{equation}
    \Xi \equiv X_{{\rm C}, L}.
\end{equation}
Knowing $X_{{\rm C}, L}$, one can derive the coefficients $c'$, which describe the nonlinear system in the $L$-th chart,
\begin{equation}\label{eq:fixed-c}
    c' = \yFc(X_{{\rm C}, L}),
\end{equation}
from which the Carleman embedding~\eqref{eq:carleman-emb} is derived:
\begin{equation}\label{eq:fixed-emb}
    A, B, u_{\rm init} = \ySc(c', x_{\rm init}).
\end{equation}
This is similar to the embedding subroutine used in Table~\ref{table:piecewise-adaptive}.

To initialize the Carleman embedding on the static linearization grid, one computes the integer vector $L_0$, which indicates the tile containing the initial condition.
The components of the vector $L_0$ are computed as
\begin{equation}
    l_0^{i} = \left\lfloor \frac{X^{i}_0 - X^{i}_{\rm GC}}{2 \xi^{i}} \right\rfloor.
\end{equation}
After this, the center of this tile, $X_{{\rm C}, L_0}$, can be found using Eq.~\eqref{eq:tile-center-vec}. 
The local coordinate of the initial condition is
\begin{equation}
    x_{{\rm init}, 0} = X_0 - X_{{\rm C}, L_0}.
\end{equation}
Using $X_{{\rm C}, L_0}$ and $x_{{\rm init}, 0}$, one constructs the initial Carleman embedding by employing Eqs.~\eqref{eq:fixed-c} and~\eqref{eq:fixed-emb}. 
Then, the vector $x$ is evolved in time, and when the local coordinates no longer satisfy condition~\eqref{eq:fixed-x-cond}, the system transitions to a neighboring chart.
In particular, for each direction $k$ such that
\begin{equation}
   |x^{k}| > \xi^{k},
\end{equation}
the components 
$l^{k}$ 
of the vector $L$ are modified as
\begin{equation}\label{eq:fixed-new-l}
    l^{k} \rightarrow l^{k} + \frac{x^{k}}{|x^{k}|}.
\end{equation}
Then, one computes the center of the new tile using Eq.~\eqref{eq:tile-center-vec} and constructs new local embedding using Eqs.~\eqref{eq:fixed-c} and~\eqref{eq:fixed-emb}, and so on.

In general, the geometry of the charts can be adapted to the topology of the system's evolution to minimize the number of shifts between linearization regions.
However, in this case, detecting the appropriate moment of the transition between neighboring charts can be more complicated than in the case of the hypercubes used in Fig.~\ref{fig:scheme-carleman-fixed}.
In principle, the construction of the linearization grid should account for the number and positions of fixed points in the nonlinear system under consideration, in order to ensure the stability of the Carleman embedding and to minimize the number of tile transitions.
The static grid can be scaled by adjusting $\xi$ or shifted by modifying $X_{\rm GC}$ to adapt to a particular nonlinear system.
For instance, each tile should include only a single fixed point.
Moreover, one should avoid placing a fixed point, such as the center of a limit cycle, near the corner of multiple tiles.
Otherwise, the number of tile transitions will be significant when the system evolves near that fixed point.
Typically, the accuracy of the linearization deteriorates near the chart boundaries.
For this reason, it is advisable to avoid scenarios where the trajectory stays close to the boundary for an extended period, especially when using large tiles. 
Otherwise, numerical errors may accumulate.

Since the position and the parameters of the anchored grid are known in advance, the described GCE technique may be more suitable for applications that require great speed.
However, one still needs to detect the moment when chart transitions are necessary.
In particular, this operation is problematic for quantum computers and, 
can either be implemented via measurement or amplitude estimation.
In the worst case, all components of the local vector $x$ must be analyzed to detect the precise moment when a chart transition becomes necessary.
This potentially requires measuring $\dim$ values, i.e. all $\dim$ components of $x$.
Ultimately, like the PCE and ACE techniques, one must shift the position by
all $\dim$ components of the vector $x_{\xi}$ to construct the new Carleman embedding in the dynamically shifted chart.

\subsection{
Comparison of Adaptive and Static Global Carleman Methods (ACE/PCE/GCE) 
\label{sec:ace_vs_pce}
}

\begin{figure}[!t]
\centering
\includegraphics[width=0.49\textwidth]{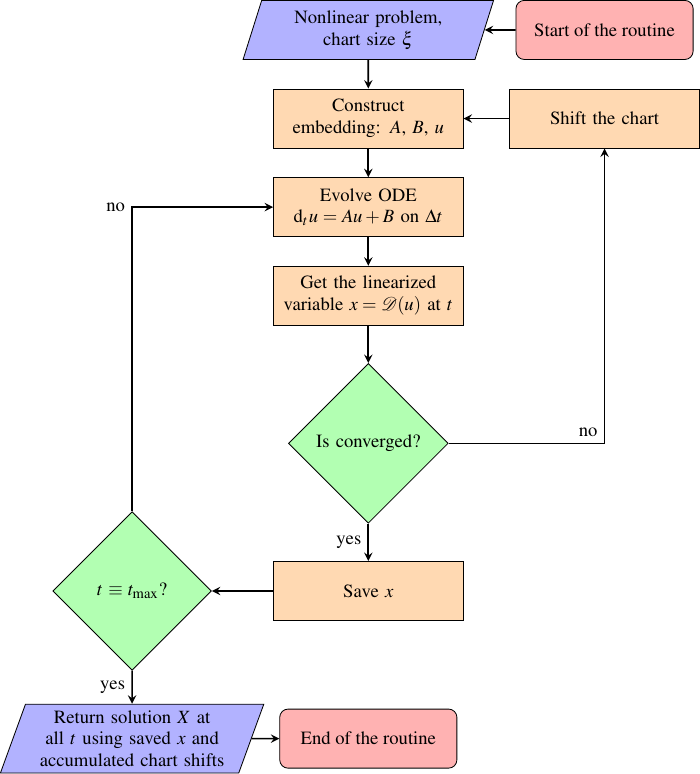}
\caption{
    \label{fig:flowchart-general}
    Flowchart showing the overall logic of the piecewise Carleman embedding methods.
    The construction of the matrices $A$ and $B$, and the vector $u$ is described in Section~\ref{sec:carleman-std}.
    The function $\mathcal{D}$ is described in Section~\ref{sec:carleman-gl}. 
    The reconstruction of the nonlinear variable $X$ is performed according to Eq.~\eqref{eq:x-reconstruction}.
    The convergence test for the PCE is $|x|\leq \xi$ at a given time $t$. 
    If the test does not pass, the embedding is modified by shifting the chart by $x$.
    If the test is satisfied, the current trajectory point $x$ is saved and one proceeds to the next time point.
    For the GCE, the convergence test checks whether 
    $|x^{k}|\leq \xi^{k}$ 
    holds for all $k = 0, 1,\dots n - 1$.
    Whenever the test fails for any $k$, the base point is moved to the new tile element in the corresponding $k$-th direction.
}
\end{figure}

\begin{figure}[!t]
\centering
\includegraphics[width=0.49\textwidth]{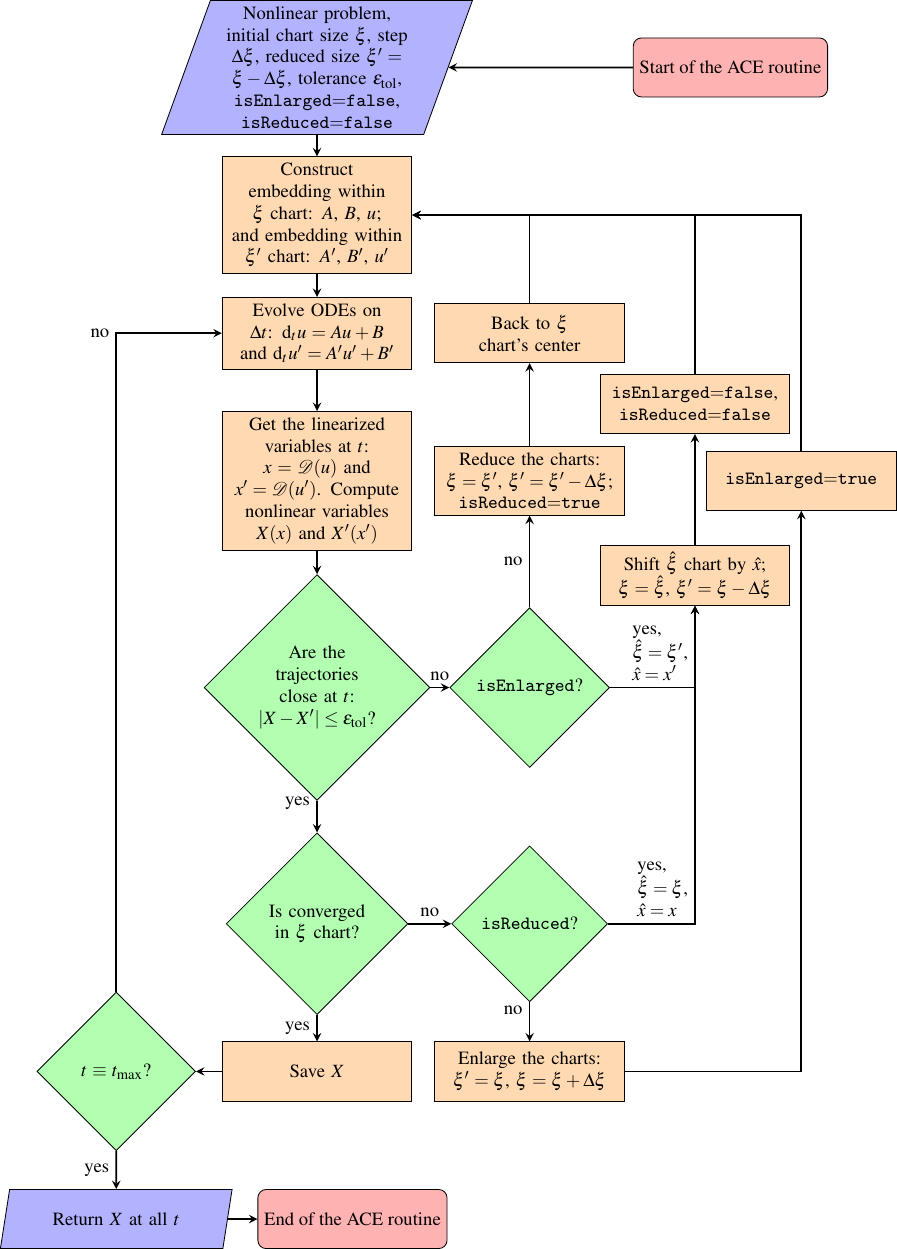}
\caption{
    \label{fig:flowchart-ACE}
    Simplified flowchart of the ACE technique where a comparison with reduced and enlarged charts is performed to improve the embedding's precision and speed.
    The \texttt{isEnlarged} flag indicates whether the comparison is performed with an enlarged chart (\texttt{true}) or a reduced chart (\texttt{false}).
    The \texttt{isReduced} flag indicates whether $\xi$ chart has already been reduced after the previous shift. 
}
\end{figure}

Flowcharts for the PCE, GCE, and ACE techniques are shown in Figs.~\ref{fig:flowchart-general} and ~\ref{fig:flowchart-ACE}, respectively.
Red rounded-corner blocks indicate the entry and exit points of each method.
Blue trapezoidal blocks indicate key input and output data.
Green diamond blocks represent decision points (i.e., if-then statements).
Orange rectangular blocks represent the main operations of each method.
The complexity of the shifting and rescaling operations is discussed in greater detail in Appendix~\ref{sec:complexity}.

The logic behind the PCE technique is quite straightforward and is the following.
The nonlinear problem of interest is embedded into a chart of size $\xi$ centered at the location $X_0$, 
i.e., $x = 0$, Fig.~\ref{fig:scheme-PCE}.
The linearly embedded solution vector, $x$, is evolved inside the chart.
If the vector $x$ reaches the boundary of the chart $\xi$ at time $t_*$, the chart is shifted by centering the new chart at the last computed point, so that one replaces $x(t_*)\rightarrow 0$. 
Then, the process repeats.
 
The logic of the GCE technique is similar to that of the PCE approach.
The main difference is that the linearization of an $n$-dimensional nonlinear problem is performed on a predefined grid of elements or tiles, each defined by its center $X_{\rm GC}$ and the fixed sizes $\xi^{(i)}$ for $i = 0, 1, ... n - 1$, Fig.~\ref{fig:scheme-carleman-fixed}.
To determine when the embedding must shift from one tile to a neighboring one, each component of the linearized variable $x$ is compared with the corresponding tile size $\xi^{(i)}$.
If a component $x^{(k)}$ exceeds $\xi^{(k)}$ for some $k$, the current tile is shifted to a neighboring tile in the $k$-th direction.
After all components of $x$ have been evaluated, the nonlinear problem is embedded into the new tile.
 
The
logic of the ACE technique is shown schematically in simplified form in flowchart~\ref{fig:flowchart-ACE}.
ACE begins by embedding a nonlinear problem into a chart of size $\xi$ and a concentric chart of smaller size, $\xi' = \xi - \Delta \xi$.
One compares the trajectories computed in the $\xi$-chart and the reduced $\xi'$-chart, which may be shifted multiple times within the $\xi$-chart.
This comparison is performed by computing the difference $\epsilon$, Eq.~\eqref{eq:diff-trajectory}, and comparing it with the user-defined tolerance $\ytol$.
If the difference exceeds the tolerance, this indicates that the charts are not sufficiently small.
Both charts are then reduced, and the trajectories are compared again within these reduced charts.
This procedure is repeated until the difference $\epsilon$ remains below the threshold $\ytol$ along the entire trajectory within the $\xi$-chart.
Similarly, if the accuracy is much smaller than the error tolerance, then the chart size is increased so that it remains close to but stays below threshold. 
As long as the accuracy remains below the error tolerance,
it is assumed that the $\xi$-chart and $\xi'$-chart are equivalent.
The trajectory computed in the $\xi$-chart is stored, $\xi$ is taken as the working chart size, and the $\xi$-chart is shifted to compute the next segment of the nonlinear trajectory.

\begin{figure*}[!t]
\centering
\includegraphics[width=0.90\textwidth]{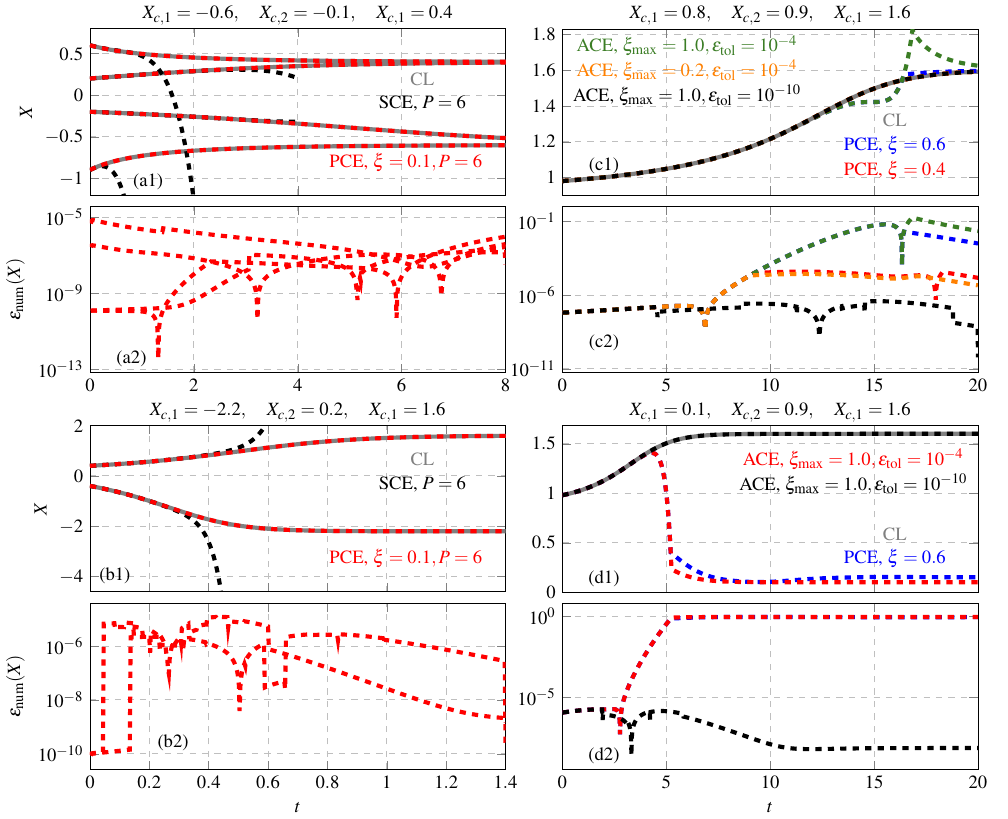}
\caption{
    \label{fig:ca-gl-oneD-case} 
    Results from simulations of the problem~\eqref{eq:1d} with $X_{c,1} = -0.6$, $X_{c,2} = -0.1$, $X_{c,3} = 0.4$ (a); $X_{c,1} = -2.2$, $X_{c,2} = 0.2$, $X_{c,3} = 1.6$ (b); $X_{c,1} = 0.8$, $X_{c,2} = 0.9$, $X_{c,3} = 1.6$ (c); $X_{c,1} = 0.1$, $X_{c,2} = 0.9$, and $X_{c,3} = 1.6$ (d).
    The CL simulations are shown by gray lines.
    (a) and (b): Comparison with the SCE (black dashed lineas) and PCE (red dashed lines) simulations for various initial conditions. 
    (c) and (d): Comparison with the ACE and PCE simulations.
    Plots (a2), (b2), (c2), and (d2) shown numerical errors in the PCE and ACE simulations.
    All PCE and ACE simulations have $\yNca = 6$.
    All ACE simulations have $\xi_{\rm init} = \xi_{\rm max}$ and $\Delta\xi = 0.02$.
}
\end{figure*}

\begin{figure}[!t]
\centering
\includegraphics[width=0.49\textwidth]{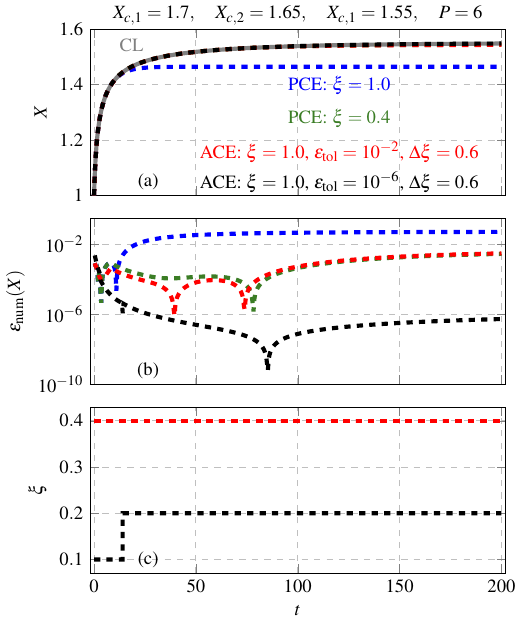}
\caption{
    \label{fig:ACE-vs-PCE}
    Results from simulations of the one-dimensional problem~\eqref{eq:1d} with $X_{c,1} = 1.70$, $X_{c,2} = 1.65$, and $X_{c,3} = 1.55$: 
    (a) nonlinear trajectories, (b) numerical error, Eq.~\eqref{eq:num-error}, in the embedded trajectories, (c) change of the chart size in the ACE simulations.
    The classical simulation is shown by the gray line.
    PCE simulations with fixed chart radius $\xi = 1.0$ and $\xi = 0.4$ are shown by the blue and green dashed lines, respectively. 
    ACE simulations with the tolerance $\ytol = 10^{-2}$ and $\ytol = 10^{-6}$ are shown by the red and black dashed lines, respectively.
    Both ACE simulations have the same initial chart size, $\xi = 1.0$, and the same step $\Delta \xi = 0.6$.
}
\end{figure}

\section{Numerical  Results 
}\label{sec:numetical-tests}

In this section, we evaluate the globalized Carleman methods proposed in the previous section on a set of nonlinear test problems and compare the results with standard classical (CL) simulations, where the nonlinear dynamical system is directly integrated in time using a Runge-Kutta method.
To compare CL and Carleman simulations, we compute the relative error
\begin{equation}\label{eq:num-error}
    \epsilon_{\rm num}\left(X^{i}\right) = \frac{|X^{i}_{\rm CL} - X^{i}_{\rm CE}|}{|X|},
\end{equation}
where 
$X^{i}_{\rm CL}$ 
is the $i$-th variable computed by the CL simulation, 
$X^{i}_{\rm CE}$ 
is the $i$-th variable computed by the PCE, ACE, or GCE techniques, and $|X|$ is 
\begin{equation}
    |X| = \sqrt{\sum_{i = 0}^{\dim-1}|X^i|^2}.
\end{equation}

The implementation of the PCE, ACE, and GCE techniques, along with the numerical simulations presented in this section, are provided in Ref.~\onlinecite{PCEcode}.

\subsection{1D Systems: One degree of freedom}\label{sec:1d}

To test the PCE and ACE methods, we first simulate a simple nonlinear problem with $\yNnl = 3$ and a single scalar variable $X$:
\begin{equation}\label{eq:1d}
    \diff_t X = (X_{c,1} - X)(X_{c,2} - X)(X_{c,3} - X),
\end{equation}
where the coefficients $X_{c,1}$, $X_{c,2}$, and $X_{c,3}$ indicate the positions of the fixed points.
One can see the simulation results in Fig.~\ref{fig:ca-gl-oneD-case} for several sets of $X_{c,k}$.
In particular, the plots (a) and (b) show standard Carleman simulations (SCE, black dashed lines).
The SCE simulations are unstable and cannot reproduce the evolutionary CL trajectories (gray lines).
On the other hand, the PCE simulations (red dashed lines) with $\xi = 0.1$ and $\yNca = 6$ correctly model the nonlinear problems and can catch all fixed points in the system starting from different initial conditions.

The comparison between PCE and ACE techniques is demonstrated in plots (c) and (d).
In both cases, if $\xi$ is not small enough, the PCE technique can either have relatively high error or fail to converge to the correct fixed point (blue dashed line in (d)).
ACE also does not find an optimal $\xi$ if the tolerance $\epsilon_{\rm tol}$ is not sufficiently small; for example, see green line in (c) or red line in (d).
In particular, in the case with $\xi_{\rm max} = 1.0$ and $\epsilon_{\rm tol} = 10^{-4}$, i.e., the green line in (c), ACE reduces the chart radius only down to $\xi \approx 0.87$.
This results in high numerical errors.

To improve the ACE precision, one can either impose a smaller $\xi_{\rm max}$ or set a lower tolerance $\epsilon_{\rm tol}$.
For instance, if the tolerance is decreased from $10^{-4}$ to $10^{-10}$  (black line in (c)), ACE reduces $\xi$ down to $\leq 0.12$ (the chart radius varies in time for different parts of the simulated trajectory), which results in significantly lower numerical errors.

One can also increase $\Delta \xi$.
When $\Delta \xi$ is small, then two charts being compared are nearly the same size, leading to only minor differences between
the trajectories simulated within them.
Thus, a sufficiently low tolerance must be imposed to ensure that the ACE algorithm is sensitive to differences between the trajectories.
If the tolerance is not small enough, then the ACE algorithm may treat the charts with $\xi$ and $\xi - \Delta\xi$ as equivalent,
resulting in no update to $\xi$ along the trajectory.
By increasing $\Delta\xi$, the difference between the two trajectories becomes more noticeable, allowing a higher tolerance to be used without reducing the maximum allowable convergence radius $\xi_{\rm max}$.
This effect is demonstrated in the numerical simulations presented below.

\subsection{1D Systems:  Comparison of PCE and ACE}\label{sec:compare_methods}
In this subsection, we compare the performance of the the PCE vs ACE global Carleman methods at solving the nonlinear problem~\eqref{eq:1d}. For this test case, we consider  three fixed points: $X_{c,1} = 1.70$, $X_{c,2} = 1.65$, and $X_{c,3} = 1.55$ where $X_{c,2}$ is  unstable point, and $X_{c,1}$ and $X_{c,3}$ are stable.
The results are summarized in Fig.~\ref{fig:ACE-vs-PCE}.

If PCE is used with $\xi = 1.0$, the simulation accuracy deteriorates as the trajectory approaches regions with concentrated fixed points.
The PCE method can achieve high accuracy provided that a sufficiently small chart radius, e.g. $\xi = 0.4$,  is used.
However, representing a nonlinear problem via a linear embedding with a fixed small chart radius along the entire trajectory leads to a large number of chart shifts and may therefore not be optimal.

In contrast, the ACE method adapts the chart size locally where needed.
The red and black dashed lines in Fig.~\ref{fig:ACE-vs-PCE} show the ACE embedding, for which the initial tentative chart size is $\xi = 1.0$.
For $\ytol = 10^{-2}$, the ACE immediately reduces the embedding size from $\xi = 1.0$ to $\xi = 0.4$, achieving accuracy comparable to the PCE embedding with fixed size $\xi = 0.4$.
When the tolerance is decreased further, the ACE reduces the chart size to $\xi = 0.1$ at the beginning of the trajectory, where it changes rapidly from $X = 1.0$ to $X \approx 1.5$, and subsequently increases the chart size to $\xi = 0.2$.
This results in a high-precision embedding.

While both methods can yield high-precision linear embeddings, for trajectories that pass close to multiple fixed points, it is advantageous to vary the chart radius, $\xi$, along the path.
This prevents the degradation of precision as neighboring fixed points are approached, while reducing the number of times the chart needs to be switched along the path. Thus, the ACE technique may be the more computationally optimal approach.

\begin{figure*}[!t]
\centering
\includegraphics[width=0.90\textwidth]{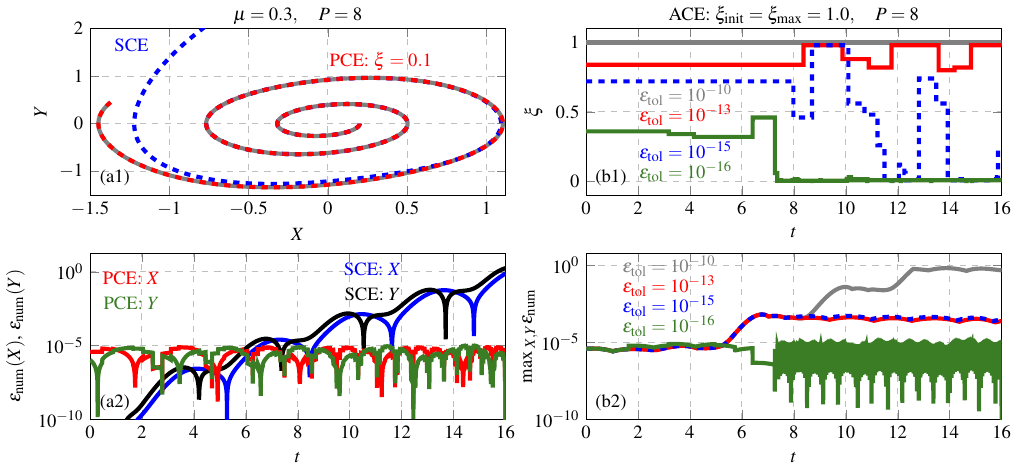}
\caption{
    \label{fig:VDP-2d} 
    Results of simulating the Van der Pol limit cycle~\eqref{eq:vdp-NORM} with the initial conditions $(X,Y) = (0.2, 0.0)$.
    (a1) SCE (blue dashed line) and PCE (red dashed line) simulations and the comparison with the CL simulation (gray line).
    (a2) Error of the SCE and PCE results.
    (b1) Evolution of the convergence radius $\xi$ in the ACE simulations under various values of the tolerance and with a fixed step size of $\Delta\xi = 0.02$.
    (b2) Maximum error of the ACE results vs. time.
}
\end{figure*}

\begin{figure*}[!t]
\centering
\includegraphics[width=0.90\textwidth]{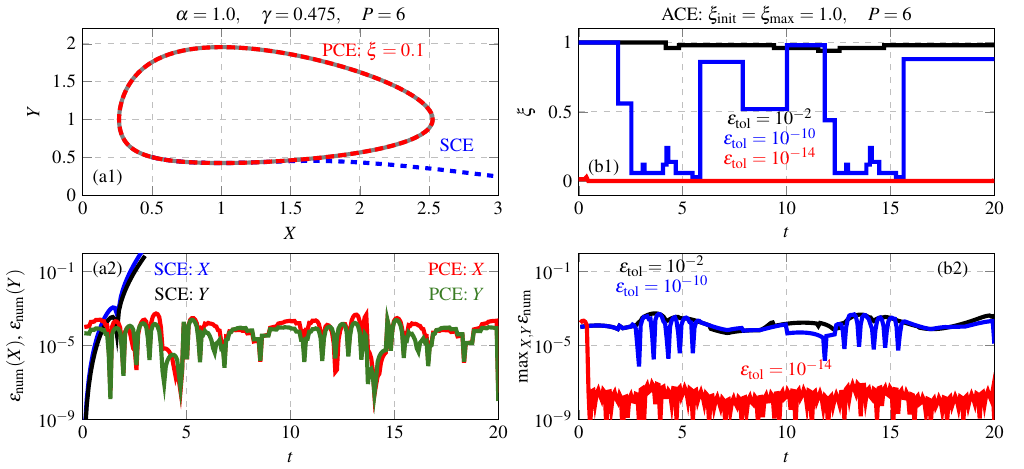}
\caption{
    \label{fig:LV-2d} 
    Results of simulating the Lotka-Volterra predator-prey model~\eqref{eq:LV2D-NORM} with the initial conditions $(X,Y) = (0.5, 0.5)$.
    (a1): SCE (blue dashed line) and PCE (red dashed line) simulations compared to the CL simulation (gray line).
    (a2): Error of the SCE and PCE results.
    (b1): Evolution of the radius $\xi$ in the ACE simulations with various values of the tolerance and with a fixed step size of $\Delta\xi = 0.02$.
    (b2): Maximum error of the ACE results vs. time.
}
\end{figure*}

\subsection{2D Limit Cycles: the Van der Pol and Lotka-Volterra predator-prey models}\label{sec:vdp-lv}

We now test the PCE and ACE techniques on two-dimensional nonlinear systems, including the Van der Pol limit cycle and the Lotka-Volterra predator-prey model.
To simulate the  Van der Pol limit cycle,
\begin{subequations}\label{eq:vdp-ORIG}
\begin{eqnarray}
    &&\diff_t X = Y,\\
    &&\diff_t Y = \mu (1 - X^2) Y - X,
\end{eqnarray}
\end{subequations}
we introduce new variables
\begin{subequations}\label{eq:xy-shift}
\begin{eqnarray}
    &&x = X - \xi_x,\\
    &&y = Y - \xi_y.
\end{eqnarray}
\end{subequations}
After substituting Eqs.~\eqref{eq:xy-shift} into the system~\eqref{eq:vdp-ORIG}, one obtains the following system of equations for the variables $x$ and $y$:
\begin{subequations}\label{eq:vdp-NORM}
\begin{eqnarray}
    \diff_t x &=& y + \xi_y,\\
    \diff_t y &=& (-\xi_x + \mu\xi_y - \mu \xi_x^2\xi_y) - (1 + 2\mu\xi_x\xi_y)x \nonumber\\
        && + \mu(1 - \xi_x^2)y - \mu\xi_y x^2 - 2\xi_x\mu x y  - \mu x^2 y.
\end{eqnarray}
\end{subequations}
The corresponding elements of the matrices $F_p$, defined in Eq.~\eqref{eq:nl}, are
\begin{subequations}\label{eq:vdp-F}
\begin{eqnarray}
    &&F_{0, 0} = \xi_y,\quad F_{0, 1} = -\xi_x + \mu\xi_y - \mu \xi_x^2\xi_y,\\
    &&F_{1, 01} = 1,\\
    &&F_{1, 10} = - (1 + 2\mu\xi_x\xi_y),\quad F_{1,11} = \mu(1 - \xi_x^2),\\
    &&F_{2, 10} = - \mu\xi_y,\quad F_{2, 11} = - 2\xi_x\mu,\\
    &&F_{3, 11} = - \mu,
\end{eqnarray}
\end{subequations}
where $F_{p,rc}$ denotes the element in the $r$-th row and $c$-th column of the $p$-th matrix $F_p$, and $F_0$ is a vector.
The variables $x$ and $y$ are considered local within a particular chart.
Thus, their values are restricted to the interval $[-\xi,\xi]$. 
Both variables are initialized to zero.
Once $x^2 + y^2 \geq \xi^2$, the embedding chart is shifted, as illustrated in Fig.~\ref{fig:scheme-PCE}.
The values of $x$ and $y$ taken immediately before the moment when the condition~\eqref{eq:emb-condition} is violated correspond to $x_{\xi}^{(0)}$ and $x_{\xi}^{(1)}$, respectively, according to the notation used in Fig.~\ref{fig:scheme-PCE}.
The coefficients $\xi_x$ and $\xi_y$ accumulate the shifts in the linearization charts, $\Xi \equiv (\xi_x, \xi_y)$.
More precisely, $\xi_x = \xi_y = 0$ for the initial linearization, i.e., within the initial chart $C_0$, Fig.~\ref{fig:scheme-PCE}.
When the chart shifts from $C_0$ to $C_1$, the values $x_{\xi,0}^{(0)}$ and $x_{\xi,0}^{(1)}$ are added to $\xi_x$ and $\xi_y$, respectively.
Then, $x$ and $y$ are reset to zero, the embedding~\eqref{eq:carleman-emb} is reconstructed using matrices~\eqref{eq:vdp-F} with the updated values of the coefficients $\xi_x$ and $\xi_y$, and the system of linear equations~\eqref{eq:carleman-emb} is evolved within the newly constructed chart $C_1$.
After $M$ chart shifts, the vector $\Xi$ is 
\begin{equation}
    \Xi \equiv (\xi_x, \xi_y) = \left(\sum_{k = 0}^{M-1}x_{\xi,k}^{(0)}, \sum_{k = 0}^{M-1}x_{\xi,k}^{(1)}\right).
\end{equation}
The vector $\Xi$ is used to reconstruct the original variables $X$ and $Y$ using Eq.~\eqref{eq:x-reconstruction}.

For $\mu>0$, the Van der Pol equations generate a limit cycle because both the origin $X=Y=0$ and the point at infinity are unstable fixed points.
The results of simulating
the Van der Pol limit cycle are presented in Fig.~\ref{fig:VDP-2d}.
The standard Carleman embedding (blue dashed line in (a1)) accurately reproduces the nonlinear dynamics for small values of $X$ and $Y$, but  becomes unstable as the trajectory grows.
In contrast, the PCE model (red dashed line in (a1)) successfully captures the full limit cycle dynamics.
Because this system is integrable, we can understand the lack of stability through the approach of Appendix~\ref{sec:higher-dflows}.
The problem is that there are two unstable fixed points within the Carleman linearization region, and, hence, the Carleman convergence radius is smaller than the radius of the trajectory.
By using a linearization radius that is sufficiently small, here $\xi=0.1$, the unstable fixed points can be avoided by the linearization chart.

We also applied the ACE technique to model the limit cycle under various tolerance values $\epsilon_{\rm tol}$, using a fixed step size of $\Delta \xi = 0.02$, as shown in plots (b).
It can be observed that lowering the tolerance prompts the ACE method to adjust the convergence radius more actively, resulting in reduced numerical errors.
However, this increased sensitivity comes at the cost of greater computational complexity, making the ACE simulations more demanding as the tolerance is lowered.

\begin{figure*}[!t]
\centering
\includegraphics[width=0.90\textwidth]{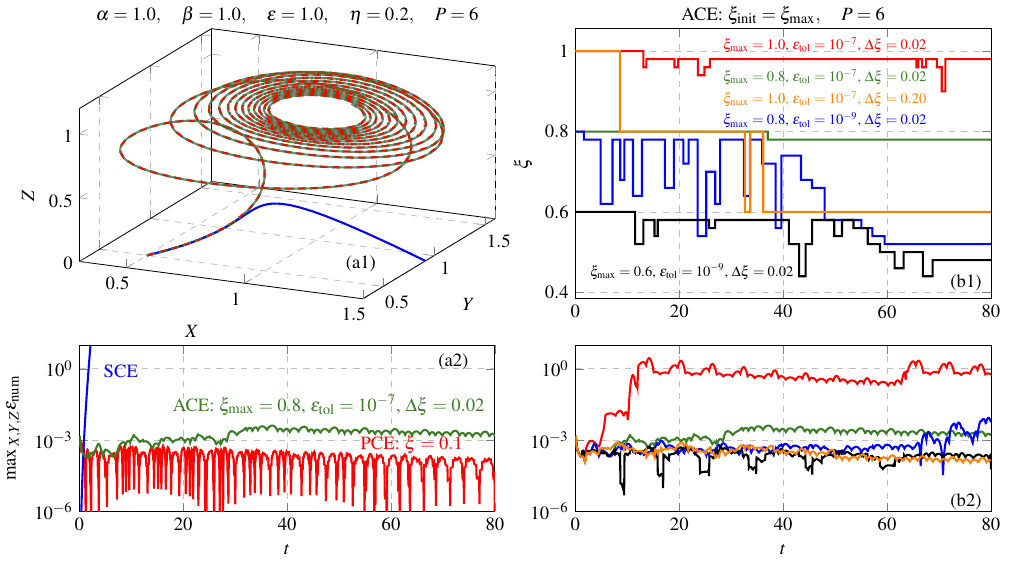}
\caption{
    \label{fig:LV-3D} 
    (a1): Simulations of the Lotka-Volterra model~\eqref{eq:LV3D-NORM} with three variables using CL (gray line, hidden behind red and green lines), SCE (blue line), PCE (red dashed line), and ACE (green dashed line).
    The initial conditions are $(X, Y, Z) = (0.5, 0.5, 0.0)$.
    (a2): Maximum numerical errors in the PCE and ACE simulations.
    (b1) and (b2): Evolution of $\xi$ and maximum numerical errors in the ACE simulations for various values of the tolerance, maximum convergence radius, and step size $\Delta\xi$.
}
\end{figure*}

Another interesting problem with two degrees of freedom is the Lotka-Volterra predator-prey model
\begin{subequations}\label{eq:LV2D-ORIG}
\begin{eqnarray}
    &&\diff_t X =   \alpha X - \beta X Y,\\
    &&\diff_t Y = - \gamma Y + \delta X Y.
\end{eqnarray}
\end{subequations}
For $\alpha,\gamma>0$, there is a hyperbolic fixed point at $X=Y=0$ and a neutrally stable fixed point at $X=\gamma/\delta$, $Y=\alpha/\beta$.
Here, we normalize the variables $X$ and $Y$ to $\gamma/\delta$ and $\alpha/\beta$, correspondingly, and introduce the shifts~\eqref{eq:xy-shift}.
The resulting system becomes
\begin{subequations}\label{eq:LV2D-NORM}
\begin{eqnarray}
    \diff_t x &=& (\alpha \xi_x - \alpha \xi_x \xi_y)  + (\alpha - \alpha \xi_y) x \nonumber\\
            && - \alpha \xi_x y - \alpha x y,\\
    \diff_t y &=& (-\gamma \xi_y + \gamma \xi_x \xi_y) + \gamma \xi_y x \nonumber\\
            && + (-\gamma + \gamma \xi_x) y + \gamma x y.
\end{eqnarray}
\end{subequations}
According to system~\eqref{eq:LV2D-NORM}, the corresponding elements of the matrices $F_p$, introduced in Eq.~\eqref{eq:nl}, are
\begin{subequations}\label{eq:LV2D-F}
\begin{eqnarray}
    &F_{0,0} = \alpha \xi_x - \alpha \xi_x \xi_y,\quad 
        &F_{0,1} = -\gamma \xi_y + \gamma \xi_x \xi_y,\\
    &F_{1, 00} = \alpha - \alpha \xi_y,\quad   &F_{1, 10} = \gamma \xi_y,\\
    &F_{1, 01} = - \alpha \xi_x,\quad          &F_{1, 11} = -\gamma + \gamma \xi_x,\\
    &F_{2, 01} = - \alpha,\quad                &F_{2, 11} = \gamma.
\end{eqnarray}
\end{subequations}

The comparison between the classical and Carleman simulations of the system~\eqref{eq:LV2D-NORM} with parameters $\alpha=1$ and $\gamma=0.475$ is shown in Fig.~\ref{fig:LV-2d}.
Again, the challenge of this model is that the hyperbolic fixed point at $X=Y=0$ is closer to the trajectory than the radius of the trajectory itself, which makes it impossible to cover this region with a single Carleman linearization chart and still retain convergence.
One can see that, while the SCE simulation is unstable, the PCE and ACE techniques accurately reproduce the nonlinear dynamics of the predatory-prey model. 
The tradeoff of the how the choice of various parameters, such as the tolerance, affect the accuracy and complexity of the globalized algorithm are explored in the subpanels of the figure.  
If the error tolerance in the ACE simulation is not sufficiently small (black lines in (b)), ACE keeps the chart size close to the maximum possible value.
Contrary to the case of the Van der Pol limit cycle, this does not result in large numerical errors.
The ACE method with $\epsilon_{\rm tol} = 10^{-10}$ (blue lines) varies the linearization size $\xi$ over a broad range of values: in some parts of the trajectory $\xi < 0.1$, while in others $\xi$ is close to one. 
Nevertheless, the resulting numerical error in the ACE simulation is similar to that of the PCE simulation with $\xi = 0.1$.
The ACE case with an even smaller tolerance (red lines) appears to overfit the nonlinear problem, resulting in much smaller but strongly oscillatory numerical errors. 
The same behavior is observed in the Van der Pol case with the tolerance $\epsilon_{\rm tol} = 10^{-16}$ (green lines in Fig.~\ref{fig:VDP-2d}b).

\subsection{The Lotka-Volterra superpredator-predator-prey model}
\begin{figure*}[!t]
\centering
\includegraphics[width=0.90\textwidth]{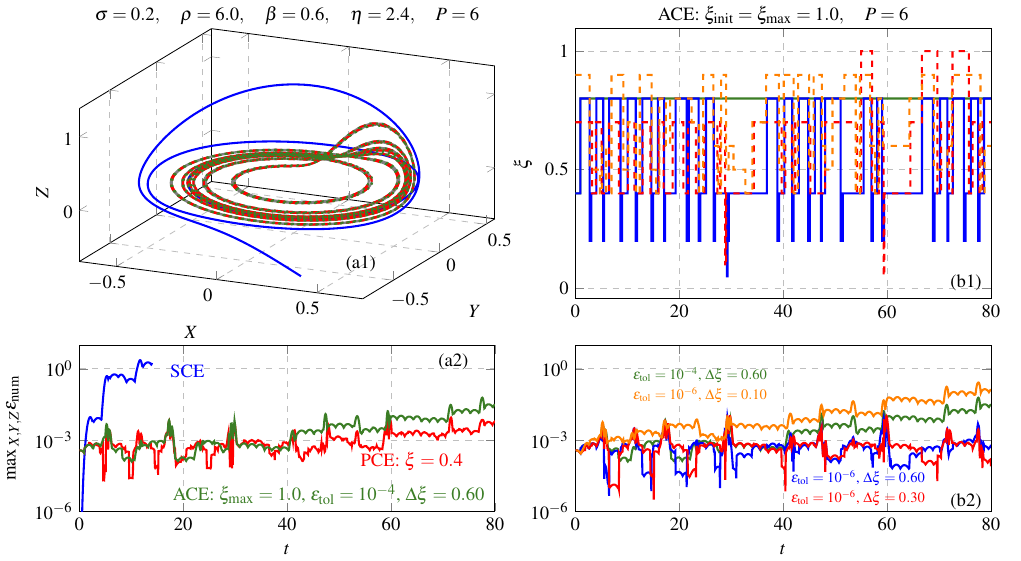}
\caption{
    \label{fig:RoA_3D} 
    (a1) Simulations of the R\"ossler attractor~\eqref{eq:Rossler-NORM} using CL (gray line), SCE (blue line), PCE (red dashed line), and ACE (green dashed line).
    The initial conditions are $(X, Y, Z) = (0.0, 0.4, 0.0)$.
    (a2): Maximum numerical errors in the PCE and ACE simulations.
    (b1) and (b2): Evolution of $\xi$ and maximum numerical errors in the ACE simulations for various values of the tolerance and step size $\Delta\xi$.
}
\end{figure*}

\begin{figure*}[!t]
\centering
\includegraphics[width=0.90\textwidth]{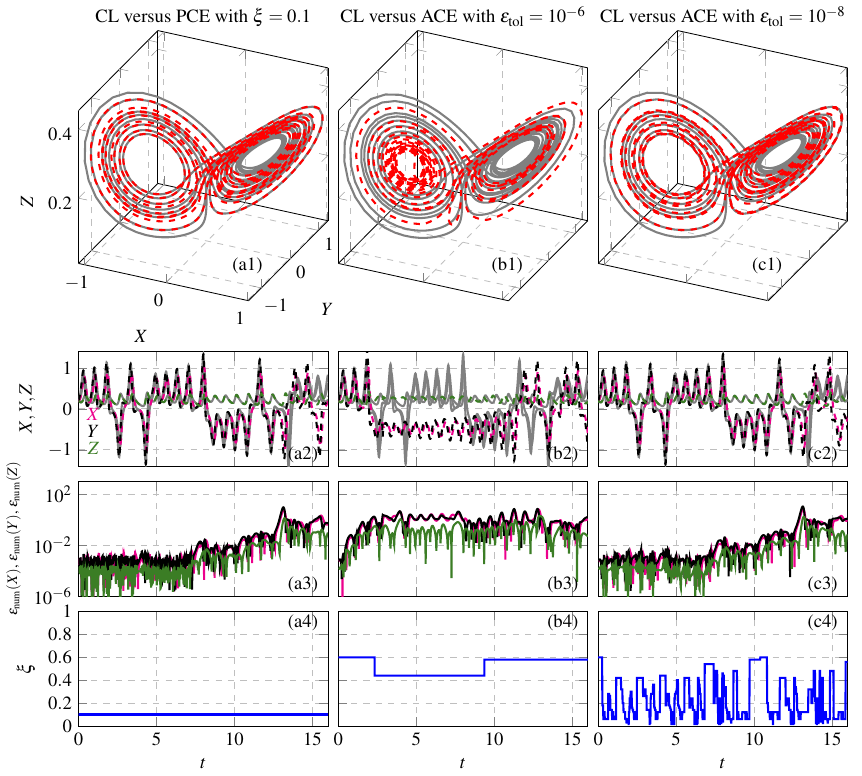}
\caption{
    \label{fig:Lorenz} 
    Simulations of the Lorenz strange attractor~\eqref{eq:LA-NORM} with $\sigma = 10$, $\rho = 28$, $\beta = 8/3$, and $c_z = 2 c_x = 4$. The initial conditions are $(X, Y, Z) = (0.2, 0.2, 0.2)$.
    The gray lines are CL simulations. The colored lines are PCE (a) or ACE (b,c) simulations.
    All Carleman simulations used $\yNca = 6$.
    All ACE simulations used $\xi_{\rm init} = \xi_{\rm max} = 0.6$ and $\Delta\xi = 0.02$.
    The bottom row presents the temporal evolution of $\xi$.
}
\end{figure*}
\begin{figure*}[!t]
\centering
\includegraphics[width=0.90\textwidth]{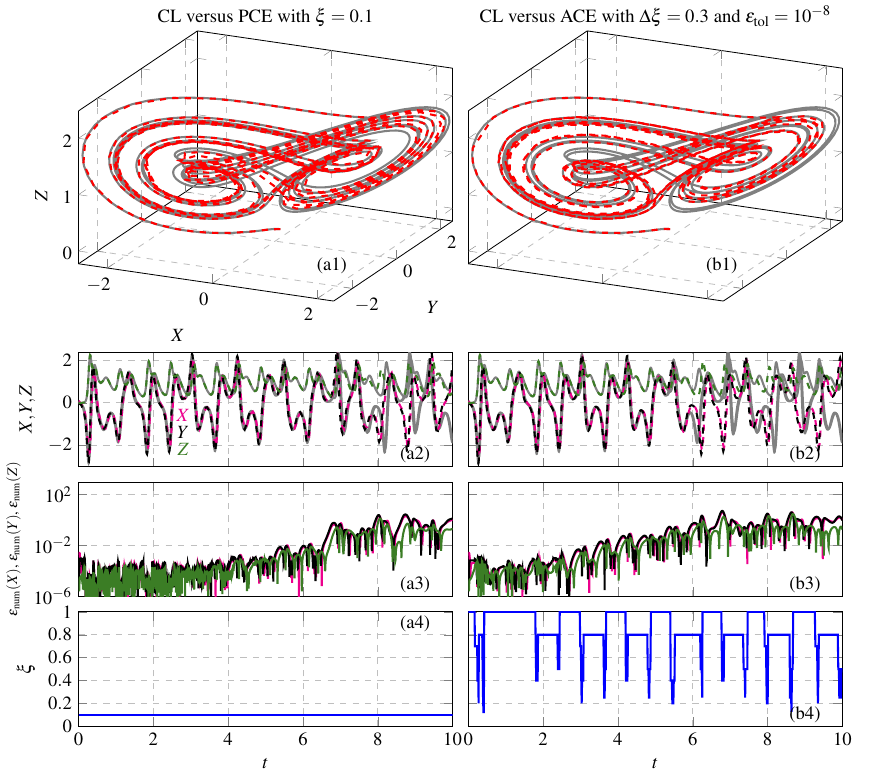}
\caption{
    \label{fig:Chen} 
    Simulations of the Chen strange attractor~\eqref{eq:Chen-NORM} with 
    $\sigma = 40$, $\rho = 28$, and $\beta = 6$.
    The initial conditions are $(X, Y, Z) = (0.1, 0.0, 0.0)$.
    The gray lines are CL simulations. The colored lines are PCE (a) or ACE (b) simulations.
    All Carleman simulations used $\yNca = 6$.
    The ACE simulation used $\xi_{\rm init} = \xi_{\rm max} = 1.0$.
    The bottom row presents the temporal evolution of $\xi$.
}
\end{figure*}

To test the PCE and ACE techniques on more complicated models, we now simulate systems with three degrees of freedom.
Let us consider the Lotka-Volterra superpredator-predator-prey (SPP) model with three independent variables:
\begin{subequations}\label{eq:LV3D-ORIG}
\begin{eqnarray}
    &&\diff_t X = \alpha X - \beta X Y,\\
    &&\diff_t Y = \epsilon XY - \epsilon YZ,\\
    &&\diff_t Z = -\eta Z  + \eta Y.
\end{eqnarray}
\end{subequations}
Similar to the previously considered cases, we introduce the shifts $\xi_x$, $\xi_y$, and $\xi_z$, which accumulate the displacements $x_{\xi}$ according to Eq.~\eqref{eq:gl-shift}.
The system becomes
\begin{subequations}\label{eq:LV3D-NORM}
\begin{eqnarray}
    \diff_t x &=& (\alpha\xi_x - \beta\xi_x\xi_y) + (\alpha - \beta\xi_y)x\nonumber\\
        &&- \beta \xi_x y - \beta xy,\\
    \diff_t y &=& (\epsilon \xi_x\xi_y - \epsilon \xi_y \xi_z) + \epsilon \xi_y x \nonumber\\
        && + (\epsilon \xi_x - \epsilon\xi_z) y - \epsilon\xi_y z\nonumber\\
        && + \epsilon xy - \epsilon yz,\\
    \diff_t z &=& \left(-\eta \xi_z + \eta\xi_y \right) + \eta y - \eta z.
\end{eqnarray}
\end{subequations}

Results from simulating the three-dimensional Lotka-Volterra model with parameters $\alpha=1$, $\beta=1$, $\epsilon=1$, and $\eta=0.2$ are presented in Fig.~\ref{fig:LV-3D}.
For these parameters, this example has a hyperbolic fixed point at $X=Y=Z=0$ and a stable fixed point at $X=Y=Z=\alpha/\beta=1$.
As can be seen from the figure, the orbit cycles towards the stable fixed point, but the initial condition is too far away for the standard Carleman embedding (SCE) to handle.
SCE
(blue line) becomes unstable very quickly, while the PCE technique with $\xi = 0.1$ successfully reproduces the nonlinear dynamics of the SPP model.
Figure~\ref{fig:LV-3D} also demonstrates that the ACE simulations with a step size of $\Delta\xi = 0.02$ require $\xi_{\rm max} \leq 0.8$ to achieve the same level of precision as the PCE simulation (compare red and green lines in (b)).
However, using a larger step size, $\Delta \xi = 0.2$ (orange line), allows ACE to detect variations in the evolutionary trajectory more easily and better adapt the convergence radius $\xi$.
As a result, one can reduce the tolerance and use $\xi_{\rm max} = 1$, while maintaining precision comparable to the PCE simulations with a small $\xi$.
While the cases with $\Delta\xi = 0.02$ (red and green lines) keep the linearization size close to its maximum value, the case with $\Delta\xi = 0.2$ decreases $\xi$ from $1.0$ to $0.6$. 
A similar effect in the case with $\Delta\xi = 0.02$ can be achieved by reducing the tolerance to $\epsilon_{\rm tol} = 10^{-9}$ (blue line).
In all cases, ACE tends to decrease $\xi$ over time, which can be attributed to the decreasing radius of the trajectory, requiring smaller charts for accurate linearization of the nonlinear problem.

\subsection{The R\"ossler, Lorenz, and Chen attractors}\label{sec:attractors-3d}
We also consider chaotic systems such as the R\"ossler, Lorenz, and Chen strange attractor.
The R\"ossler attractor is described by the following system of equations: 
\begin{subequations}\label{eq:Rossler-ORIG}
\begin{eqnarray}
    &&\diff_t X = - Y - Z,\\
    &&\diff_t Y = X + \sigma Y,\\
    &&\diff_t Z = \frac{\beta}{\eta\rho} -\rho Z + \eta \rho X Z.
\end{eqnarray}
\end{subequations}
The R\"ossler attractor has two hyperbolic fixed points at 
\begin{align}
    X_\pm=- \sigma Y_\pm= \sigma Z_\pm =\left(1\pm \sqrt{1-4\beta \sigma/ \rho^2}\right)/2\eta.
\end{align}
For the parameters $\sigma=0.2$, $\rho=6$, $\beta=0.6$, $\eta=2.4$, there is a fixed point near the origin, at $X=1.39\times 10^{-3}$, $Z=-Y=6.97\times 10^{-3}$, that is strongly attracting in the $Z$-direction but weakly unstable in the $X-Y$ plane. 
There is another fixed point near $X\simeq 0.415$, $Z=-Y\simeq 2.08$, that has an unstable direction that pushes trajectories back towards the origin.

Introducing the shifts $\xi_x$, $\xi_y$, and $\xi_z$ recasts the above system of equations as
\begin{subequations}\label{eq:Rossler-NORM}
\begin{eqnarray}
    \diff_t x &=& - (\xi_y + \xi_z) - y - z,\\
    \diff_t y &=& (\xi_x + \sigma \xi_y) + x + \sigma y,\\
    \diff_t z &=& \left(\frac{\beta}{\eta\rho} - \rho\xi_z + \eta \rho \xi_x \xi_z \right)\nonumber\\
        && + \eta \rho \xi_z x + (\eta \rho\xi_x - \rho)z + \eta \rho x z.
\end{eqnarray}
\end{subequations}
Results from simulating
the R\"ossler attractor~\eqref{eq:Rossler-NORM} are shown in Fig.~\ref{fig:RoA_3D} after the rescaling procedure is applied.
The SCE technique (blue line) fails to reproduce the nonlinear dynamics 
and eventually goes unstable. Clearly, the issue is that there are only hyperbolic fixed points in this system, and, in fact, there is an unstable hyperbolic fixed point at the center of the attractor.
This means that any Carleman linearization atlas that seeks to cover the entire attractor must deal with a hyperbolic fixed point near the center of the domain.

The PCE method with a small $\xi$ and the ACE technique manage to successfully simulate
the attractor.
One can see from Fig.~\ref{fig:RoA_3D}.b2 that, similarly to the previous cases, a low step size $\Delta\xi$ can result in higher numerical errors (for example, compare orange and red lines).
In addition, ACE requires a sufficiently small tolerance for the chosen step $\Delta\xi$ to minimize the numerical error (compare the green and blue lines).
The spikes in the numerical error observed in both PCE and ACE simulations occur during the time intervals when the trajectories rapidly grow in the $z$ direction.
ACE also appears to reduce $\xi$ during these spikes (for example, compare the blue line in (b1) with that in (b2)).
This may indicate that the crest in (a1) is the most demanding part of the attractor for linearization and requires the largest number of linearization regions.

\begin{figure*}[!t]
\centering
\includegraphics[width=0.90\textwidth]{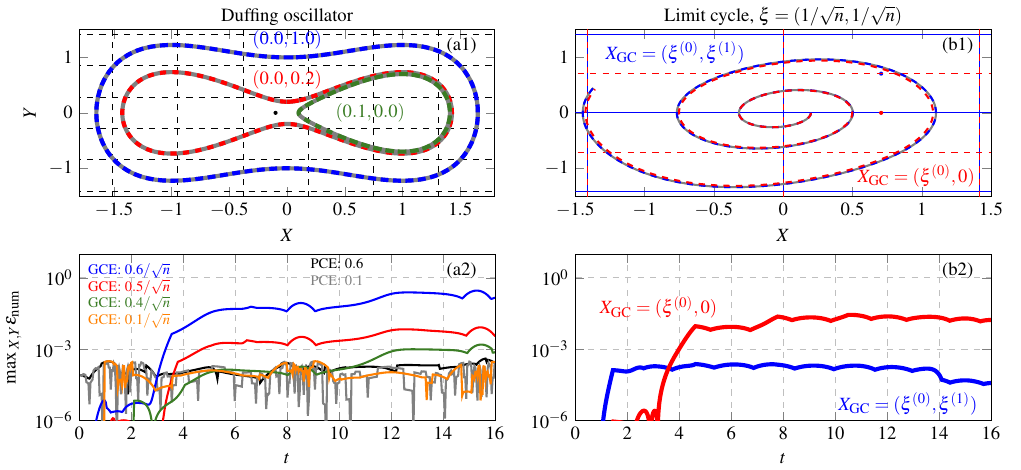}
\caption{
    \label{fig:GCE-2D} 
    (a1): Simulations of the Duffing oscillator with various initial conditions indicated by the colored text.
    The CL and GCE simulations are indicated by solid gray and dashed colored lines, respectively.
    All GCE simulations are performed in the same grid (black lines): $X_{\rm GC} = (-0.1, 0.0)$ (black dot), $\xi^{(i)} = 0.4/\sqrt{\dim}$, with $i = 0,1$.
    (a2): Numerical errors in the GCE and PCE simulations of the Duffing oscillator with the initial condition $(0.5, 0.5)$, the GCE grid center $X_{\rm GC} = (-0.1,0.0)$, and various sizes of the linearization regions (indicated by the colored text).
    (b1): Simulations of the Van der Pol limit cycle with the initial condition $(0.2, 0.0)$.
    The CL and GCE simulations are indicated by solid gray and dashed colored lines, respectively.
    The GCE simulations have the same tile sizes, $|\xi| = 1$, the grid centers are $X_{\rm GC} = \xi$ (blue dot) and $X_{\rm GC} = (\xi^{(0)}, 0)$ (red dot).
    The corresponding GCE grids are indicated by the solid blue and dashed red horizontal and vertical lines.
    (b2): Numerical errors in the GCE simulations of the limit cycle.
}
\end{figure*}
\begin{figure*}[!t]
\centering
\includegraphics[width=0.90\textwidth]{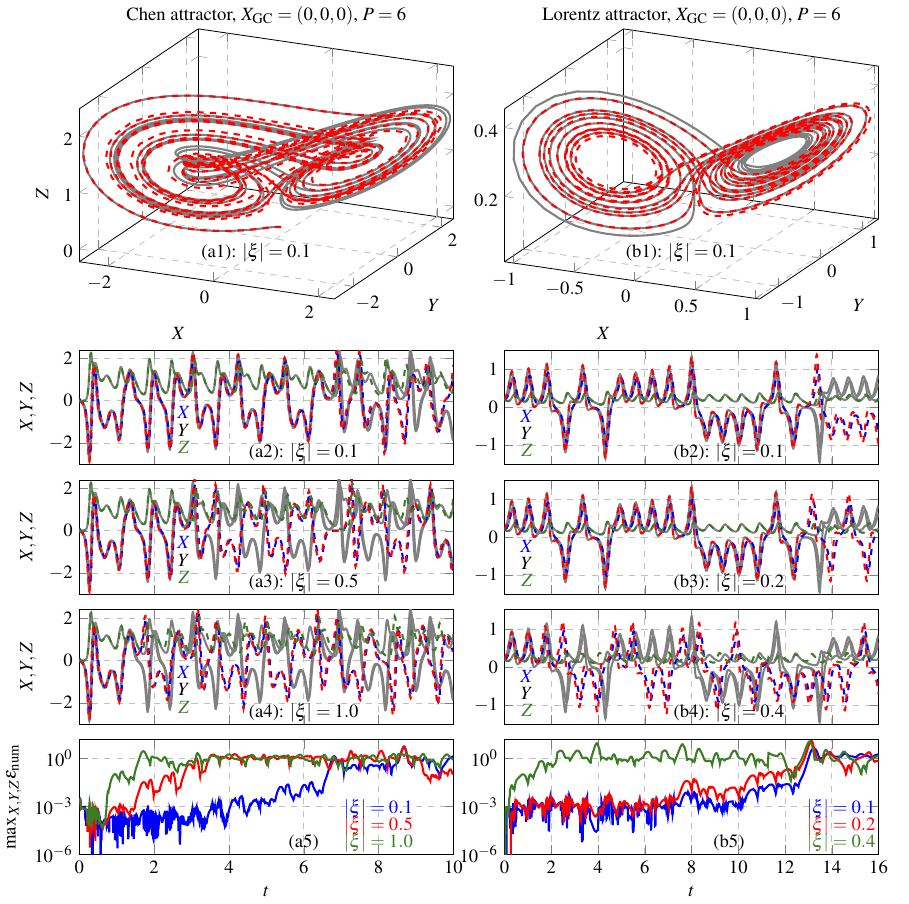}
\caption{
    \label{fig:GCE-3D} 
    (a1): Classical (gray line) and GCE (red line) simulations of the Chen strange attractor with 
    $\sigma = 40$, $\rho = 28$, $\beta = 6$, and the initial condition $(0.1, 0.0, 0.0)$.
    (a2)-(a4): Comparison of the time evolution of individual variables ($X$, $Y$, $Z$) in classical simulations (gray lines) and GCE simulations for various tile sizes $\xi$ (colored lines).   
    (a5): Numerical error in the GCE simulations of the Chen attractor.
    (b1): Simulations of the Lorenz strange attractor with $\sigma = 10$, $\rho = 28$, $\beta = 8/3$, $c_z = 2 c_x = 4$, and the initial condition $(0.2, 0.2, 0.2)$. 
    (b2)-(b4): Comparison of the time evolution of individual variables in classical simulations (gray lines) and GCE simulations for various tile sizes $\xi$ (colored lines). 
    (b5): Numerical error in the GCE simulations of the Lorenz attractor.
    In all GCE simulations shown in plots (a) and (b), all components of $\xi$ are set to $|\xi|/\sqrt{\dim}$, and the grid center is located at $(0,0,0)$.
}
\end{figure*}

The Lorenz 1963 equations are given by 
\begin{subequations}\label{eq:LA-ORIG}
\begin{eqnarray}
    &&\diff_t X = \sigma(Y - X),\\
    &&\diff_t Y = X (\rho - Z) - Y,\\
    &&\diff_t Z = XY - \beta Z.
\end{eqnarray}
\end{subequations}
This system has a fixed point at the origin that becomes unstable if $4\rho \sigma>-(\sigma-1)^2$. 
There are another two hyperbolic fixed points at the location
\begin{align}
    X_\pm=Y_\pm=\pm\sqrt{\beta Z_\pm} &&
    Z_\pm = \rho-1
\end{align}
that appear for $\rho>1$.
For the Lorenz attractor, the parameters $\sigma=10$, $\rho = 28$, and $\beta = 8/3$ are chosen, which places the pair of hyperbolic fixed points at $X_\pm=Y_\pm=\pm6\sqrt{2}$ and $Z_\pm=27$.
The Lorenz attractor has two lobes that are formed in the parameter regime where the two hyperbolic fixed points are strongly attracting along one axis, but weakly unstable in a 2D plane.
In the region where the two lobes nearly meet, it is difficult to predict whether a trajectory will continue circling the same lobe or switch to circling the other lobe.

To prepare the system of equations describing the Lorenz attractor so that they are suitable for the Carleman embedding,
we normalize the variables $X$, $Y$, and $Z$ to $\eta_{x}$, $\eta_{y}$, and $\eta_{z}$, respectively, where $\eta_x = \eta_y = c_x\sqrt{\beta(\rho - 1)}$ and $\eta_z = c_z (\rho - 1)$.
We also introduce the shifts $\xi_x$, $\xi_y$, and $\xi_z$, as was done in all previous cases.
As a result, the system is recast as:
\begin{subequations}\label{eq:LA-NORM}
\begin{eqnarray}
    \diff_t x &=& \sigma (\xi_y - \xi_x) - \sigma x + \sigma y,\\
    \diff_t y &=& (\rho \xi_x - \xi_y - \eta_z \xi_x \xi_z)  + (\rho - \eta_z \xi_z)x \nonumber\\
    &&- y - \eta_z \xi_x z - \eta_z xz,\\
    \diff_t z &=& (\eta \xi_x \xi_y - \beta \xi_z) + \eta \xi_y x + \eta \xi_x y - \beta z + \eta x y,
\end{eqnarray}
\end{subequations}
where $\eta = \eta_x^2 / \eta_z$.
The elements of the matrices $F_p$ can be derived from the above system.

We also consider
the Chen equations: 
\begin{subequations}\label{eq:Chen-ORIG}
\begin{eqnarray}
    &&\diff_t X = \sigma(Y - X),\\
    &&\diff_t Y = (\rho - \sigma) X - X Z + \rho Y,\\
    &&\diff_t Z = XY - \beta Z.
\end{eqnarray}
\end{subequations}
This system has a fixed point at the origin that can become unstable when $ 3 > \rho/2\sigma-\sqrt{3} >  -3 $ and two hyperbolic fixed points at 
\begin{align}
    X_\pm=Y_\pm=\pm\sqrt{\beta Z_\pm} && Z_\pm=2\rho-\sigma.
\end{align}
Here, the parameters $\sigma=40$, $\rho=28$, and $\beta=6$ are chosen which places the pair of fixed points at $X_\pm=Y_\pm=4\sqrt{6}\simeq \pm 9.8$ and $Z_\pm=16$.
Like the Lorenz attractor, the Chen attractor has two lobes that are formed when the two hyperbolic fixed points are attracting along one axis and repelling in the 2D plane.
However, it has different topological structure for the pathways that connect the two lobes of the attractor.

For linearization, the variables $X$ and $Y$ are normalized to $\sqrt{\beta(2\rho - \sigma)}$, and the variable $Z$ is normalized to $2\rho - \sigma$.
After introducing the shifts $\xi_x$, $\xi_y$, and $\xi_z$, we rewrite the system as
\begin{subequations}\label{eq:Chen-NORM}
\begin{eqnarray}
    \diff_t x &=& \sigma (\xi_y - \xi_x) - \sigma x + \sigma y,\\
    \diff_t y &=& [(\rho - \sigma)\xi_x + \rho \xi_y - (2\rho - \sigma) \xi_x \xi_z] \nonumber\\
        && + [(\rho - \sigma) - (2\rho - \sigma)\xi_z] x \nonumber\\
        && + \rho y - (2\rho - \sigma) \xi_x z - (2\rho - \sigma) xz,\\
    \diff_t z &=& (-\beta \xi_z + \beta \xi_x \xi_y)\nonumber\\
        && + \beta \xi_y x + \beta \xi_x y - \beta z + \beta x y.
\end{eqnarray}
\end{subequations}

The results from simulating
the Lorenz and Chen attractors are shown in Fig.~\ref{fig:Lorenz} and Fig.~\ref{fig:Chen}, respectively.
The standard embedding, not shown here, proves to be unstable for computing these systems.
In 
contrast, the PCE and ACE techniques succeed in bounding the modeling error in these chaotic systems.
For the Lorenz and Chen attrators, the Carleman method must deal with two hyperbolic fixed points that located within the central region of the attractor.
Thus, it is not possible to cover the entire attractor with a single Carleman chart that retains convergence for the entire trajectory.

We note that, due to the \emph{butterfly effect}, which is expected from the theory of chaotic dynamical systems,  for times longer than a few Lyapunov times, the numerical error using different numerical techniques becomes significant.
Since these systems have nonzero Lyapunov exponents, even small differences in initial conditions or numerical algorithms will lead to a divergence in the simulated trajectories.
The divergence  manifests not only locally, but also globally, when at a given time instant two simulated trajectories are found on different loops of the attractor, i.e. near different fixed points.
For instance, this can be observed in the ACE simulation with $\epsilon_{\rm tol} = 10^{-8}$ of the Lorenz attractor shown in Fig.~\ref{fig:Lorenz}c2 where the $X$ and $Y$ signals modeled by ACE have an opposite sign compared to the CL signals starting at $t\approx 15$.
This behavior is unavoidable, even when comparing standard methods, because differences in floating-point precision alone can cause a chaotic dynamical system to follow a different trajectory.

For this reason, it may be reasonable to primarily use the PCE technique without attempting to adapt the convergence radius when modeling chaotic systems, as such adaptation might not significantly reduce the numerical error.
However, ACE may be preferable in scenarios involving multiple fixed points, particularly when some are closely spaced and others are not, as discussed in Section~\ref{sec:ace_vs_pce}.

\subsection{Testing the GCE technique}\label{sec:test-gce}

To test the GCE method, we start from two-dimensional systems such as the Van-der Pol limit cycle considered in Sec.~\ref{sec:vdp-lv} and the Duffing oscillator.
The Duffing oscillator,
\begin{subequations}\label{eq:duffing-orig}
\begin{eqnarray}
    \diff_t x &=& y,\\
    \diff_t y &=& x - x^3,
\end{eqnarray}
\end{subequations}
has two centers at $(\pm 1, 0)$ and a saddle point at $(0,0)$. 
After introducing the shifts $\xi_x$ and $\xi_y$, one recasts the system as
\begin{subequations}\label{eq:duffing-sim}
\begin{eqnarray}
    \diff_t x &=& \xi_y + y,\\
    \diff_t y &=& (\xi_x - \xi_x^3) + (1 - 3\xi_x^2) x - 3\xi_x x^2 - x^3.
\end{eqnarray}
\end{subequations}

The results from the GCE simulations of these 2D systems are shown in Fig.~\ref{fig:GCE-2D}.
As seen from Fig.~\ref{fig:GCE-2D}a2, the GCE approach requires smaller linearization regions than the PCE method.
For instance, the GCE simulation with $|\xi| = 0.6$ has numerical error $\oO(1)$, while the PCE simulation with $\xi = 0.6$ has much smaller error around $10^{-2}$.
One possible reason for the stricter requirements on the chart size in the GCE method is that, during chart transitions, the PCE technique dynamically creates a new chart centered at the trajectory point just before it crosses the linearization boundary.
According to the analysis \onlinecite{Weber16, Sanchez25}, this procedure generates a much higher accuracy solution.
In contrast, during a chart transition, the GCE technique moves from a point near the boundary of one tile to a point near the boundary in the neighboring tile, where the error is greatest.
This can potentially lead to higher error accumulation in the GCE simulation, as the trajectory spends more time near the linearization horizon than in the PCE method.

One can also see from Figs.~\ref{fig:GCE-2D}b1 and~\ref{fig:GCE-2D}b2 that one can change the GCE accuracy by shifting the alignment of the trajectory and the center or edges of the linearization grid.
In particular, when the grid center is shifted from $X_{\rm GC} = (\xi^{(0)}, 0)$ (red grid) to $X_{\rm GC} = (\xi^{(0)}, \xi^{(1)})$ (blue grid), the precision improves by several orders of magnitude.
It is difficult to pinpoint the exact reason for this improvement.
One possible explanation is that, in the case of the grid center $X_{\rm GC} = (\xi^{(0)}, 0)$ (red grid), the trajectory of the limit cycle appears to spend more time near the top and bottom boundaries of the tiles $(0,0)$ (the central tile) and $(-1,0)$ (the tile to the left).
On the other hand, while the trajectory does spend some time near the corners of the tiles of the blue grid, it crosses the boundaries while moving rapidly in a direction that is nearly perpendicular to the grid rather than spending a long time moving parallel to the edges where the error is large.

We also test the GCE method on simulations of the Chen and Lorenz attractors for the same parameters described in Sec.~\ref{sec:attractors-3d}.
The results of the GCE simulations are shown in Fig.~\ref{fig:GCE-3D}.
The error of the GCE simulations of the Chen attractor remains bounded for the maximum allowable tile size equal to $|\xi| = 1.0$, with all components of the vector $\xi$ set equal.
Comparing Fig.~\ref{fig:Chen}a3 with Fig.~\ref{fig:GCE-3D}a5 shows that the PCE and GCE simulations of the Chen attractor exhibit similar numerical errors for the same chart size, $|\xi| = 0.1$.
Similarly, a comparison between Fig~\ref{fig:Lorenz}a3 and Fig.~\ref{fig:GCE-3D}b5 reveals that the PCE and GCE simulations of the Lorenz attractor yield comparable numerical errors at the same chart size $|\xi| = 0.1$.

The Lorenz attractor appears to be a more challenging nonlinear system to model using Carleman linearization, as it requires smaller tiles than those used in the GCE simulations of the Chen attractor.
Although the numerical error remains bounded even for $|\xi| = 1.0$, the trajectories of the Lorenz attractor simulated with such a GCE grid differ drastically from the classical trajectories for the chosen initial conditions.
\section{Complexity of Global Carleman Embedding Methods}
\label{sec:complexity}

\subsection{Global Classical Carleman Algorithms}
\label{sec:classical-global-carleman}

In this section, we consider the cost of the globalized Carleman embeddings in greater detail.
Note, however, that the relatively low-dimensional ODE examples that we study in this work generated matrices that were small enough that we did not need to focus on optimizing the classical complexity. 
In Appendix~\ref{sec:complexity-remap}, we consider the cost of the remapping operation that requires shifting and rescaling the initial conditions and the velocity matrices.
We refer to this remapping operation as a \textit{diagonal affine mapping} because it has the form $x'=S^{-1}x+b$, where the scaling matrix $S^{-1}$ is required to be diagonal (and non-singular).
In Appendix~\ref{sec:homogeneous}, we discuss the classical complexity associated with using a homogeneous polynomial basis.
Finally, in Section~\ref{sec:quantum-global-carleman}, we discuss the complexity associated with potential quantum algorithm formulations.

First, we note that there are a few different factors that control the complexity of the Carleman embedding method.
One important factor is how well the equations of motion (EOM) are represented by the polynomial basis.
In the simplest situations, the equations of motion might be known analytically and might already be given in the form of polynomials or trigonometric series, in which case there is a natural associated function library.
However, other situations might require methods for polynomial approximation such as interpolation or projection, which could increase the cost over the estimates given below.
Hence, in this work, we assume that the form of $\Velocity_k$ is  known in advance and can be queried efficiently.

Next, the cost of testing for convergence of the solution and for switching between charts can be estimated from a combination of theoretical and empirical resuts.
There are several factors that enter the cost: the cost of the logical tests and the number of times that they are performed, the cost of switching to a different chart and the number of times that a switch is performed, and, for ACE, the computational overhead required to evolve multiple trajectories simultaneously.
Classically, the number of logical tests for the radius is $\oO(1)$ for PCE and ACE and $\oO(\dim)$ for GCE, while the convergence test for ACE is $\oO(\dim)$.
If the radius can be extracted directly from the data without computing the distance using components, e.g.  by either evolving or directly extracting the radius data, then the computational cost can be kept to $\oO(1)$ for the radius test.

However, as shown in Section~\ref{sec:complexity-remap}, the cost of switching the trajectory to another chart is still $\oO(\dim)$.
If optimized,  the cost of allowing an arbitrary remapping of the Carleman matrix has a constant factor $\oO(5\yNnl)\times$ overhead over that of calculating it once and, hence, scales similarly to the cost of performing a single time step.
Thus, as long as the number of logical testing and chart switching operations is a fraction of the total number of time steps, the overhead is manageable
Note that, for GCE, the Carleman matrices can be calculated before run time, so that this cost can then be amortized over the total number of trajectories.

In our actual implementation, we did not attempt to perform these classical calculations in an optimal manner.
For example, we perform logical tests on every timestep and we did not implement the recursive construction of the $A$ and $B$ Carleman matrix elements in Appendix~\ref{sec:multidim-remap}.
In practice, we found that the required frequency of switching per time step is typically on the order of  $f\sim $ 0.01 (Van der Pol) -- 0.1 (Chen attractor).
While the time it takes to reconstruct the Carleman  matrix elements was $m=30\times$ the time to take a single timestep, we believe this can be reduced by using the recursive form in Section~\ref{sec:multidim-remap}. 
Yet, this still only leads to an overall constant factor  $\oO(5\yNnl)$ cost increase of order 0.3-3 relative to SCE, for a total cost of $1.3-4\times$ the cost of SCE.

For our current implementation of ACE, the need to simultaneously follow two trajectories and  to try additional trajectories during the testing stage multiplies the total cost by a constant factor of order $2-4\times$, depending on how many trajectories need to be compared.
However, ACE can reduce the switching frequency by $3-10\times$ or more, so that, in principle, the total cost can be less that of PCE.
In practice, we find that at low precision, ACE can outperform PCE, but as the precision is increased, it eventually becomes more expensive.
Once the tolerance becomes small enough, ACE must use a radius that is as small as PCE/GCE, and, hence, it becomes roughly $2-4\times$ more expensive because there is no longer potential for speedup.
We believe that, in the future, ACE can be improved by determination of the radius of convergence found directly from the convergence of the series via the Cauchy-Hadamard Theorem.~\cite{Shabat1992book}
This could reduce the costs associated with the ACE algorithm significantly, making it competitive with PCE at high precision.

\subsection{On Global Quantum Carleman Algorithms}
\label{sec:quantum-global-carleman}
A complete understanding of the complexity of a quantum implementation of the global Carleman algorithms is beyond the scope of this work.
Yet, the fact that a single Carleman linearization chart is not sufficient to cover the phase space implies that nonlinear operations are always required for the Carleman approach to faithfully simulate the trajectories of a generic nonlinear dynamical system.
For the global Carleman methods, the nonlinearity is only ever introduced at the remapping stage when the trajectory crosses the boundary between charts.
From this perspective, it is clear that a fully quantum (non-hybrid) implementation of the present version of the global Carleman algorithms would require exponential computational cost due to the nonlinear operations at the remapping stage.
This implies that any quantum algorithm for nonlinear dynamics that claims to achieve an exponential speedup can only do so for special classes of nonlinear dynamical systems, e.g. attracted to a single stable fixed point,  neutrally stable limit cycle, completely integrable systems, etc.
\textit{\textbf{This represents another fundamental no-go theorem~\cite{Liu21,Brustle2025quantum,Ameri2026quantum} on the ability of quantum computers to speed up the simulation of the individual trajectories of a generic nonlinear dynamical system.}}

The standard Carleman embedding is well suited for methods that exploit linear systems well, such as quantum computing and tensor networks, as it only requires solving a linear initial-value problem~\eqref{eq:carleman-emb}. 
For SCE, amplitude encoding can be used to achieve an exponential speedup in memory and in operations per time step. 
For low-dimensional ODEs, the case $\oO(\yNca)\gg \oO(\dim)\sim \oO(\yNnl)\sim \oO(1)$ is of interest.
The classical tensor product encoding would require $\oO(\Npoly^\dim)\geq \Binomial^{\Npoly+\dim}_{\Npoly}$ terms.
If an exponentially efficient block encoding of the velocity, $\Velocity(x)$, can be found, then a quantum computer could use amplitude encoding to reduce the required storage to  $\oO(\dim\log(\Npoly))$.
For PDEs or high-dimensional ODEs, where $\oO(\dim)\gg\oO(\yNca)\gg\oO(\yNnl)\sim\oO(1)$, the number of independent terms becomes $\oO((\dim/e)^{\Npoly}/\Npoly!)$.
While a classical tensor product encoding would require $\oO( \dim^{\Npoly})\geq \Binomial^{\dim+\Npoly}_{\dim}$ terms, if an efficient block encoding exists, then a quantum computer could use amplitude encoding to compress the required memory exponentially to $\oO(\log(\dim^{\Npoly}))=\oO(\Npoly\log(\dim))$.
Note, however, that, in order to achieve the Courant-Friedrich-Lewy (CFL) stability/accuracy criterion for stable and accurate time integration, the time step, $\Delta t<\tau/\yNca$, must scale inversely with the number of polynomials, $\yNca$.
For ODEs where $\yNca\gg\dim$, this reduces the overall quantum speedup of the Carleman method to polynomial at best.
For PDEs, where $\dim \gg\yNca$, one must consider the question of how $\dim$ and  $\yNca$ scale with $1/h$, where $h$ is the grid spacing. 
In the PDE case, $\dim\sim 1/h^d$ where $d$ is the spatial dimension.
As explained in Appendix~\ref{sec:higher-dflows}, $\yNca$ must be chosen to accurately capture the dependence of $x(t,x_0)$.
Hence, if this requires the resolution necessary to capture small radial variations, $\yNca\sim 1/h$, then, once again, the CFL condition implies that the overall quantum speedup will be polynomial at best.

For global quantum Carleman algorithms, the key challenge lies in the fact that the position of the orbit, $x(t)$, must be ultimately be read out and copied from the previous linearization chart in order to determine the starting point in the next linearization chart and perform the required remapping, and this requires multiple nonlinear operations. 
Moreover, the globalized Carleman methods  require  careful detection of  
boundary crossings  between charts. 
This requires various logical tests for the position  and, potentially, for convergence along the way.
Many of these tests require nonlinear operations.
This tends to imply that the dynamic adaptive methods are less suitable for a fully quantum implementation.
Nonetheless, a complete understanding of which approaches are optimal and a complete accounting of their associated costs has yet to be developed.

If a fully quantum algorithm wants to reduce the exponential amount of required memory and/or work, then one strategy is to reduce the number of nonlinear operations as much as possible. 
This can potentially be achieved if the number of tests and the number of chart-changing operations is kept sufficiently small.
This tends to imply that the static grid-based GCE method is best, since the Carleman matrices are precomputed in advance and the logical tests are linear.
Unfortunately, the exponential growth in required memory would restrict the quantum algorithm to a relatively small number of chart changes.

Another strategy is to recognize that all the nonlinear operations can be performed efficiently if the position at the chart boundary is first read out to generate a single copy that is stored in the form of a tensor product of bitstrings.
The most efficient way to do this is using amplitude estimation.
Since these states are classical tensor product states that can be copied efficiently, the subsequent nonlinear operations are benign classical operations that do not incur costs that grow exponentially in time.

This approach is equivalent to implementing the globalized Carleman algorithm as a hybrid classical-quantum algorithm where the classical computer performs the nonlinear work required for remapping at the boundaries.
The strategy is to solve the system~\eqref{eq:carleman-emb} on a quantum computer for a fixed time interval, then measure the local variables $x$ and classically analyze whether the system remains within the initial linearization chart.
If this is the case, a new chart is constructed at the measured point, and quantum Hamiltonian simulation is performed within this newly created chart for the next fixed time interval, after which the process repeats.
If the measured variables fall outside the chart, the results are discarded, and the system~\eqref{eq:carleman-emb} is solved within the same chart over a shorter time interval. 

For a hybrid algorithm, the primary limiting factor for achieving speedup will be the data exchange between the classical and quantum registers.
However, the only data that needs to be exchanged during each chart shift are the displacements $x_{\xi}$ or the components of the local $\dim$-dimensional vector $x$. This is always much smaller than size of the full Carleman vector 
$\yNu=\Binomial^{\dim+\yNca}_\dim$.
Thus, the relatively small amount of data exchange required 
is potentially compatible with the constraints of an efficient hybrid quantum–classical algorithm.
In the hybrid setting, PCE may be preferable over GCE because it tends to be more accurate. 
Yet, ACE will be preferable over PCE if it can significantly reduce the number of chart-changes, and, hence, the number of times that data must be transferred between classical and quantum registers.

\section{Conclusion}\label{sec:conclusion}

We have demonstrated that the Carleman embedding can be globalized to significantly enhance the accuracy and bound the maximum error when simulating nonlinear systems with multiple fixed points, even in cases with chaotic behavior. 
A common problem with the nonlinear dynamical systems for which the standard Carleman embedding fails is that there are multiple typically hyperbolic or unstable fixed points that have a smaller distance to the trajectory than the overall size of the trajectory.
By partitioning the state space into multiple linearization charts, we successfully stabilize the Carleman method for many cases that are otherwise intractable for the standard Carleman linearization.
This resolves the issue that the radius of convergence of the Carleman solution is smaller than the radius of the orbit (see Appendix~\ref{sec:higher-dflows}).

To avoid large errors in cases where multiple nearby fixed points appear within the same linearization chart, we designed an adaptive technique that dynamically adjusts the chart radius $\xi$ by analyzing the nonlinear terms in the Carleman embedding and by comparing two embeddings with different convergence radii.
This technique either reduces the convergence radius to minimize numerical error or enlarges the embedding charts to accelerate simulation speed. 
This adaptive adjustment enables the identification of more optimal linearization chart sizes for different segments of the simulated trajectory. 
However, numerical tests indicate that, in the case of chaotic systems, the non-adaptive technique with a small fixed linearization radius 
is faster at very high precision, particularly when prior knowledge of the minimal required distance between fixed points in the simulated system is available.

The adaptive methods do not require prior knowledge of the locations of the fixed points, but they do require careful detection of the times when it is necessary to switch between linearization charts. 
The accuracy of the method depends not only on the truncation of the maximum monomial order, but also on the imposed convergence radius of the linearization regions in the non-adaptive version and on the tolerance specified in the adaptive version. 
Through various numerical simulations, we have demonstrated that the numerical error in this globalized Carleman approach remains bounded even for chaotic dynamical systems, including strange attractors of various types.

The fact that the global Carleman embedding methods work well can be interpreted as a proof that some nonlinear operations are always required to accurately simulate individual trajectories.
Unfortunately, quantum algorithms for the global Carleman embedding methods lose their potential for exponential speedup in compression and work during the remapping between charts.
This is because changing charts requires decompressing the amplitude encoding to classical bitstrings, copying the data, and then recompressing.
Thus, the exponential gains in compression and work that the standard quantum Carleman algorithms promise for simple integrable dynamical systems are  lost at the shift and scale remapping stage.

In contrast, the KvN method~\cite{Joseph20, Joseph23} already provides a global embedding of nonlinear dynamics, for which a highly efficient quantum algorithm has already been provided.~\cite{Novikau24KvN,Novikau25Opt}
Yet, the success 
of the global Carleman methods motivates the exploration of finite element and spectral element methods for the KvN approach to generating a linear embedding of the Koopman and Perron-Frobenius evolution operators.

\section*{Acknowledgments}
We would like to thank Matthew Kvalheim, Phillip Morrison, and Hong Qin for discussions regarding the minimal number of charts needed to linearize a vector field on a manifold.
We would like to thank Matteo Lostaglio and Yi\u{g}it Suba\c{s}i for pointing out that the minimal size of the Carleman vector is equal to the number of distinct homogeneous polynomials up to a given order.
We would like to thank Ala Shayegi for discussions leading to the understanding that the need to copy the location of the transition point in order to change linearization charts is the key bottleneck for quantum computation.
This work, LLNL-JRNL-2010809, was supported by the U.S. Department of Energy (DOE) Office of Fusion Energy Sciences “Quantum Leap for Fusion Energy Sciences” Project No. DE-SCL0000070 (FWP-SCW1680) at Lawrence Livermore National Laboratory (LLNL).
Work was performed under the auspices of the U.S. DOE under LLNL Contract DE-AC52–07NA27344.

\appendix
\def\pj{{\nu_j}}
\def\pinfty{{\nu_\infty}}
\def\Velocity{F}
\section{
Carleman embedding stability \& convergence for autonomous 1D flows }
\label{sec:1dflows}

Consider an autonomous 1D vector field $\Velocity(x)$ that can be expressed as a $\yNnl$-th order polynomial in $x$.
The Carleman procedure constructs the coordinate transformation $x(t,x_0)$ as a power series in $x_0$ that can be numerically integrated in time.
However, the Carleman procedure must fail whenever this relationship is beyond the radius of convergence of the power series. 
This is determined by the radius in the complex plane where the coordinate transformation is no longer analytic in $x_0$.

    Consider a 1D autonomous flow for $x\in \Reals$
\begin{align} \label{eq:dxdtF}
    dx/dt=\Velocity(x).
\end{align}
    The Carleman embedding constructs a solution in the form $x(t,x_0)$ that is a power series in $x_0$.
    The Carleman embedding will not converge as additional polynomials are added 
    whenever the true solution is beyond the radius of convergence of the Taylor series of $x$ as a function of $x_0$. The radius of convergence of the Taylor series is determined by the nearest singularity of $x(t,x_0)$ as a function in the complex plane.

For simplicity, first consider the case where $\Velocity(x)$ has simple zeros and $\yNnl\geq 1$.
The solution to the differential equation \ref{eq:dxdtF}
can be determined 
by dividing by the velocity and integrating to find the time as a function of $x$,
\begin{align}
t-t_0 &= \int_{\xinit}^{x_t}dx/\Velocity(x).
\end{align}
Using the partial fraction decomposition of $1/\Velocity(x)$ allows one to explicitly write the solution as
\begin{align}
t-t_0&=\int_{\xinit}^{x_t} \sum_{j=1}^n \frac{\tau_j dx}{(x-x_j)}=\sum_{j=1}^{n} \tau_j\log\abs{\frac{x_t - x_j}{x_0 - x_j}}, 
\label{eq:1d_polynomial_solution}
\end{align}
where the fixed points $x_j$ for $j = 1, 2, \dots n$ are ordered by their distance to the initial condition, $x_0$.

For a single fixed point, the solution is simply 
\begin{align}\label{eq:1d_polynomial_solution-one-fixed}
    x(t)=x_1+(\xinit-x_1)e^{(t-t_0)/\tau_1}.
\end{align}
The Carleman method constructs an approximate  Taylor series of Eq.~\ref{eq:1d_polynomial_solution-one-fixed}, and, hence, the Carleman procedure can converge over the entire orbit, as long as the Carleman radius $\xi$ is larger than $\abs{\xinit-x_1}$.
This is because, if $\tau_1<0$ ($>0$) the future (past) trajectory is guaranteed to remain within the Carleman disk.
However, as soon as there is more than one fixed point, the radius of convergence of Eq.~\ref{eq:1d_polynomial_solution} will be limited by the second 
closest fixed point to the origin, $x_2$.

These arguments can be generalized to the case where $\Velocity(x)$ is a rational function with a polynomial numerator of order $n$ and a polynomial denominator of order $m$.
In this case, the partial fraction expansion of $1/\Velocity(x)$ yields the result
\begin{align}
t-t_0 &= \int_{x_0}^{x_t}dx/\Velocity(x) =\int_{x_0}^{x_t} dx \sum_{j} \tau_j (x-x_j)^{-\pj} +  c_j x^j
\nonumber\\
&= \sum_{j}\tau_j \frac{\left( (x_t-x_j)^{1-\pj}-(x_0-x_j)^{1-\pj}\right)}{(1-\pj)}+ c_j  \frac{ x_t^{j+1}-x_0^{j+1}}{(j+1)},
\label{eq:1d_rational_solution}
\end{align}
where $n=\sum_j\pj$.
Clearly, at each $x_j$, the series has poles where $\pj\geq 2$ and branch cuts where $\pj=1$ or is not an integer.
For a single fixed point, the solution is constructed correctly.
However, when there are two fixed points, the radius of convergence is asymptotically limited to the 
second nearest fixed point to the origin, $x_2$.

Consider the logistic equation 
\begin{align}
    dx/dt = \gamma (x-x_1)(x-x_2)
\end{align}
whose solution is
\begin{align}
     x_t  = \frac{ e^{\gamma (x_2-x_1)t} x_1(x_2-x_0) + x_2(x_0 - x_1) }
     {e^{\gamma(x_2-x_1) t}(x_2-x_0)+(x_0-x_1)}.
\end{align}
For $x_2>x_1$ and  $\gamma>0$ $(<0)$, the fixed point $x_1$ ($x_2$) is stable and the fixed point $x_2$ ($x_1$) is unstable.
As required by Painlev\'e theory,~\cite{Hille1997ordinary} the solution has a moveable pole singularity given by the location where the denominator vanishes.
When Taylor expanded around the basepoint, $x=a$, so that the initial series expands in terms of $(x_0-a)^k$, the radius of convergence is $r_a(t)=|x_a^*(t)|$ where
\begin{align}
    x_a^*(t)=x_0^*(t)-a=\frac{  e^{\gamma(x_2-x_1) t}(x_2-a)-(x_1-a) }{ e^{\gamma(x_2-x_1) t}-1 }.
\end{align}
At first, the radius of convergence is infinite but it decreases in time and asymptotically approaches $|x_{1/2}-a|$ as $ t\rightarrow +/-s\infty$ where $s={\rm sign}\,(\gamma(x_2-x_1))$.
In fact, there is always a complex time when the radius of convergence vanishes
\begin{align}
     (x_2-x_1)\gamma t^*=\log((x_2-a)/(x_1-a)).
\end{align}
The radius can vanish at some point along the trajectory if $t^*$ is real, i.e. if $x_2-a\neq x_1-a$ have the same sign, but not if they have opposite signs.

First, consider the case $\gamma>0$ where $x_1$ is stable and $x_2$ is unstable.
For an initial condition of the form $x_1< x_0<x_2$, the trajectory is attracted to $x_1$.
Convergence is worst near $x_0$, but improves as $x$ travels towards $x_1$.  
For $x_0<x_1$, the trajectory is still attracted to $x_1$, but convergence is worst where $x$ is furthest from the origin.
However, for $x_0>x_2$, convergence cannot be achieved as the trajectory flies off towards $x\rightarrow\infty$ and, in fact, the solution diverges from the true solution as $x\rightarrow 1$.

Second, consider the case $\gamma<0$ where $x_1$ is unstable and $x_2$ is stable.
Now, for an initial condition of the form $x_1<x_0<x_2$, the trajectory is repelled from $x_1$ and attracted to $x_2$.
However, if the number of polynomials is fixed, then the method always fails once the trajectory gets close enough to $x_2$.
If $x_0<0$, then the trajectory flies off towards $x\rightarrow -\infty$, but the evolution in time abruptly accumulates large errors once $x<-x_2$.
After this point, convergence can no longer be achieved and the solution civerges from the true solution as $x\rightarrow -1$.
Clearly, convergence also cannot be achieved in the case where $x_0>x_2$ and the solution diverges from the true solution for $x\rightarrow 1$.

We have performed a series of tests that exhibit the behavior predicted for
all of these cases.  
For example, numerical evidence in Sec.~\ref{sec:carleman-gl} shown in Fig.~\ref{fig:ca-gl-oneD-case}(a) (black dashed lines) demonstrates that convergence is difficult to achieve even for the region $x_1<x<x_2$. In this case, convergence can be achieved if more polynomials are retained; we found $\yNca=20$ and above was sufficient.

\section{Linearization for higher-order 1D fixed points}
\label{sec:higher-order-fixed-points}
Even for a single fixed point of order $\pj>1$, the standard Carleman method fails unless one modifies the formulation. 
In this case, one can linearize the region near the fixed point using the non-analytic change of variables
\begin{align}
(\pj-1)\log\abs{z}&= 
\dx/\abs{\dx}^{\pj}
\\
  \abs{z}^{\pj-1}&=\exp{\left( 
  \dx/\abs{\dx}^{\pj}
  \right) },
\end{align}
where, here, $\Delta x =x-x_j$.
In this formula, we used the choice of sign conventions to handle
the region below, $x<x_j$, and above, $x_j<x$, the fixed point to correspond to $z<1$ and $1<z$, respectively. 
 
This formula also holds when $\pj<1$ and even for a singular fixed point where $\pj<0$.
In this case, the linearizing transformation for $\pj<1$ can be rewritten as 
\begin{align}
(1-\pj)\log\abs{z}&=
\dx \abs{\dx}^{-\pj},
\\
 \abs{z}^{1-\pj}&=\exp{\left(   
 \dx \abs{\dx}^{-\pj}
 \right) }.
\end{align}
Hence, it is not as singular as the case of $\pj>1$, and, can even be chosen to be analytic when $\pj$ is a negative even integer.

If the partial fraction expansion of $\Velocity(x)$ includes a polynomial of order $\pinfty\geq 1$, then the singular point at infinity can be handled in a similar fashion.
The change of variables $y=1/x$ repositions the singular point at $x=\pm\infty$ to the point $y=0$ and changes the effective order of the polynomial to $\dot y \propto   y^{2-\pinfty}$. Hence, this can be handled as a singular point of order $2-\pinfty$, which leads to 
the linearizing transformation
\begin{align}
(\pinfty+1)\log\abs{z} &=-y\abs{y}^{\pinfty}=-1/x\abs{x}^{\pinfty},  
\\
\abs{z}^{\pinfty+1}
&=\exp{\left(
- y \abs{y}^{\pinfty}
\right) }
\\
&=
\exp{\left(-1/x\abs{x}^{ \pinfty)}
\right) }.
\end{align}
In this case, our choice of sign conventions handles the regions  $-\infty<x<0$ and $0<x<\infty$ with the regions $z<1$ and $z>1$, respectively.

\def\Bz{\boldsymbol{z}}
\def\tcoord{t}
\def\constantcoord{p}
\def\qcoord{{q}}
\def\rcoord{r}

\def\jindex{j}
\def\kindex{k}
\def\dimindex{\dim}
\def\oneindex{1}
\def\symmetry{ independent }

\section{Carleman embedding stability \& convergence in higher dimensions}
\label{sec:higher-dflows}
The Carleman embedding with time dependent coefficients
yields the linear differential equation
\begin{align}
    du_t/dt= A(t)u_t+B(t).
\end{align}
The general solution to this linear equation is simply
\begin{align}
    u_t=e^{\int_0^t A(s)ds}u_0+\int_0^t e^{\int_s^t A(s')ds'}B(s) ds
\end{align}
where $u_0$ is the initial condition at time $t=0$.
Hence, the solution of the set of ODEs at any point in time, $x_t$, which is given by the first block of the Carleman vector, $u_t$, is represented as a linear combination of the components of $u_0$, with coefficients that depend on time.
Since the components of $u_0$ are monomials of the initial conditions, $x_0$,
the Carleman embedding assumes that, in the chosen coordinate system, the solution, $x_t$, can be written as a convergent Taylor series of the initial conditions, $x_0$, with coefficients that depend on time.
\emph{Clearly, in the case of higher dimensions, the radius of convergence must be checked for each coordinate, $x^j$, independently, and convergence or lack thereof can change over time.}

When the Taylor series for the solution does not converge, this causes the Carleman solution to become singular in the limit that an infinite number of terms are retained, i.e. in the limit $\yNca\rightarrow\infty$.
However, it is important to remember that, under rather mild assumptions on the behavior of $A(t)$ and $B(t)$, any finite dimensional truncation of the Carleman embedding will lead to a finite result for finite times.
For example, when $A$ and $B$ are constant in time, the solution is simply determined by the generalized eigenvectors of $A$, and consists of a sum of terms  with the time dependence $t^ke^{\lambda t}$.
When the Taylor series does not converge, finite truncations become more and more singular as more polynomials are retained.
Although numerical overflow can occur, since no finite time singularity appears for the truncated solution, we should be careful to say that the truncated Carleman solution diverges from the true solution rather than to say that the truncated solution blows up.

The question of where the solution to a set of nonlinear ODEs has a convergent Taylor series is directly connected with the question of where the solution, $x_t$, can become singular for any complex values of it's parameters, $x_0$, and for any complex values of time, $t$.~\cite{Hille1997ordinary, Arnold1992odes}
In fact, proving the existence of the solution beyond the point at which the Taylor series stops converging requires the notion of analytical continuation of the solution.
Hence, this question is also intimately connected to the existence and uniqueness of the solutions.

First, there are \emph{fixed singularities} that can arise at a subset of any of the singularities of $\Velocity(t,x)$ itself.
Fixed singularities clearly limit the useful extent of the Carleman embedding in time.
Yet, since their potential locations can be predetermined, one can predetermine a lower bound for the radius of convergence in time, which would allow one to predict an upper bound for how many Carleman charts are required by the fixed singularities.

Second, there are \emph{moveable singularities} that arise when either $x_t$ or $dx_t/dt$ becomes singular at locations determined by the initial conditions rather than the singularities of $\Velocity(t,x)$.
Moveable singularities are responsible for determining the radius of convergence in the initial condition, $r(t)$, and for how this radius changes in time.
Unfortunately, it is more challenging to predict the locations of the moveable singularities.

For first order ODEs in a single variable, $dx/dt=\Velocity(t,x)$, a great deal is known about the singularities of the solutions.~\cite{Hille1997ordinary}
For example, Painlev\'e theory proves that, in the case where $\Velocity(t,x)=P(t,x)/Q(t,x)$ is algebraic in $t$ and rational in $x$,
the only moveable singularities are poles and algebraic branch points.~\cite{Hille1997ordinary}
Algebraic branch points arise whenever $dx/dt\rightarrow\infty$ at the set of roots in $x$ for which the denominator vanishes, $Q(t,x)=0$, and are moveable if $Q$ has nontrivial dependence on $t$.
If the numerator, $P(t,x)$, is order $p$  in $x$ and the denominator, $Q(t,x)$, is order $q$ in $x$, then, when $x\rightarrow\infty$,  moveable algebraic branch points arise for $p>q+2$ and moveable poles arise for $p=q+2$, but not for $p<q+2$.
If the only singularities are poles, then $q=0$ and $p=2$, so $\Velocity(t,x)$ must be quadratic in $x$, i.e. a Ricatti equation.
If $\Velocity(t,x)$ is linear in $x$, then the ODE is linear and the solution only has fixed singularities.
Hence, whenever the Carleman embedding is applied to the case of a nonlinear 1D ODE with polynomial $\Velocity(t,x)$, there are moveable singularities to contend with.

In fact, in 1D, the fixed points completely determine the asymptotic region of convergence.
Even if the solution is holomorphic for any finite time, it must asymptotically approach the stable/unstable fixed points as $t\rightarrow +/- \infty$. 
Since the solution must be asymptotic to a discontinuous function, the radius of convergence must  asymptotically limit towards the location of the nearest unstable/stable fixed point.
Clearly, the radius of convergence asymptotically  vanishes at an unstable/stable fixed point as $t\rightarrow +/-\infty$.

This argument extends to higher dimensions for any system that has topologically distinct asymptotic basins of attraction, e.g. determined by a Lyapunov function.
If the flow must asymptotically approach distinct basins of attraction, e.g. with distinct stable/unstable fixed points or limit cycles, then the solution must be asymptotic to a discontinuous function.
Thus, an upper bound on the asymptotic radius of convergence is given by the distance in parameter space to the separatrices that determine the boundaries between different basins of convergence.

Even in the case of non-dissipative, e.g. Hamiltonian flows, there are typically topologically distinct basins, e.g. determined by conserved quantities such as the Hamiltonian and other action invariants. 
The flow will have distinct neutrally stable O-points in the different basins and the angles that specify the motion around each O-point have different winding numbers.
This implies that the solution cannot be holomorphic across the separatrices that determine the boundaries of these topologically distinct basins.
Thus, in any dimension, the presence of multiple fixed points implies that there are nontrivial upper bounds on the region of convergence of the Taylor series of the solution.
It should also be clear that, in this context, proximity should be measured in terms of the canonical coordinates that determine the distinct basins rather than some other notion of distance.

\def\ConstCoord{J}
\def\VaryCoord{\theta}
\def\TotalCoord{{x^\dim}}
\def\Transf{\mathsf{T}}
\def\Tinv{\mathsf{T}^{-1}}
\def\xcoord{x}

The location of the singularities is easiest to understand in the case where the dynamical system is completely integrable by quadratures.~\cite{Olver1993book,Bluman1989book} 
In this case, the solution can be expressed through a series of autonomous 1D quadratures that can each be analyzed as in Appendix~\ref{sec:1dflows}.
For example, whenever the ODE system has a set of symmetries that forms a solvable Lie algebra of the same dimension or higher, then it is completely integrable by quadratures.~\cite{Olver1993book,Bluman1989book}
In fact, the flow of any vector field that is an element of a transitive finite-dimensional solvable Lie algebra is completely integrable by quadratures.~\cite{Carinena2016solvability, Kozlov2024solvable}
Finding the explicit solution requires a two-stage process of first finding a set of canonical coordinates that expose the solvable structure of the Lie algebra and then performing quadratures to solve for the flow itself.~\cite{Grabowska2020solvable, Kozlov2024solvable}
Since both the first and second stages require performing quadratures, each quadrature is subject to the same issues discussed for the autonomous 1D case. 
The first stage, which sets up the canonical coordinate system itself, yields fixed singularities and places asymptotic restrictions on moveable singularities, and, hence, convergence due to coordinate topology.
The second stage considers the velocity itself and controls the moveable singularities, i.e. how the radius of convergence depends on time.

For example, in a sufficiently small neighborhood of a generic initial condition, one can always use the flowbox rectification theorem~\cite{Arnold1992odes} to straighten out the vector field.
On an $\dim$-dimensional region of phase space, this process determines $\dim-1$ constants of the motion, $\ConstCoord=\{\ConstCoord^j\}$, so that the set of ODEs reduces to the 1D case, $d\TotalCoord/dt= \Velocity^\TotalCoord(\ConstCoord,\TotalCoord)$, where $\TotalCoord$ is the remaining coordinate that parameterizes motion along the trajectory. 
Hence, the equation of motion for coordinate $\TotalCoord$ can be solved by quadrature $t=\int d\TotalCoord/\Velocity^\TotalCoord(\ConstCoord,\TotalCoord)$,
which reduces to the autonomous 1D case.
If the velocity vanishes more than once, then the transformation will suffer from the same convergence issues studied in Appendix~\ref{sec:1dflows}.
For example, this happens whenever there are turning points in the original coordinate system or whenever there is a separatrix that connects fixed points together.
Although generic separatrices do not intersect multiple fixed points, these situations do occur for many completely integrable dynamical systems of interest.

\newcommand\sn[1]{{\rm sn}\,#1}
\newcommand\cn[1]{{\rm cn}\,#1}
\newcommand\dn[1]{{\rm dn}\,#1}
\newcommand\am[1]{{\rm am}\,#1}
\def\Integers{\mathbb{Z}}
Due to the movable singularities, the presence of fixed points only sets an upper bound on the asymptotic radius of convergence and the radius at finite times can differ than this.
The example of the logistic equation in Appendix~\ref{sec:1dflows} clearly shows that, while the radius may initially be large, it typically gets smaller in time, can even vanish, and is asymptotically determined by the location of the fixed points.
As another example, consider the  Hamiltonian motion of $\{x,p\} \in \Reals^2$ for  Hamiltonian
\begin{align}
    H=p^2/2+U(x)
\end{align} 
where $U(x)$ is a potential that is either 4th order in $x$ or of the pendulum form where $U(x)=-\cos(x)$.
Clearly, there are fixed singularities at any value of $H$ that corresponds to an X-point or O-point.
The solutions~\cite{Lichtenberg2013book, NIST:DLMF:22.19} can be expressed in terms of the Jacobian elliptic functions, $\sn(\phi,k)$, $\cn(\phi,k)$ and $\dn(\phi,k)$, where $\phi=\omega_0 t$.
Here, $k(H)$ and $\omega_0(H)$ are constants of the motion that depend on the Hamiltonian, but have different functional forms in topologically distinct regions.  
Elliptic functions are doubly periodic in the complex plane and must have poles within the fundamental torus at~\cite{NIST:DLMF:22.19}
\begin{align}
  \phi^*_{n,m}=  \omega_0 t^*_{n,m} = 2n K(k)+i(2m+1)K'(k)
\end{align} 
for $n,m\in\Integers$. 
These are moveable singularities that depend on $\omega_0(H)$ and on $k(H)$  via  the complete Legendre elliptic integral, $K(k)$, and it's derivative $K'(k)$.
They limit the radius of convergence of the solution in $x$ both as a function of time, $t$, and as a function of the initial conditions, through the energy, $H=p_0^2/2+U(x_0)$.

Poles in the complex plane are necessary because the only entire functions on a compact surface, the torus here, are constant.
An important point for the Carleman algorithm is that, on the torus, periodicity implies that distance to the nearest singularity varies periodically on every cycle.
In fact, this same argument applies to any integrable system defined on a higher-dimensional torus.
Poles must be present and the distance to the poles varies quasiperiodically on every cycle.

Thus, we see that, even for completely integrable systems in higher dimensions, a similar but more complicated set of stability and convergence issues arise as for the 1D case  of Appendix~\ref{sec:1dflows}.
In the case of higher dimensions, both completely unstable and partially unstable hyperbolic fixed points can generate an unstable Carleman embedding.
For example, for generic initial conditions, the Carleman embedding will not remain stable over the entire trajectory unless there is a completely stable critical point that is closest to the basepoint of the Carleman linearization region and the initial condition is sufficiently close to the stable fixed point/limit cycle.
For this reason, even stable limit cycles pose issues because they are generated by a set of unstable fixed points on both the inside and the outside of the attractor.
This poses the geometric issue that the overall radius of the limit cycle is larger than the distance to the next critical point, so that a single Carleman chart covering the entire attractor must include at least one unstable fixed point.  
Perhaps the issues with this scenario can be avoided by choosing an annular coordinate system that avoids these other fixed points.
Clearly, this requires knowing the general structure and shape of the attractor.


\def\Poly{F}
\def\Coeff{F}
\def\Xpoly{x}
\def\Binomial{C}
\def\MPOLY{M}

\def\Npoly{\yNnl}
\def\dPoly{\dim}

\def\Mpoly{\yNca}

\def\Xscale{s}
\def\Xshift{a}
\def\Nt{N_t}
\def\Nrk{n_r}
\def\Nsteps{N_{rt}}
\def\Nadapt{N_A}

\section{Complexity of Carleman embedding methods}
\label{sec:complexity-remap}

\subsection{Classical complexity of tensor product encoding}
\label{sec:classical_complexity}

The classical complexity of the Carleman algorithm can vary widely depending on how the Carleman vector and matrix are stored because the number of elements is quite large but there is a lot of symmetry and sparsity that one can use to reduce the cost. 
For example, one can use a sparse format for the matrix or, even better, one can store the matrix in the form of the sum of tensor products in Eq.~\ref{eq:ca-matrices-A}. 
There are also other important optimizations such as using the symmetric form of the tensors discussed in Appendix~\ref{sec:homogeneous} rather than a tensor product form which has many repeated entries.

In the following, let’s assume that the Carleman vector is stored as a direct sum of tensor products of the $\dim$-dimensional vector $x$.
Since we will need to evaluate the elements of the vector to compute the matrix-vector product, we can assume that all elements must be evaluated on each time step.
Each tensor product of order $k$ has $\dim^k$ elements, so the total number of elements in the Carleman vector is
\begin{align}\label{eq:NC_tensor-product}
    N_C = \sum_{k=1}^\yNca \dim^k = (\dim^\yNca-1)/(1-1/\dim) = \oO(\dim^\yNca).
\end{align}
The classical cost of recursively computing all components of the Carleman vector from a tensor product, e.g. from the initial conditions, is also $N_C$.
Hence, remapping the initial conditions also requires $\oO(N_C)$ operations.
Using the symmetric format in Appendix~\ref{sec:homogeneous} would reduce the value but does not change the way that $N_C$ enters the other estimates.
Let’s also assume that the $\Velocity_k$ matrices are stored as $(k+1)$-index tensors with $\dim$ dimensions per index. This implies that each $\Velocity_k$ has $\dim^{k+1}$ elements and the largest $\Velocity_k$ has $\dim^{\yNnl+1}$ elements, so the total number of velocity elements is
\begin{align} \label{eq:NF_tensor-product}
    N_F = \sum_{j=1}^{\yNnl+1} \dim^j = (\dim^{\yNnl+1}-1)/(1-1/\dim) = \oO(\dim^{\yNnl+1}),
\end{align}
where the last form assumes $p+1\gg 1$.
Again, using the symmetric format in Appendix~\ref{sec:homogeneous} would reduce this value.
The Carleman matrix $A$ has $N_C^2 = \oO(\dim^{2\yNca})$ total elements, and the most expensive parts are the largest blocks, which already have $\oO(\dim^{2\yNca})$ elements. However, there are only $\oO(N_F N_C) = \oO(\dim^{\yNca+\yNnl+1})$ nonzero elements, so the overall complexity can be reduced to this amount by using a sparse matrix format.

However, storing the matrix in the sum of tensor product form of Eq.~\ref{eq:ca-matrices-A} greatly reduces the cost. 
This only requires $N_F$ elements for all the $\Velocity_k$'s and, since there are $P(P+1)/2$ total terms for  all the terms in the sum for all the blocks in Eq.~\ref{eq:ca-matrices-A}, the total cost is only
\begin{align}
N_A = \oO(P^2) \oO( N_F)   = \oO(\yNca^2\dim^{ \yNnl+1}).
\end{align}
The cost of the matrix-vector product in this format is only
\begin{align}
N_{AC} = \oO(N_AN_C)=\oO(\yNca^2 N_F N_C)  = \oO(\yNca^2\dim^{\yNca+\yNnl+1}).
\end{align}
So, it’s far better to store the matrix in the form of Eq.~\ref{eq:ca-matrices-A}. Now, we don’t need recompute the dense or even the sparse form of $A$.
Instead, we just need to determine how to update the $\Velocity_k$ matrices.

In Appendices~\ref{sec:1d-remap} and \ref{sec:multidim-remap}, we show that the complexity of simply shifting and rescaling the $\Velocity_k$ matrices is only $\oO(\yNnl N_F)$. 
Naively, one would expect the scale to cost $\oO(\yNnl)$ steps and the shift to cost $\oO(\yNnl^2)$ steps, but using the recursive form in Appendix~\ref{sec:multidim-remap} reduces the cost of shifting to $\oO(\yNnl)$ steps.
In 1D, the total cost is only $\oO(\yNnl \yNca )$. 
For $\dim$ dimensions, if we use the sum of tensor product form in Eq.~\ref{eq:ca-matrices-A}, then updating all terms costs
\begin{align}
M_{A} =  \oO(p P^2)  \oO( N_F ) = \oO(\yNnl)  \oO( N_{AC}).
\end{align}
So, this is just a constant factor of $\oO(\yNnl)$ increase over the cost of building the matrix. 
Performing the matrix-vector product using the updated matrix has a similar form
\begin{align}
M_{AC}  =\oO(\yNnl  \yNca^2) \oO(  N_F   N_C)   = \oO(\yNnl)  \oO(N_{AC}).
\end{align}
Thus, using the sum of tensor products format of Eq.~\ref{eq:ca-matrices-A} only leads to a constant factor $\oO(\yNnl)$ increase over the cost of a single evaluation of the matrix-vector product, which is needed for every time step.

\subsection{1D diagonal affine remapping}
\label{sec:1d-remap}
Let us begin by considering the 1D case where the coefficients of the polynomial are known.
Consider an $\Npoly$-th order polynomial expressed in the form
\begin{align}
\Poly(\Xpoly) = \sum_{k=0}^\Npoly \Coeff_k \Xpoly^k = \left(\left(\Coeff_{\Npoly} \Xpoly + \Coeff_{\Npoly-1}\right)\Xpoly +\dots  +\Coeff_1\right)\Xpoly+\Coeff_0.
\end{align}
Evaluation of the value of the polynomial from the coefficients requires $\Npoly$ multiplications and $\Npoly$ additions.  
Rescaling the basepoint by a factor $s$ generates a transformation of the coefficients
\begin{align}
\bar \Xpoly^k = (\Xpoly/\Xscale)^k && \bar\Coeff_k = \Coeff_k \Xscale^k
\end{align}
that costs $p$ multiplications of $x/s$ to form the initial conditions and $2\Npoly$ multiplications to form $\Xscale^k$  
and multiply by $\Coeff_k$.
Shifting the basepoint generates a recursive transformation of the coefficients starting from $\bar \Coeff_{\Npoly}=\Coeff_\Npoly$ going downward via
\begin{align}
\bar \Xpoly^k =  (\Xpoly-\Xshift)^k && \bar\Coeff_k = \Coeff_k+ \bar \Coeff_{k+1} \Xshift
\end{align}
that costs $2\Npoly$ multiplications and additions. 
Performing both at the same time
\begin{align}
\bar \Xpoly^k =  \left((\Xpoly-\Xshift)/\Xscale\right)^k && \bar\Coeff_k = (\Coeff_k+ \bar \Coeff_{k+1} \Xshift)\Xscale^k
\end{align}
simply adds the costs of both for a total of $\Npoly$ multiplications for the initial conditions and $3\Npoly$ multiplications and $\Npoly$ additions for the velocity, i.e. $4\Npoly$ operations for the velocity.
Thus, the total cost is $\oO(5\Npoly)$ operations. 
Since the cost of rescaling and shifting the coefficients is similar to that of polynomial evaluation, which must happen many times for stable numerical integration along the trajectory, this cost is negligible if the number of tests can be kept to a small fraction of the number of time integration points.

In 1D, the Carleman velocity can be computed using $dx^k/dt=kx^{k-1}\Velocity_1(x)$.
Thus, the cost of computing the velocity for all components is simply the cost of recursively computing $x^k$ and then multiplying by the  velocity $\Velocity_1$, for a total cost of $\oO(2\yNca+5\yNnl)$ operations.

\subsection{Higher dimensional diagonal affine remapping}
\label{sec:multidim-remap}
\newcommand\yAbar[2]{\bar A^{[#2]}_{#1}}

Consider a multidimensional shift, $a=\{a^\jindex\}$, and scale, $s=\{s^\jindex\}$, operation. First, one must form the new initial conditions from the tensor product $\left(\otimes[(x-a)/s]\right)^k$, where  $k=1,\dots,\yNca$ ranges over all polynomial orders, where $\yNca$ is the number of monomials in the Carleman embedding according to the notation of Sec.~\ref{sec:carleman-std}.
If stored in tensor product format, one only needs a single copy of the initial conditions after the shift and scale factors are applied.
However, full evaluation of all elements is required for computing the matrix-vector product and this is required on the next time step.
This requires the $\yNca$ tensor products to be explicitly computed.
The classical cost of storing and recursively computing all components of the initial condition tensor is $N_C\sim\oO(\dim^{\yNca })$, which is equivalent to the cost of manipulating the Carleman vector on each time step.

The shifted and rescaled Carleman matrix can be computed recursively in a form similar to that of Eq.~\ref{eq:ca-matrices-A} 
\begin{align}
    \yAbar{j+\yk}{j} =\sum_{i=0}^{j-1} \yIN{i}\otimes  \bar F_{\yk+1} \otimes \yIN{j-i-1},
\end{align}
for $j=1,\dots,\yNca$ and $k=-1,0,\dots,\yNnl$ (aside from the first row and the last $\yNnl-1$ rows).
In order to shift and scale these blocks, the coefficients can be recursively computed from
\begin{align}
\bar F_{\yNnl} &=  F_{\yNnl} s^{\otimes\yNnl}, \\
\bar F_{\yk} &=  (F_{\yk} +  \bar F_{\yk+1} a\otimes  )s^{\otimes k},
\end{align}
which now requires $\oO(3\yNnl)$ tensor products and $\oO(\yNnl)$ tensor additions. 
Hence, the cost of computing the $\bar \Velocity_k$ matrices is $\oO(4\yNnl)$ and, since it is $\oO(\yNnl)$, it is equivalent to the cost of evaluating the first row of the Carleman matrix. 
The cost of evaluating the standard Carleman matrix is given in Appendix~\ref{sec:classical_complexity}.
The shifted and rescaled matrices have $\oO(5\yNnl)$ additional terms in the sum. 
Thus, this multiplies the overall cost of forming the matrix and, hence, the matrix-vector product by $\oO(5\yNnl)$.
Since $\yNnl$ is not typically very large, often $\sim 2-5$, this leads to a manageable $\sim 10-25\times$ constant factor increase.

\subsection{Classical complexity of homogeneous polynomial basis}
\label{sec:homogeneous}
The multi-dimensional Carleman embedding is usually assumed to use homogeneous polynomials up to a given order $\yNca$.
Homogeneous polynomials of order $\yNca$ have the form of a sum of monomials, $\Xpoly_1^{j_1}\Xpoly_2^{j_2} \cdots \Xpoly_\dPoly^{j_\dPoly}$, subject to the constraint $\sum_i j_i=\yNca$.
The number of independent homogeneous polynomials in $\dPoly$ variables up through order $\yNca$ is 
\begin{align}
   \MPOLY= \Binomial^{\dPoly+\yNca}_\dPoly=(\dPoly+\yNca)!/\dPoly!\yNca!.
\end{align}
For PDEs, where $\dPoly\gg \yNca$, using Stirling's formula for the asymptotic expansion of the Gamma function yields
\begin{align}
    \lim_{\dPoly/\yNca\rightarrow\infty}
    \MPOLY &\sim 
    e^{(\dPoly+1/2)\log{(1+\yNca/\dPoly)}
    +\yNca\log{(\dPoly+\yNca)-\yNca}} /\yNca! 
    \\
    &\sim  (\dPoly/e)^\yNca/\yNca!.
\end{align}
When both $\dPoly\gg \yNca\gg 1$, then this becomes
\begin{align} \lim_{\dPoly/\yNca,\yNca\rightarrow\infty}
\MPOLY &\sim 
   \left(\frac{\dPoly+\yNca}{2\pi \dPoly \yNca} \right)^{1/2} e^{\dPoly\log{(1+\yNca/\dPoly)}+\yNca\log{(1+\dPoly/\yNca)}} 
    \\
    &\sim  (\dPoly/\yNca)^{\yNca}/ (2\pi \yNca)^{1/2}.
\end{align}
For ODEs, where the polynomial order can be much larger than the dimension, $\yNca\gg\dPoly$, exchanging $\dPoly$ and $\yNca$ in these formulas yields the appropriate limit.
Thus, this has the form 
\begin{align}
   \lim_{\yNca/\dPoly\rightarrow\infty} \MPOLY 
   &\sim 
   (\yNca/e)^\dPoly/\dPoly!,
    \\
    \lim_{\yNca/\dPoly,\dPoly\rightarrow\infty} \MPOLY 
    &\sim
    (\yNca/\dPoly)^{\dPoly}/(2\pi \dPoly)^{1/2} .
\end{align}
Thus,  if the homogeneous polynomial basis is used directly, then the overall classical complexity is smaller than direct sum of tensor product encodings, e.g. assumed in Eqs.~\ref{eq:NC_tensor-product} and ~\ref{eq:NF_tensor-product}. 

We note that the using the homogeneous polynomial basis is a strategy for significantly reducing the number of required basis functions.
However, it is not always possible to obtain an accurate approximation of an arbitrary function using this strongly reduced number of degrees of freedom.
Whether it is possible depends on how well the function is represented by homogeneous polynomials.
This can potentially cause practical restrictions on convergence that are more stringent than the stability issues derived in Appendices~\ref{sec:1dflows}  and ~\ref{sec:higher-dflows}.

When a dynamical system is attracted to a single fixed point, then it is usually possible to make this approximation.
If one chooses an appropriate coordinate system and linearization region, the same can be true for attraction to a single limit cycle.
However, this is never the case when the system evolves towards a strange attractor because there is no well-defined coordinate system that will allow such a reduction. 
In this case, one might expect that one would need to keep $\oO(\yNca^{d_A})$ degrees of freedom where $d_A$ is the Hausdorff dimension of the attractor.

\bibliography{main}

@InProceedings{Ambainis12,
  author =	{Andris Ambainis},
  title =	{{Variable time amplitude amplification and quantum algorithms for linear algebra problems}},
  booktitle =	{29th International Symposium on Theoretical Aspects of Computer Science (STACS 2012)},
  pages =	{636--647},
  series =	{Leibniz International Proceedings in Informatics (LIPIcs)},
  ISBN =	{978-3-939897-35-4},
  ISSN =	{1868-8969},
  year =	{2012},
  volume =	{14},
  editor =	{Christoph D{\"u}rr and Thomas Wilke},
  publisher =	{Schloss Dagstuhl--Leibniz-Zentrum fuer Informatik},
  address =	{Dagstuhl, Germany},
  URL =		{http://drops.dagstuhl.de/opus/volltexte/2012/3426},
  URN =		{urn:nbn:de:0030-drops-34261},
  doi =		{10.4230/LIPIcs.STACS.2012.636}
}

@article{Joseph23,
   title={Quantum computing for fusion energy science applications},
   volume={30},
   ISSN={1089-7674},
   url={http://dx.doi.org/10.1063/5.0123765},
   DOI={10.1063/5.0123765},
   number={1},
   journal={Physics of Plasmas},
   publisher={AIP Publishing},
   author={Joseph, I. and Shi, Y. and Porter, M. D. and Castelli, A. R. and Geyko, V. I. and Graziani, F. R. and Libby, S. B. and DuBois, J. L.},
   year={2023},
   month=jan }

@article{Joseph20,
  title = {{Koopman--von Neumann} approach to quantum simulation of nonlinear classical dynamics},
  author = {Joseph, Ilon},
  journal = {Phys. Rev. Res.},
  volume = {2},
  issue = {4},
  pages = {043102},
  numpages = {17},
  year = {2020},
  month = {Oct},
  publisher = {American Physical Society},
  doi = {10.1103/PhysRevResearch.2.043102},
  url = {https://link.aps.org/doi/10.1103/PhysRevResearch.2.043102}
}

@misc{Lin22Koopman,
      title={{Koopman--von Neumann} mechanics and the {Koopman} representation: A perspective on solving nonlinear dynamical systems with quantum computers}, 
      author={Yen Ting Lin and Robert B. Lowrie and Denis Aslangil and Yi\u{g}it Suba\c{s}ı and Andrew T. Sornborger},
      year={2022},
      eprint={2202.02188},
      archivePrefix={arXiv},
      primaryClass={quant-ph}
}

@article{An23,
  title = {Linear Combination of Hamiltonian Simulation for Nonunitary Dynamics with Optimal State Preparation Cost},
  author = {An, Dong and Liu, Jin-Peng and Lin, Lin},
  journal = {Phys. Rev. Lett.},
  volume = {131},
  issue = {15},
  pages = {150603},
  numpages = {6},
  year = {2023},
  month = {Oct},
  publisher = {American Physical Society},
  doi = {10.1103/PhysRevLett.131.150603},
  url = {https://link.aps.org/doi/10.1103/PhysRevLett.131.150603}
}

@Article{Liu23,
    author={Liu, Jin-Peng
    and An, Dong
    and Fang, Di
    and Wang, Jiasu
    and Low, Guang Hao
    and Jordan, Stephen},
    title={Efficient Quantum Algorithm for Nonlinear Reaction--Diffusion Equations and Energy Estimation},
    journal={Communications in Mathematical Physics},
    year={2023},
    month={Dec},
    day={01},
    volume={404},
    number={2},
    pages={963-1020},
    issn={1432-0916},
    doi={10.1007/s00220-023-04857-9},
    url={https://doi.org/10.1007/s00220-023-04857-9}
}

@misc{An23impr,
      title={Quantum algorithm for linear non-unitary dynamics with near-optimal dependence on all parameters}, 
      author={Dong An and Andrew M. Childs and Lin Lin},
      year={2023},
      eprint={2312.03916},
      archivePrefix={arXiv},
      primaryClass={quant-ph}
}

@article{Koopman31,
    title={{Hamiltonian} Systems and Transformations in {Hilbert} Space}, 
    author={B. O. Koopman},
    year={1931},
    url={https://www.pnas.org/doi/10.1073/pnas.17.5.315},
    journal={Proceedings of the National Academy of Sciences of the United States of America},
    pages={315-318},
    doi={10.1073/pnas.17.5.315},
    volume={17},
    number={5}
}

@article{Koopman32,
author = {Koopman, B. O. and v. Neumann, J.  },
title = {Dynamical Systems of Continuous Spectra },
journal = {Ann. Math.},
volume = {18},
year = {1932},
pages={255},
}

@article{Liu21,
   title={Efficient quantum algorithm for dissipative nonlinear differential equations},
   volume={118},
   ISSN={1091-6490},
   url={http://dx.doi.org/10.1073/pnas.2026805118},
   DOI={10.1073/pnas.2026805118},
   number={35},
   journal={Proceedings of the National Academy of Sciences},
   publisher={Proceedings of the National Academy of Sciences},
   author={Liu, Jin-Peng and Kolden, H. and Krovi, Hari K. and Loureiro, Nuno F. and Trivisa, Konstantina and Childs, Andrew M.},
   year={2021},
   month=aug 
}

@Article{Akiba23,
    author={Akiba, Takaki
    and Morii, Youhi
    and Maruta, Kaoru},
    title={{C}arleman linearization approach for chemical kinetics integration toward quantum computation},
    journal={Scientific Reports},
    year={2023},
    month={Mar},
    day={09},
    volume={13},
    number={1},
    pages={3935},
    issn={2045-2322},
    doi={10.1038/s41598-023-31009-9},
    url={https://doi.org/10.1038/s41598-023-31009-9}
}

@article{Carleman32,
    title = {Application de la th\'eory des \'equations int\'egrales lin\'eaires aux syst\`emes d'\'equations diff\'erentielles non lin\'eaires},
    author = {Torsten Carleman},
    journal = {Acta Mathematica},
    volume = {59},
    pages = {63-87},
    year = {1932},
    url = {https://doi.org/10.1007/BF02546499},
    doi = {10.1007/BF02546499}
}

@article{Jin22Sch,
  title = {Quantum Simulation of Partial Differential Equations via Schr\"odingerization},
  author = {Jin, Shi and Liu, Nana and Yu, Yue},
  journal = {Phys. Rev. Lett.},
  volume = {133},
  issue = {23},
  pages = {230602},
  numpages = {6},
  year = {2024},
  month = {Dec},
  publisher = {American Physical Society},
  doi = {10.1103/PhysRevLett.133.230602},
  url = {https://link.aps.org/doi/10.1103/PhysRevLett.133.230602}
}

@Article{Wootters82,
    author={Wootters, W. K.
    and Zurek, W. H.},
    title={A single quantum cannot be cloned},
    journal={Nature},
    year={1982},
    month={Oct},
    day={01},
    volume={299},
    number={5886},
    pages={802-803},
    issn={1476-4687},
    doi={10.1038/299802a0},
    url={https://doi.org/10.1038/299802a0}
}

@misc{Leyton08,
      title={A quantum algorithm to solve nonlinear differential equations}, 
      author={Sarah K. Leyton and Tobias J. Osborne},
      year={2008},
      eprint={0812.4423},
      archivePrefix={arXiv},
      primaryClass={quant-ph},
      url={https://arxiv.org/abs/0812.4423}, 
}

@article{Krovi23,
   title={Improved quantum algorithms for linear and nonlinear differential equations},
   volume={7},
   ISSN={2521-327X},
   url={http://dx.doi.org/10.22331/q-2023-02-02-913},
   DOI={10.22331/q-2023-02-02-913},
   journal={Quantum},
   publisher={Verein zur Forderung des Open Access Publizierens in den Quantenwissenschaften},
   author={Krovi, Hari},
   year={2023},
   month=feb, pages={913} 
}

@article{Joseph23JPA,
   title={Semiclassical theory and the {Koopman-van Hove} equation},
   volume={56},
   ISSN={1751-8121},
   url={http://dx.doi.org/10.1088/1751-8121/ad0533},
   DOI={10.1088/1751-8121/ad0533},
   number={48},
   journal={Journal of Physics A: Mathematical and Theoretical},
   publisher={IOP Publishing},
   author={Joseph, Ilon},
   year={2023},
   month=nov, pages={484001} 
}

@misc{Vaszary25,
      title={Solving the Nonlinear {Vlasov} Equation on a Quantum Computer}, 
      author={Tamas Vaszary and Animesh Datta and Tom Goffrey and Brian Appelbe},
      year={2025},
      eprint={2411.19310},
      archivePrefix={arXiv},
      primaryClass={quant-ph},
      url={https://arxiv.org/abs/2411.19310}, 
}

@article{Conde25,
  title = {Quantum {C}arleman linearization efficiency in nonlinear fluid dynamics},
  author = {Gonzalez-Conde, Javier and Lewis, Dylan and Bharadwaj, Sachin S. and Sanz, Mikel},
  journal = {Phys. Rev. Res.},
  volume = {7},
  issue = {2},
  pages = {023254},
  numpages = {16},
  year = {2025},
  month = {Jun},
  publisher = {American Physical Society},
  doi = {10.1103/PhysRevResearch.7.023254},
  url = {https://link.aps.org/doi/10.1103/PhysRevResearch.7.023254}
}

@article{Sanavio24,
   title={Explicit Quantum Circuit for Simulating the Advection–Diffusion–Reaction Dynamics},
   volume={6},
   ISSN={2689-1808},
   url={http://dx.doi.org/10.1109/TQE.2025.3544839},
   DOI={10.1109/tqe.2025.3544839},
   journal={IEEE Transactions on Quantum Engineering},
   publisher={Institute of Electrical and Electronics Engineers (IEEE)},
   author={Sanavio, Claudio and Mauri, Enea and Succi, Sauro},
   year={2025},
   pages={1–12} 
}

@article{Sanavio24Three,
    author = {Sanavio, C. and Scatamacchia, R. and de Falco, C. and Succi, S.},
    title = {Three {C}arleman routes to the quantum simulation of classical fluids},
    journal = {Physics of Fluids},
    volume = {36},
    number = {5},
    pages = {057143},
    year = {2024},
    month = {05},
    issn = {1070-6631},
    doi = {10.1063/5.0204955},
    url = {https://doi.org/10.1063/5.0204955},
    eprint = {https://pubs.aip.org/aip/pof/article-pdf/doi/10.1063/5.0204955/19963101/057143\_1\_5.0204955.pdf},
}

@article{Li25,
  title = {Potential quantum advantage for simulation of fluid dynamics},
  author = {Li, Xiangyu and Yin, Xiaolong and Wiebe, Nathan and Chun, Jaehun and Schenter, Gregory K. and Cheung, Margaret S. and M\"ulmenst\"adt, Johannes},
  journal = {Phys. Rev. Res.},
  volume = {7},
  issue = {1},
  pages = {013036},
  numpages = {18},
  year = {2025},
  month = {Jan},
  publisher = {American Physical Society},
  doi = {10.1103/PhysRevResearch.7.013036},
  url = {https://link.aps.org/doi/10.1103/PhysRevResearch.7.013036}
}

@misc{Vaszary24,
  title={{C}arleman Linearization of Partial Differential Equations}, 
  author={Tamas Vaszary},
  year={2024},
  eprint={2412.00014},
  archivePrefix={arXiv},
  primaryClass={math.GM},
  url={https://arxiv.org/abs/2412.00014}, 
}

@article{Novikau24KvN,
    title = {Quantum algorithm for the advection-diffusion equation and the {Koopman-von Neumann} approach to nonlinear dynamical systems},
    journal = {Computer Physics Communications},
    volume = {309},
    pages = {109498},
    year = {2025},
    issn = {0010-4655},
    doi = {https://doi.org/10.1016/j.cpc.2025.109498},
    url = {https://www.sciencedirect.com/science/article/pii/S0010465525000013},
    author = {I. Novikau and I. Joseph},
}

@misc{Forets17,
      title={Explicit Error Bounds for {C}arleman Linearization}, 
      author={Marcelo Forets and Amaury Pouly},
      year={2017},
      eprint={1711.02552},
      archivePrefix={arXiv},
      primaryClass={math.NA},
      url={https://arxiv.org/abs/1711.02552}, 
}

@article{Belabbas23,
    title = {A sufficient condition for the super-linearization of polynomial systems},
    journal = {Systems \& Control Letters},
    volume = {179},
    pages = {105588},
    year = {2023},
    issn = {0167-6911},
    doi = {https://doi.org/10.1016/j.sysconle.2023.105588},
    url = {https://www.sciencedirect.com/science/article/pii/S0167691123001354},
    author = {Mohamed-Ali Belabbas and Xudong Chen},
}

@article{Sanchez25,
    doi = {10.1088/1402-4896/add65a},
    url = {https://dx.doi.org/10.1088/1402-4896/add65a},
    year = {2025},
    month = {may},
    publisher = {IOP Publishing},
    volume = {100},
    number = {6},
    pages = {065231},
    author = {Mu\~noz-S\'anchez, Juan C and Elena, Santiago F and Oteo, Jos\'e-\'Angel},
    title = {{C}arleman approximants for non-linear differential systems},
    journal = {Physica Scripta},
}

@misc{Motee25,
      title={{C}arleman-{Fourier} linearization of nonlinear real dynamical systems with quasi-periodic fields}, 
      author={Nader Motee and Qiyu Sun},
      year={2025},
      eprint={2503.01498},
      archivePrefix={arXiv},
      primaryClass={math.DS},
      url={https://arxiv.org/abs/2503.01498}, 
}

@Article{Weber16,
    AUTHOR = {Weber, H. and Mathis, W.},
    TITLE = {Adapting the range of validity for the {C}arleman linearization},
    JOURNAL = {Advances in Radio Science},
    VOLUME = {14},
    YEAR = {2016},
    PAGES = {51--54},
    URL = {https://ars.copernicus.org/articles/14/51/2016/},
    DOI = {10.5194/ars-14-51-2016}
}

@article{Lin20,
  doi = {10.22331/q-2020-11-11-361},
  url = {https://doi.org/10.22331/q-2020-11-11-361},
  title = {Optimal polynomial based quantum eigenstate filtering with application to solving quantum linear systems},
  author = {Lin, Lin and Tong, Yu},
  journal = {{Quantum}},
  issn = {2521-327X},
  publisher = {{Verein zur F{\"{o}}rderung des Open Access Publizierens in den Quantenwissenschaften}},
  volume = {4},
  pages = {361},
  month = nov,
  year = {2020}
}

@misc{Low24QLSA,
      title={Quantum linear system algorithm with optimal queries to initial state preparation}, 
      author={Guang Hao Low and Yuan Su},
      year={2024},
      eprint={2410.18178},
      archivePrefix={arXiv},
      primaryClass={quant-ph},
      url={https://arxiv.org/abs/2410.18178}, 
}

@misc{Dalzell24,
      title={A shortcut to an optimal quantum linear system solver}, 
      author={Alexander M. Dalzell},
      year={2024},
      eprint={2406.12086},
      archivePrefix={arXiv},
      primaryClass={quant-ph},
      url={https://arxiv.org/abs/2406.12086}, 
}

@misc{PCEcode,
    title = {{P}iecewise {C}arleman {E}mbedding},
    howpublished = {\url{https://github.com/QuCF/QuCF/wiki/Piecewise-Carleman-embedding}},
    year={2025},
}

@article{Succi23,
    doi = {10.1209/0295-5075/acfdc7},
    url = {https://dx.doi.org/10.1209/0295-5075/acfdc7},
    year = {2023},
    month = {oct},
    publisher = {EDP Sciences, IOP Publishing and Società Italiana di Fisica},
    volume = {144},
    number = {1},
    pages = {10001},
    author = {Succi, Sauro and Itani, W. and Sreenivasan, K. and Steijl, R.},
    title = {Quantum computing for fluids: Where do we stand?},
    journal = {Europhysics Letters}
}

@article{Brunton22,
    author = {Brunton, Steven L. and Budi\v{s}i\'{c}, Marko and Kaiser, Eurika and Kutz, J. Nathan},
    title = {Modern {K}oopman Theory for Dynamical Systems},
    journal = {SIAM Review},
    volume = {64},
    number = {2},
    pages = {229-340},
    year = {2022},
    doi = {10.1137/21M1401243},
    URL = {https://doi.org/10.1137/21M1401243},
    eprint = {https://doi.org/10.1137/21M1401243},
}

@article{Schmid10, 
    title={Dynamic mode decomposition of numerical and experimental data}, 
    volume={656}, 
    DOI={10.1017/S0022112010001217}, 
    journal={Journal of Fluid Mechanics}, 
    author={Schmid, Peter J.}, 
    year={2010}, 
    pages={5–28}
}

@book{Mauroy2020koopman,
  title={{K}oopman operator in systems and control},
  author={Mauroy, Alexandre and Susuki, Y and Mezic, Igor},
  volume={7},
  year={2020},
  publisher={Springer}
}

@Article{Dongwei24,
    AUTHOR = {Shi, Dongwei and Yang, Xiu},
    TITLE = {{K}oopman Spectral Linearization vs. {C}arleman Linearization: A Computational Comparison Study},
    JOURNAL = {Mathematics},
    VOLUME = {12},
    YEAR = {2024},
    NUMBER = {14},
    ARTICLE-NUMBER = {2156},
    URL = {https://www.mdpi.com/2227-7390/12/14/2156},
    ISSN = {2227-7390},
    DOI = {10.3390/math12142156}
}

@Article{Mezic05,
    author={Mezi{\'{c}}, Igor},
    title={Spectral Properties of Dynamical Systems, Model Reduction and Decompositions},
    journal={Nonlinear Dynamics},
    year={2005},
    month={Aug},
    day={01},
    volume={41},
    number={1},
    pages={309-325},
    issn={1573-269X},
    doi={10.1007/s11071-005-2824-x},
    url={https://doi.org/10.1007/s11071-005-2824-x}
}

@article{Colbrook2024,
  title={Limits and powers of {K}oopman learning},
  author={Colbrook, Matthew J and Mezi{\'c}, Igor and Stepanenko, Alexei},
  journal={arXiv preprint arXiv:2407.06312},
  year={2024}
}

@article{Dudkovskaia25,
    doi = {10.1088/1361-6587/addb73},
    url = {https://dx.doi.org/10.1088/1361-6587/addb73},
    year = {2025},
    month = {jun},
    publisher = {IOP Publishing},
    volume = {67},
    number = {6},
    pages = {065033},
    author = {Dudkovskaia, A V and Li, D and Candy, J and Belli, E A and Yang, C},
    title = {Dynamic mode decomposition for gyrokinetic eigenmode analysis},
    journal = {Plasma Physics and Controlled Fusion},
}

@article{Kaptanoglu20,
    author = {Kaptanoglu, A. A. and Morgan, K. D. and Hansen, C. J. and Brunton, S. L.},
    title = {Characterizing magnetized plasmas with dynamic mode decomposition},
    journal = {Physics of Plasmas},
    volume = {27},
    number = {3},
    pages = {032108},
    year = {2020},
    month = {03},
    issn = {1070-664X},
    doi = {10.1063/1.5138932},
    url = {https://doi.org/10.1063/1.5138932},
    eprint = {https://pubs.aip.org/aip/pop/article-pdf/doi/10.1063/1.5138932/13806412/032108\_1\_online.pdf},
}

@article{Sasaki19,
doi = {10.1088/1361-6587/ab471b},
url = {https://dx.doi.org/10.1088/1361-6587/ab471b},
year = {2019},
month = {oct},
publisher = {IOP Publishing},
volume = {61},
number = {11},
pages = {112001},
author = {Sasaki, M and Kawachi, Y and Dendy, R O and Arakawa, H and Kasuya, N and Kin, F and Yamasaki, K and Inagaki, S},
title = {Using dynamical mode decomposition to extract the limit cycle dynamics of modulated turbulence in a plasma simulation},
journal = {Plasma Physics and Controlled Fusion},
}

@article{Taylor18,
    author = {Taylor, Roy and Kutz, J. Nathan and Morgan, Kyle and Nelson, Brian A.},
    title = {Dynamic mode decomposition for plasma diagnostics and validation},
    journal = {Review of Scientific Instruments},
    volume = {89},
    number = {5},
    pages = {053501},
    year = {2018},
    month = {05},
    issn = {0034-6748},
    doi = {10.1063/1.5027419},
    url = {https://doi.org/10.1063/1.5027419},
    eprint = {https://pubs.aip.org/aip/rsi/article-pdf/doi/10.1063/1.5027419/15631032/053501\_1\_online.pdf},
}

@article{Nayak21,
    title = {Detection and prediction of equilibrium states in kinetic plasma simulations via mode tracking using reduced-order dynamic mode decomposition},
    journal = {Journal of Computational Physics},
    volume = {447},
    pages = {110671},
    year = {2021},
    issn = {0021-9991},
    doi = {https://doi.org/10.1016/j.jcp.2021.110671},
    url = {https://www.sciencedirect.com/science/article/pii/S0021999121005660},
    author = {Indranil Nayak and Mrinal Kumar and Fernando L. Teixeira},
}

@article{Kusaba22,
    doi = {10.35848/1347-4065/ac1c3c},
    url = {https://dx.doi.org/10.35848/1347-4065/ac1c3c},
    year = {2021},
    month = {dec},
    publisher = {IOP Publishing},
    volume = {61},
    number = {SA},
    pages = {SA1011},
    author = {Kusaba, Akira and Kuboyama, Tetsuji and Shin, Kilho and Sasaki, Makoto and Inagaki, Shigeru},
    title = {A new combination of Hankel and sparsity-promoting dynamic mode decompositions and its application to the prediction of plasma turbulence},
    journal = {Japanese Journal of Applied Physics},
}

@article{LoVerso25,
    title = {Enhancing computational efficiency in nuclear fusion through reduced order modelling: Applications in magnetohydrodynamics},
    journal = {Fusion Engineering and Design},
    volume = {216},
    pages = {115080},
    year = {2025},
    issn = {0920-3796},
    doi = {https://doi.org/10.1016/j.fusengdes.2025.115080},
    url = {https://www.sciencedirect.com/science/article/pii/S0920379625002777},
    author = {Matteo {Lo Verso} and Stefano Riva and Carolina Introini and Eric Cervi and Luciana Barucca and Marco Caramello and Matteo {Di Prinzio} and Francesca Giacobbo and Laura Savoldi and Antonio Cammi},
}

@article{Indranil24,
  title = {Accelerating particle-in-cell kinetic plasma simulations via reduced-order modeling of space-charge dynamics using dynamic mode decomposition},
  author = {Nayak, Indranil and Teixeira, Fernando L. and Na, Dong-Yeop and Kumar, Mrinal and Omelchenko, Yuri A.},
  journal = {Phys. Rev. E},
  volume = {109},
  issue = {6},
  pages = {065307},
  numpages = {15},
  year = {2024},
  month = {Jun},
  publisher = {American Physical Society},
  doi = {10.1103/PhysRevE.109.065307},
  url = {https://link.aps.org/doi/10.1103/PhysRevE.109.065307}
}

@article{Pascuale22,
    author = {De Pascuale, Sebastian and Green, David L. and Lore, Jeremy D.},
    title = {Data-driven linear time advance operators for the acceleration of plasma physics simulation},
    journal = {Physics of Plasmas},
    volume = {29},
    number = {11},
    pages = {113903},
    year = {2022},
    month = {11},
    issn = {1070-664X},
    doi = {10.1063/5.0110393},
    url = {https://doi.org/10.1063/5.0110393},
    eprint = {https://pubs.aip.org/aip/pop/article-pdf/doi/10.1063/5.0110393/16778809/113903\_1\_online.pdf},
}

@article{Brunton16,
  title={Discovering governing equations from data by sparse identification of nonlinear dynamical systems},
  author={Brunton, Steven L and Proctor, Joshua L and Kutz, J Nathan},
  journal={Proceedings of the national academy of sciences},
  volume={113},
  number={15},
  pages={3932--3937},
  year={2016},
  publisher={National Academy of Sciences},
  doi={10.1073/pnas.1517384113},
  url={https://doi.org/10.1073/pnas.1517384113}
}

@article{Brunton17,
   title={Chaos as an intermittently forced linear system},
   volume={8},
   ISSN={2041-1723},
   url={http://dx.doi.org/10.1038/s41467-017-00030-8},
   DOI={10.1038/s41467-017-00030-8},
   number={1},
   journal={Nature Communications},
   publisher={Springer Science and Business Media LLC},
   author={Brunton, Steven L. and Brunton, Bingni W. and Proctor, Joshua L. and Kaiser, Eurika and Kutz, J. Nathan},
   year={2017},
   month=may 
}

@article{DiGrazia24,
    title = {A data-driven Vertical Stabilization system for the ITER tokamak based on Dynamic Mode Decomposition},
    journal = {Journal of the Franklin Institute},
    volume = {361},
    number = {2},
    pages = {816-833},
    year = {2024},
    issn = {0016-0032},
    doi = {https://doi.org/10.1016/j.jfranklin.2023.12.027},
    url = {https://www.sciencedirect.com/science/article/pii/S0016003223007974},
    author = {Luigi Emanuel {di Grazia} and Massimiliano Mattei and Adriano Mele and Alfredo Pironti},
}

@article{Faraji24Part1,
    doi = {10.1088/1361-6463/ad0910},
    url = {https://dx.doi.org/10.1088/1361-6463/ad0910},
    year = {2023},
    month = {nov},
    publisher = {IOP Publishing},
    volume = {57},
    number = {6},
    pages = {065201},
    author = {Faraji, F and Reza, M and Knoll, A and Kutz, J N},
    title = {Dynamic mode decomposition for data-driven analysis and reduced-order modeling of E × B plasmas: I. Extraction of spatiotemporally coherent patterns},
    journal = {Journal of Physics D: Applied Physics},
}

@article{Faraji24Part2,
doi = {10.1088/1361-6463/ad0911},
url = {https://dx.doi.org/10.1088/1361-6463/ad0911},
year = {2023},
month = {nov},
publisher = {IOP Publishing},
volume = {57},
number = {6},
pages = {065202},
author = {Faraji, F and Reza, M and Knoll, A and Kutz, J N},
title = {Dynamic mode decomposition for data-driven analysis and reduced-order modeling of E × B plasmas: II. Dynamics forecasting},
journal = {Journal of Physics D: Applied Physics},
}

@article{Budisic12,
    author = {{Budi\v{s}i\'{c}}, Marko and Mohr, Ryan and {Mezi\'{c}}, Igor},
    title = {Applied {K}oopmanism},
    journal = {Chaos: An Interdisciplinary Journal of Nonlinear Science},
    volume = {22},
    number = {4},
    pages = {047510},
    year = {2012},
    month = {12},
    issn = {1054-1500},
    doi = {10.1063/1.4772195},
    url = {https://doi.org/10.1063/1.4772195},
    eprint = {https://pubs.aip.org/aip/cha/article-pdf/doi/10.1063/1.4772195/13471578/047510\_1\_online.pdf},
}

@misc{Breunung24,
      title={Learning Global Linear Representations of Nonlinear Dynamics}, 
      author={Thomas Breunung and Florian Kogelbauer},
      year={2024},
      eprint={2408.03437},
      archivePrefix={arXiv},
      primaryClass={math.DS},
      url={https://arxiv.org/abs/2408.03437}, 
}

@misc{Rozwood24,
      title={{K}oopman-Assisted Reinforcement Learning}, 
      author={Preston Rozwood and Edward Mehrez and Ludger Paehler and Wen Sun and Steven L. Brunton},
      year={2024},
      eprint={2403.02290},
      archivePrefix={arXiv},
      primaryClass={cs.AI},
      url={https://arxiv.org/abs/2403.02290}, 
}

@misc{Shi25koopman,
      title={{K}oopman Operators in Robot Learning}, 
      author={Lu Shi and Masih Haseli and Giorgos Mamakoukas and Daniel Bruder and Ian Abraham and Todd Murphey and Jorge Cortes and Konstantinos Karydis},
      year={2025},
      eprint={2408.04200},
      archivePrefix={arXiv},
      primaryClass={cs.RO},
      url={https://arxiv.org/abs/2408.04200}, 
}

@article{Korda18,
    title = {Linear predictors for nonlinear dynamical systems: {K}oopman operator meets model predictive control},
    journal = {Automatica},
    volume = {93},
    pages = {149-160},
    year = {2018},
    issn = {0005-1098},
    doi = {https://doi.org/10.1016/j.automatica.2018.03.046},
    url = {https://www.sciencedirect.com/science/article/pii/S000510981830133X},
    author = {Milan Korda and Igor Mezić},
}

@inproceedings{Abraham17, series={RSS2017},
   title={Model-Based Control Using {K}oopman Operators},
   url={http://dx.doi.org/10.15607/RSS.2017.XIII.052},
   DOI={10.15607/rss.2017.xiii.052},
   booktitle={Robotics: Science and Systems XIII},
   publisher={Robotics: Science and Systems Foundation},
   author={Abraham, Ian and de la Torre, Gerardo and Murphey, Todd},
   year={2017},
   month=jul, 
   collection={RSS2017} 
}

@article{Rauh09,
    title = {{C}arleman Linearization for Control and for State and Disturbance Estimation of Nonlinear Dynamical Processes},
    journal = {IFAC Proceedings Volumes},
    volume = {42},
    number = {13},
    pages = {455-460},
    year = {2009},
    note = {14th IFAC Conference on Methods and Models in Automation and Robotics},
    issn = {1474-6670},
    doi = {https://doi.org/10.3182/20090819-3-PL-3002.00079},
    url = {https://www.sciencedirect.com/science/article/pii/S1474667015304821},
    author = {Andreas Rauh and Johanna Minisini and Harald Aschemann},
}

@article{Otto21,
   author = "Otto, Samuel E. and Rowley, Clarence W.",
   title = "{K}oopman Operators for Estimation and Control of Dynamical Systems", 
   journal= "Annual Review of Control, Robotics, and Autonomous Systems",
   year = "2021",
   volume = "4",
   number = "Volume 4, 2021",
   pages = "59-87",
   doi = "https://doi.org/10.1146/annurev-control-071020-010108",
   url = "https://www.annualreviews.org/content/journals/10.1146/annurev-control-071020-010108",
   publisher = "Annual Reviews",
   issn = "2573-5144",
   type = "Journal Article",
  }

@article{Taylor21,
    title = {Active learning in robotics: A review of control principles},
    journal = {Mechatronics},
    volume = {77},
    pages = {102576},
    year = {2021},
    issn = {0957-4158},
    doi = {https://doi.org/10.1016/j.mechatronics.2021.102576},
    url = {https://www.sciencedirect.com/science/article/pii/S0957415821000659},
    author = {Annalisa T. Taylor and Thomas A. Berrueta and Todd D. Murphey},
}

@article{Bevanda21,
   title={{K}oopman operator dynamical models: Learning, analysis and control},
   volume={52},
   ISSN={1367-5788},
   url={http://dx.doi.org/10.1016/j.arcontrol.2021.09.002},
   DOI={10.1016/j.arcontrol.2021.09.002},
   journal={Annual Reviews in Control},
   publisher={Elsevier BV},
   author={Bevanda, Petar and Sosnowski, Stefan and Hirche, Sandra},
   year={2021},
   pages={197–212} 
}

@ARTICLE{Carrier24,
  author={Carrier, Matthew J. and Farmer, William A. and Srinivasan, Bhuvana},
  journal={IEEE Transactions on Plasma Science}, 
  title={Deep {K}oopman Neural Network for Analyzing High-Energy-Density Simulations of Electrical Wire Explosions}, 
  year={2024},
  volume={52},
  number={10},
  pages={4916-4932},
  doi={10.1109/TPS.2024.3440255}
}

@article{Brunton20,
   author = "Brunton, Steven L. and Noack, Bernd R. and Koumoutsakos, Petros",
   title = "Machine Learning for Fluid Mechanics", 
   journal= "Annual Review of Fluid Mechanics",
   year = "2020",
   volume = "52",
   number = "Volume 52, 2020",
   pages = "477-508",
   doi = "https://doi.org/10.1146/annurev-fluid-010719-060214",
   url = "https://www.annualreviews.org/content/journals/10.1146/annurev-fluid-010719-060214",
   publisher = "Annual Reviews",
   issn = "1545-4479",
   type = "Journal Article",
}

@article{Ji23,
    title = {Signal propagation in complex networks},
    journal = {Physics Reports},
    volume = {1017},
    pages = {1-96},
    year = {2023},
    note = {Signal propagation in complex networks},
    issn = {0370-1573},
    doi = {https://doi.org/10.1016/j.physrep.2023.03.005},
    url = {https://www.sciencedirect.com/science/article/pii/S0370157323001321},
    author = {Peng Ji and Jiachen Ye and Yu Mu and Wei Lin and Yang Tian and Chittaranjan Hens and Matjaž Perc and Yang Tang and Jie Sun and Jürgen Kurths},
}

@INPROCEEDINGS{Han20,
  author={Han, Yiqiang and Hao, Wenjian and Vaidya, Umesh},
  booktitle={2020 59th IEEE Conference on Decision and Control (CDC)}, 
  title={Deep Learning of {K}oopman Representation for Control}, 
  year={2020},
  volume={},
  number={},
  pages={1890-1895},
  doi={10.1109/CDC42340.2020.9304238}
}

@misc{Azencot20,
  title={Forecasting Sequential Data using Consistent {K}oopman Autoencoders}, 
  author={Omri Azencot and N. Benjamin Erichson and Vanessa Lin and Michael W. Mahoney},
  year={2020},
  eprint={2003.02236},
  archivePrefix={arXiv},
  primaryClass={physics.comp-ph},
  url={https://arxiv.org/abs/2003.02236}, 
}

@article{Fukami20,
    author = {Fukami, Kai and Nakamura, Taichi and Fukagata, Koji},
    title = {Convolutional neural network based hierarchical autoencoder for nonlinear mode decomposition of fluid field data},
    journal = {Physics of Fluids},
    volume = {32},
    number = {9},
    pages = {095110},
    year = {2020},
    month = {09},
    issn = {1070-6631},
    doi = {10.1063/5.0020721},
    url = {https://doi.org/10.1063/5.0020721},
    eprint = {https://pubs.aip.org/aip/pof/article-pdf/doi/10.1063/5.0020721/14710996/095110\_1\_online.pdf},
}

@misc{Arathoon23,
  title={{K}oopman Embedding and Super-Linearization Counterexamples with Isolated Equilibria}, 
  author={Philip Arathoon and Matthew D. Kvalheim},
  year={2023},
  eprint={2306.15126},
  archivePrefix={arXiv},
  primaryClass={math.DS},
  url={https://arxiv.org/abs/2306.15126}, 
}

@misc{Kvalheim24flows,
      title={Linearizability of flows by embeddings}, 
      author={Matthew D. Kvalheim and Philip Arathoon},
      year={2024},
      eprint={2305.18288},
      archivePrefix={arXiv},
      primaryClass={math.DS},
      url={https://arxiv.org/abs/2305.18288}, 
}

@article{Kvalheim21,
    title = {Existence and uniqueness of global {K}oopman eigenfunctions for stable fixed points and periodic orbits},
    journal = {Physica D: Nonlinear Phenomena},
    volume = {425},
    pages = {132959},
    year = {2021},
    issn = {0167-2789},
    doi = {https://doi.org/10.1016/j.physd.2021.132959},
    url = {https://www.sciencedirect.com/science/article/pii/S0167278921001160},
    author = {Matthew D. Kvalheim and Shai Revzen},
}

@article{Kvalheim25,
    title = {Global linearization of asymptotically stable systems without hyperbolicity},
    journal = {Systems \& Control Letters},
    volume = {203},
    pages = {106163},
    year = {2025},
    issn = {0167-6911},
    doi = {https://doi.org/10.1016/j.sysconle.2025.106163},
    url = {https://www.sciencedirect.com/science/article/pii/S0167691125001458},
    author = {Matthew D. Kvalheim and Eduardo D. Sontag},
}

@misc{Korda20,
      title={Optimal construction of {K}oopman eigenfunctions for prediction and control}, 
      author={Milan Korda and Igor Mezi\'{c}},
      year={2020},
      eprint={1810.08733},
      archivePrefix={arXiv},
      primaryClass={math.OC},
      url={https://arxiv.org/abs/1810.08733}, 
}

@article{Liu25Koopman,
    title = {Properties of immersions for systems with multiple limit sets with implications to learning {K}oopman embeddings},
    journal = {Automatica},
    volume = {176},
    pages = {112226},
    year = {2025},
    issn = {0005-1098},
    doi = {https://doi.org/10.1016/j.automatica.2025.112226},
    url = {https://www.sciencedirect.com/science/article/pii/S0005109825001189},
    author = {Zexiang Liu and Necmiye Ozay and Eduardo D. Sontag},
}

@article{Liu23Koopman,
    title = {On the Non-Existence of Immersions for Systems with Multiple Omega-Limit Sets},
    journal = {IFAC-PapersOnLine},
    volume = {56},
    number = {2},
    pages = {60-64},
    year = {2023},
    note = {22nd IFAC World Congress},
    issn = {2405-8963},
    doi = {https://doi.org/10.1016/j.ifacol.2023.10.1408},
    url = {https://www.sciencedirect.com/science/article/pii/S2405896323018165},
    author = {Zexiang Liu and Necmiye Ozay and Eduardo D. Sontag},
}

@misc{Novikau25Opt,
      title={An efficient explicit implementation of a near-optimal quantum algorithm for simulating linear dissipative differential equations}, 
      author={Ivan Novikau and Ilon Joseph},
      year={2025},
      eprint={2501.11146},
      archivePrefix={arXiv},
      primaryClass={quant-ph},
      url={https://arxiv.org/abs/2501.11146}, 
}

@article{Steeb83,
  title={Embedding of nonlinear finite dimensional systems in linear infinite dimensional systems and Bose operators},
  author={Steeb, WH},
  journal = {Hadronic J.},
  volume = {6},
  pages = {68-76},
  year={1983}
}

@article{Kowalski87,
    title = {Hilbert space description of classical dynamical systems {I}},
    journal = {Physica A: Statistical Mechanics and its Applications},
    volume = {145},
    number = {3},
    pages = {408-424},
    year = {1987},
    issn = {0378-4371},
    doi = {https://doi.org/10.1016/0378-4371(87)90003-3},
    url = {https://www.sciencedirect.com/science/article/pii/0378437187900033},
    author = {Krzysztof Kowalski},
}

@book{Cornea2003Lusternik-Schnirelmann,
  title={Lusternik-Schnirelmann category},
  author={Cornea, Octavian and Lupton, Gregory and Oprea, John and Tanr\'e, Daniel},
  series={Mathematical Surveys and Monographs},
  number={103},
  year={2003},
  publisher={American Mathematical Soc.}
}

@article{Ostrand1971covering,
  title={Covering dimension in general spaces},
  author={Ostrand, Phillip A},
  journal={General Topology and its applications},
  volume={1},
  number={3},
  pages={209--221},
  year={1971},
  publisher={Elsevier},
  doi={10.1016/0016-660X(71)90093-6},
  url={https://doi.org/10.1016/0016-660X(71)90093-6}
}

@misc{Jennings25,
    title={Quantum algorithms for general nonlinear dynamics based on the {C}arleman embedding}, 
    author={David Jennings and Kamil Korzekwa and Matteo Lostaglio and Andrew T Sornborger and Yigit Subasi and Guoming Wang},
    year={2025},
    eprint={2509.07155},
    archivePrefix={arXiv},
    primaryClass={quant-ph},
    url={https://arxiv.org/abs/2509.07155}, 
}

@article{Endo2026,
  title={Divergence-free algorithms for solving nonlinear differential equations on quantum computers},
  author={Endo, Katsuhiro and Takahashi, Kazuaki Z},
  journal={Physical Review Research},
  volume={8},
  number={1},
  pages={013057},
  year={2026},
  publisher={APS}
}

@article{Wang2026quantum,
  title={Quantum Algorithms for Nonlinear Differential Equations via Pivot-Shifted {C}arleman Linearization},
  author={Wang, Ke and Jia, Zikang and Veerapaneni, Shravan and Ding, Zhiyan},
  journal={arXiv preprint arXiv:2605.20071},
  year={2026}
}

@book{Arnold1992odes,
  title={Ordinary differential equations},
  author={Arnold, Vladimir I},
  year={1992},
  publisher={Springer Science \& Business Media}
}

@book{Lichtenberg2013book,
  title={Regular and chaotic dynamics},
  author={Lichtenberg, Allan J and Lieberman, Michael A},
  year={2013},
  publisher={Springer Science \& Business Media}
}

@book{Olver1993book,
  title={Applications of Lie groups to differential equations},
  author={Olver, Peter J},
  volume={107},
  year={1993},
  publisher={Springer Science \& Business Media}
}

@book{Shabat1992book,
  title={Introduction to complex analysis Part II: Functions of several variables},
  author={Shabat, Boris Vladimirovich},
  publisher={American Mathematical Society},
  year={1992}
}

@book{Bluman1989book,
  title={Symmetries and differential equations},
  author={Bluman, George W and Kumei, Sukeyuki and Bluman, GW},
  volume={81},
  year={1989},
  publisher={Springer}
}

@article{Brustle2025quantum,
  title={Quantum and classical algorithms for nonlinear unitary dynamics},
  author={Brustle, Noah and Wiebe, Nathan},
  journal={Quantum},
  volume={9},
  pages={1741},
  year={2025},
  publisher={Verein zur F{\"o}rderung des Open Access Publizierens in den Quantenwissenschaften}
}

@article{Ameri2026quantum,
  title={Quantum lower bounds for simulating fluid dynamics},
  author={Ameri, Abtin and Carolan, Joseph and Childs, Andrew M and Krovi, Hari},
  journal={arXiv preprint arXiv:2603.12161},
  year={2026}
}

@article{Carinena2016solvability,
doi = {10.1088/1751-8113/49/42/425202},
url = {https://doi.org/10.1088/1751-8113/49/42/425202},
year = {2016},
month = {sep},
publisher = {IOP Publishing},
volume = {49},
number = {42},
pages = {425202},
author = {Cariñena, J F and Falceto, F and Grabowski, J},
title = {Solvability of a Lie algebra of vector fields implies their integrability by quadratures},
journal = {Journal of Physics A: Mathematical and Theoretical}
}

@article{Grabowska2020solvable,
  title={Solvable Lie algebras of vector fields and a Lie's conjecture},
  author={Grabowska, Katarzyna and Grabowski, Janusz and others},
  journal={SIGMA. Symmetry, Integrability and Geometry: Methods and Applications},
  volume={16},
  pages={065},
  year={2020},
  publisher={SIGMA. Symmetry, Integrability and Geometry: Methods and Applications}
}

@article{Kozlov2024solvable,
  title={Solvable algebras and integrable systems},
  author={Kozlov, Valery V},
  journal={Regular and Chaotic Dynamics},
  volume={29},
  number={5},
  pages={717--727},
  year={2024},
  publisher={Springer}
}

@book{Hille1997ordinary,
  title={Ordinary differential equations in the complex domain},
  author={Hille, Einar},
  year={1997},
  publisher={Courier Corporation}
}

@misc{NIST:DLMF:22.19,
         key = "{{DLMF}}",
       title = "{NIST Digital Library of Mathematical Functions, Ch. 22 Jacobian Elliptic Functions, Sec. 19 Applications}",
howpublished = {\url{https://dlmf.nist.gov/}, Release 1.2.8 of 2026-09-15},
         url = {https://dlmf.nist.gov/22.19},
        note = {F.~W.~J. Olver, A.~B. {Olde Daalhuis}, D.~W. Lozier, B.~I. Schneider,
                R.~F. Boisvert, C.~W. Clark, B.~R. Miller, B.~V. Saunders,
                H.~S. Cohl, and M.~A. McClain, eds.}
}

\end{document}